\documentclass[useAMS,usenatbib,fleqn]{mnras}

\usepackage{times}
\usepackage{graphicx}
\usepackage{subfigure}
\usepackage{amsmath}
\usepackage{amssymb}
\usepackage{comment}
\newcommand{\rd}{{\rm d}}
\newcommand{\gsim}{\mathrel{\hbox{\rlap{\lower.55ex \hbox {$\sim$}}
                   \kern-.3em \raise.4ex \hbox{$>$}}}}
\newcommand{\lsim}{\mathrel{\hbox{\rlap{\lower.55ex \hbox {$\sim$}}
                   \kern-.3em \raise.4ex \hbox{$<$}}}}

\newcommand{\St}{\mbox{St}}

\title[Modelling dust growth \& dynamics]{Modelling dust coagulation, dynamical drag and turbulent mixing during star and disc formation}
\author[M. R. Bate, M. A. Hutchison \& D. Elsender]{Matthew R. Bate$^{1}$\thanks{E-mail:
M.R.Bate@exeter.ac.uk (MRB)}, Mark A. Hutchison$^{2,3}$ \& Daniel Elsender$^{1}$\\
$^{1}$ Department of Physics and Astronomy, University of Exeter, Stocker
Road, Exeter EX4 4QL \\
$^{2}$ Universit$\bar{\mbox{a}}$ts-Sternwarte, Ludwig-Maximilians-Universit$\bar{\mbox{a}}$t M\"unchen, Scheinerstr. 1, 81679 M\"unchen, Germany \\
$^{3}$ Hochschule f\"ur angewandte Wissenschaften M\"unchen, Lothstra$\beta$e 34, 80335 M\"unchen, Germany 
}

\date{Submitted to MNRAS, revised after first referee's report}

\begin{document}
\maketitle
\begin{abstract}
Planet formation in the discs around young stars involves the coagulation of sub-micron sized dust grains into much larger grains that may be mixed by turbulence and migrate through the disc.  In this paper, we describe how we have combined a method for modelling the coagulation of a population of dust grains with the {\sc multigrain} algorithm for modelling the dynamical evolution of a population of dust grains that are subject to strong gas drag.  We solve the dynamical evolution of the dust grains due to gas drag using a recently-developed implicit integration method, and we introduce a new implicit method to model the diffusion of the dust due to unresolved hydrodynamic turbulence.  The resulting smoothed particle hydrodynamics (SPH) code allows us, for the first time, to model the growth, mixing and migration of dust grain populations during the early stages of star formation and the formation, growth and evolution of a young protoplanetary disc using three-dimensional hydrodynamical simulations. In doing so, we find that including turbulent dust diffusion within the disc provides a substantial enhancement of the rate of dust grain growth due to the fact that the turbulent diffusion provides a source of small and intermediate dust grains to the regions in which the largest dust grains are growing.
\end{abstract}

\section{Introduction}
\label{introduction}

Dust plays crucial roles in both star formation and planet formation. In present-day star formation, dust provides the dominant source of opacity and the cooling of dense molecular gas (hydrogen number densities $n_{\rm H} \gsim 10^4$~cm$^{-3}$) is dominated by the thermal emission of dust grains \citep[e.g.,][]{Goldsmith2001}. Dust grains also play an important role in setting the degree of ionisation in dense molecular clouds and protoplanetary discs by absorbing electrons and ions, with small grains being more effective \citep[e.g.,][]{Wardle2007, Bai2011, Guillet_etal2020, TsuOku2022, Lebreuilly_etal2023}.  Thus, variation of the dust grain population may alter the coupling between the magnetic field and the gas, thereby affecting the dynamics of molecular cloud collapse and the evolution of magnetised discs. For planet formation, dust coagulation into pebbles is believed to be necessary both to form planetesimals (e.g., via the streaming instability; \citealt{YouGoo2005, LiYou2021}, although note that a broad dust size distribution can hinder its effectiveness;  \citealt{Paardekooper_etal2021, McNally_etal2021})  and also to allow the efficient growth of planetary cores \citep[i.e., via pebble accretion;][]{JohLac2010, OrmKla2010, JohLam2017}.

Despite the importance of dust, until recently little attention has been paid to its evolution in studies of star formation.  To determine the dust opacity and/or degree of ionisation, star formation calculations usually assume a constant interstellar grain size distribution \citep*[e.g., a MRN size distribution with a maximum grain size, $s \sim 0.1~\mu$m;][]{MatRumNor1977}, with an abundance that is either appropriate for solar metallicity (Z$_\odot$), or that is scaled by a constant factor \citep[e.g.,][]{Myers_etal2011, Bate2014, Bate2019}.  This is reasonable for the early, low-density phases of star formation since, for example, \cite{Ormel_etal2009} found that large grains could not grow within molecular clouds unless they were both dense and long-lived.  However, as soon as protoplanetary discs are formed, their much higher densities allow rapid grain growth \citep[e.g.,][]{WeiCuz1993}.

Dust grain growth at the centre of a collapsing molecular cloud core has been studied by \cite{HirOmu2009}, for varying initial metallicities.  Beginning with typical MRN size distributions, they showed that there was little dust coagulation during the early phase of the collapse (densities $n_{\rm H} \lesssim 10^7$~cm$^{-3}$) in any of their calculations, but as the density continued to increase, small grains became depleted for metallicities above $\sim 10^{-4} ~{\rm Z}_\odot$.  Nevertheless, because of the relatively short timescales, both high densities ($n_{\rm H} \gtrsim 10^{11}$~cm$^{-3}$, found within the first hydrostatic core; \citealt{Larson1969}) and high metallicities were required for the peak grain size (in terms of dust mass) to increase beyond its initial value of 0.25~$\mu$m, and the peak remained at sub-micron sizes.

\cite{Bate2022} was the first to model the coagulation of a dust grain population in three-dimensional hydrodynamical simulations of star formation, by adapting the method of \cite{HirOmu2009} to follow the evolution of the grain population associated with each fluid element.  In this way, he was able to investigate both the temporal and spatial distributions of dust particle sizes during molecular cloud collapse and the formation and evolution of first hydrostatic cores \citep{Larson1969} or pre-stellar discs \citep*[rapidly-rotating first hydrostatic cores that have a disc-like morphology and may have radii of several tens of au;][]{Bate1998, Bate2011, SaiTom2006, SaiTomMat2008, MacInuMat2010, MacMat2011}.  In agreement with previous solar metallicity results \citep*{HirOmu2009, WonHirLi2016}, \citeauthor{Bate2022} found that the maximum grain size did not increase substantially within the collapsing envelope.  However, once the first hydrostatic core formed, he found that rapid dust growth to sizes in excess of 100 $\mu$m occurred within the core, even before stellar core formation. Due to the strong dependence of the grain growth rate on density, he found that progressively larger grains were produced at smaller distances from the centre of the first hydrostatic core. Moreover, in pre-stellar discs that rotate so quickly that they become bar unstable \citep{Durisenetal1986, Bate1998} and form spiral arms, larger grains were preferentially produced in the spiral arms due to the higher densities.  Thus, \cite{Bate2022} found that the grain size distribution can vary substantially in the very earliest stages of star formation.  Subsequent studies using other codes have obtained similar results \citep[e.g.,][who also included magnetic fields and the effects of grain growth on the level of ionisation]{Marchand_etal2023}.  The general results are also in agreement with earlier three-dimensional calculations of star formation that did not model the full grain size distribution, but modelled grain growth using a single-size approximation \citep*{TsuMacInu2021}.
More recently, \citep{NavarroAlmaida_etal2024} have shown that grain growth within the first hydrostatic core can affect its chemical evolution.

A fundamental limitation of the method of \cite{Bate2022} was that the dust could not move relative to the gas.  The dust population associated with each fluid element was evolved independently and dust could not move between fluid elements.  This approximation was valid as far as the calculations were evolved (up until stellar core formation) due to the fact that the dust was well coupled to the gas --- the low-density regions never produced large grains, and large grains were only produced at high (gas) densities. However, to follow the evolution of the dust on longer timescales, the dynamical evolution of the dust needs to be taken into account.

In this paper, we present a new combination of methods that allow a dust grain population to be evolved due to coagulation, dynamical dust-gas drag and turbulent mixing, in three-dimensional hydrodynamical simulations that follow the collapse of a molecular cloud core to form a protostar and the formation, growth and dynamical evolution of its disc.  Such calculations are timely because they will allow the wealth of millimetre (e.g., ALMA), scattered-light (e.g. VLT-SPHERE, GPI), and recent infrared (2-21 $\mu$m: JWST) observations of the dust component of young discs that have been, and continue to be, collected to be compared with simulations.  For example, observations of edge-on discs allow the processes of dust settling and turbulent stirring within protoplanetary discs to be constrained \citep[e.g.,][]{Villenave_etal2020, Duchene_etal2024}.

This paper is arranged as follows.  In Section \ref{sec:method} we describe each of the individual methods for modelling the dynamics and size evolution of the dust, and how these methods are combined together and implemented the {\sc sphNG} code.  In Section \ref{sec:tests}, we present a number of tests of the code related to dust growth, settling and turbulent stirring within a protoplanetary disc.  In Section \ref{sec:results}, we present results from several star formation calculations that include dust coagulation, turbulent mixing and settling/migration, demonstrating the potential of the new code for studying dust evolution during star and disc formation. Finally, in Section \ref{sec:conclusions} we present our conclusions.

\section{Method}
\label{sec:method}

This paper brings together several existing methods, and introduces a new method, to model various aspects of dust evolution in discs around young stars. We combine the dust coagulation of \citep{Bate2022}, the {\sc multigrain} algorithm for modelling the dynamics of a population of dust grains subject to strong gas drag \citep*{HutPriLai2018}, and the implicit integration method developed by \cite{ElsBat2024} to solve for the evolution of the dust subject to gas drag.  In addition, we present a new method for modelling turbulent diffusion of dust.  We therefore begin with overviews of each of the existing methods, followed by descriptions of how they are combined. We then describe the new turbulent diffusion method, and finish with a brief summary of the main elements of the smoothed particle hydrodynamics (SPH) code,  {\sc sphNG}, in which the methods are implemented.

\subsection{The grain growth model}
\label{sec:coagulation}

The growth of a local distribution of dust grains can be described by the deterministic
mean-field Smoluchowski equation assuming binary collisions \citep{Smoluchowski1916}.
\cite{Bate2022} described a method to model the coagulation of a 
discretised population of dust grains within an SPH code.  
The grain growth method closely followed the method developed by \cite{HirYan2009}, which was used by  \cite{HirOmu2009} to study dust coagulation at the centre of collapsing molecular clouds.  
The original method modelled grain growth in a single zone using $N$ discrete bins to represent a range of grain sizes.  \cite{Bate2022} implemented an almost identical method, but with each individual SPH gas particle acting as a separate `zone' containing a population of dust grains.  Thus, the grain growth in the fluid element represented by each SPH particle (or `zone') occurred independently of neighbouring SPH particles.  This approximation is good as long as the physical flux of grains between `zones' is negligible (or neighbouring zones have similar populations of grains).  This assumption limited the calculations of \cite{Bate2022} to the early stages of star formation, covering only the evolution of the first hydrostatic core phase \citep{Larson1969} up to the formation of the stellar core.

We note that this discretised method uses a pair-wise or point-based approximation
for modelling the dust coagulation.
Alternative methods for modelling grain growth can be based around finite element or 
finite volume methods.  A recent example is the finite-volume coagulation scheme
of \cite{LomLai2021}, which can achieve comparable accuracy with fewer dust species
compared to some other schemes -- an appealing feature for computationally demanding 3D simulations. However, this method represents the dust size distribution differently from hydrodynamic solvers: the finite-volume scheme uses a high-order reconstruction of each mass or size bin, whereas hydrodynamic models typically evolve only bin-averaged quantities. Converting between these representations at every time step is non-trivial.

Here we give a brief summary of the main elements of the grain growth method that we
use for this paper.  Further details may be found in \cite{Bate2022}, and in the original papers by \cite{HirYan2009} and \cite{HirOmu2009}.  In addition, in Appendix \ref{appendixA} 
we compare the performance of our chosen grain growth method with analytic solutions to
the Smoluchowski equation.  We note that the method of \cite{HirYan2009} can be used to study 
both coagulation and fragmentation of dust grains.  In this paper we have chosen only to 
consider coagulation, and we have chosen a coagulation threshold (see below) that tends not 
to produce very large grains in our simulations of protoplanetary disc formation 
(Section \ref{sec:results}). Such growth tends to produce monodisperse grain populations
in the long term.  The use of different coagulation thresholds that result in the production of larger
grains that are less well-coupled to the gas may also mean that grain fragmentation becomes
important.  Thus, care should be taken when attempting to draw general conclusions from the
results presented in this paper.

For simplicity, we assume spherical grains with a material density $\rho_{\rm gr}$ that is constant, such that the mass of a dust grain of radius, $a$, is given by
\begin{equation}
m(a)= \frac{4 \pi}{3} a^3 \rho_{\rm gr}.
\label{eq:mass_grain}
\end{equation}
We consider a single grain material density of $\rho_{\rm gr}=3.0$~g~cm$^{-3}$.

For each SPH particle, we create $N$ discrete bins that are logarithmically spaced in grain radius from $a_{\rm min}$ to $a_{\rm max}$, so that radii of grains in the $i$th bin lies between $a^{\rm (b)}_{i-1}$ and $a^{\rm (b)}_i$ where $a^{\rm (b)}_i = \delta a^{\rm (b)}_{i-1}$ and $\log(\delta) = \log(a_{\max}/a_{\min})/N$ (the super-script (b) denotes the grain radius boundaries of the bin).
The number density of grains with radii between $a$ and $a+\rd a$ is given by $n(a)~\rd a$.  The method of \cite{HirYan2009} considers the distribution function of grain mass rather than grain size to ensure conservation of the total grain mass.  Thus, they denote the number density of grains with masses between $m$ and $m+{\rm d}m$ as $\widetilde{n}(m)~{\rm d}m$, and for the two distributions $n(a)~\rd a = \widetilde{n}(m)~\rd m$.  The quantities $n(a)~\rd a$ or $\widetilde{n}(m)~\rd m$ give the number density of grains in some size or mass interval, respectively.

\cite{HirYan2009} defined the grain radius and the mass in the $i$th bin as $a_i \equiv (a^{(b)}_{i-1} + a^{(b)}_{i})/2$ and $m_i \equiv m(a_i)$, respectively, and the upper boundary of the mass bin is defined as $m^{\rm (b)}_i \equiv m(a^{\rm (b)}_i)$.  \cite{Bate2022} also tried defining the grain radius using a geometric mean rather than an arithmetic mean and found no significant differences in the results for numbers of bins of 10 per decade in radius or greater.  It is interesting to note, however, that if the number of bins is small enough, the algorithm completely fails to grow the grain population and this limit depends on which form of averaging is used.  This is because for dust grains to populate bins with larger sizes the sum of the masses must lie within the mass range of the higher-mass bin.  When using the geometric mean to define the average grain size in the $i$th bin, this results in a condition that $2 (a_i)^3 > (a_i \sqrt{\delta})^3$ in order for the sum of the masses of two grains in the $i$th bin to exceed the lower mass boundary of the $(i+1)$th bin.  This requires $\delta<2^{2/3} \approx 1.587$ (or at least 5 bins per decade in radius).  When using the arithmetic mean, this condition becomes $2[(a^{\rm (b)}_i/\delta +  a^{\rm (b)}_i)/2]^3 > (a^{\rm (b)}_i)^3$, which reduces to $\delta<1/(2^{2/3}-1) \approx 1.702$ (or at least $\approx 4.3$ bins per decade in radius).  So using the arithmetic mean, somewhat fewer bins can be used before the algorithm fails completely.  In practice, we always use at least 10 bins per decade in radius for science calculations, and the testing performed by \cite{Bate2022} demonstrated that at these dust resolutions the type of mean does not significantly affect the grain growth (as long as the above limits on $\delta$ are satisfied).  We note, however, that because \cite{HutPriLai2018} defined the grain size associated with their bins using the geometric mean, when we use their 10-bin dust settling test in Section \ref{sec:tests} we also use the geometric mean.  For all our other calculations we use the arithmetic mean.

The mass density of grains in the $i$th bin, $\widetilde{\rho}_i$, is defined as
\begin{equation}
\widetilde{\rho}_i \equiv m_i \widetilde{n}(m_i) \left[ m^{(b)}_i - m^{(b)}_{i-1} \right],
\end{equation} 
and the time evolution of $\widetilde{\rho}_i$ can be expressed as
\begin{equation}
\frac{ \rd \widetilde{\rho}_i }{\rd t} = - m_i \widetilde{\rho}_i \sum^N_{k=1} \alpha_{ki} \widetilde{\rho}_k + \frac{1}{2} \sum^N_{j=1} \sum^N_{k=1} \alpha_{kj} \widetilde{\rho}_k \widetilde{\rho}_j m^{kj}_{\rm coag}(i),
 \label{eq:coag}
\end{equation}
where we only consider coagulation.  Shattering can also be included \citep[see][]{HirYan2009}, but for the present we are only concerned with early grain growth during which the relatively small grains are reasonably well coupled to the gas and high relative speeds do not occur.  In the above equation
\begin{equation}
 \alpha_{kj} = \left\{  \begin{array}{ll} 
\displaystyle \frac{\beta \sigma_{kj} v_{kj}}{m_j m_k}  & {\rm if}~~ v_{kj} < v^{kj}_{\rm coag}, \\ 
\\
 0 & {\rm otherwise,} 
 \end{array} \right.
 \label{eq:alpha}
\end{equation}
and \cite{Bate2022} use $m^{kj}_{\rm coag}(i) = m_k+m_j$ if the mass of the coagulated particle $m_k + m_j$ is within the mass range of the $i$th bin, otherwise $m^{kj}_{\rm coag}(i) = 0$.  Coagulation only occurs if the relative velocity, $v_{kj}=| \mbox{\boldmath{$v$}}_k - \mbox{\boldmath{$v$}}_j|$, is less than the coagulation threshold velocity $v^{kj}_{\rm coag}$ (equation \ref{eq:alpha}).  The coagulation cross section is $\sigma_{kj} = \pi (a_k+a_j)^2$ and $\beta$ is the sticking probability which we set to unity \citep[][show that the sticking probability is of order unity below the coagulation threshold velocity]{Blum2000}.  Typically, we use the coagulation threshold velocity \citep*{ChoTieHol1993,DomTie1997,YanLazDra2004,HirOmu2009}
\begin{equation}
v^{kj}_{\rm coag} = 21.4 \left[ \frac{a_k^3 + a_j^3}{(a_k+a_j)^3} \right]^{1/2} \frac{\gamma_{\rm SE}^{5/6}}{E^{1/3} R_{kj}^{5/6} \rho_{\rm gr}^{1/2}},
\label{eq:coagulation_barrier}
\end{equation}
where $\gamma_{\rm SE}$ is the surface energy per unit area, $R_{kj} \equiv a_k a_j/(a_k+a_j)$ is the reduced radius of the grains, and $E$ is related to Poisson's ratios ($\nu_k$ and $\nu_j$) and Young's modulus ($E_k$ and $E_j$) by $1/E \equiv (1-\nu_k)^2/E_{k} + (1-\nu_j)^2/E_{j}$.  For our calculations we use the same values for these parameters as \cite{HirYan2009}, appropriate for graphite grains: $\gamma_{\rm SE}=12$~erg~cm$^{-2}$, $E=3.4\times 10^{10}$~dyn~cm$^{-2}$.  For two identical particles, these parameters result in $v^{ii}_{\rm coag} \approx 228~ (a_i/\mbox{0.1 $\mu$m})^{-5/6}$~cm~s$^{-1}$.

As discussed by \cite{Bate2022}, in the SPH code, rather than evolving $\widetilde{\rho}_i$ directly, we set $\widetilde{\rho}_i = \varrho_i  \rho$, where $\rho$ is the total SPH density (gas and dust) of the particle, and we evolve $\varrho_i$ (i.e., the mass density of grains in the $i$th bin is evolved as a fraction of the total mass density).  This naturally allows for changes in the size of a Lagrangian `zone' (i.e., changes in the effective size of an SPH particle), for example during collapse of a cloud in which the density of grains increases simply due to the compression of the `zone'.  In  the {\sc multigrain} method described by \cite{HutPriLai2018} (see the next section), they use the symbol $\varepsilon_i \equiv \varrho_i$.  So from this point on, we use $\varepsilon_i$ to denote the fractional mass density of the grains in the $i$th bin rather than $\varrho_i$.  Then the total dust mass fraction of any SPH particle is $\varepsilon = \sum_i \varepsilon_i$.

A typical interstellar grain size distribution is often specified as a power-law distribution for sizes smaller than some maximum, cut-off size (e.g., $n(a) \propto a^{-3.5}$ for $a<a_{\rm cutoff}$).  Therefore, in some of our calculations we initially set $\varepsilon_i$ such that
\begin{equation}
\varepsilon_i = \frac{ \varepsilon }{\chi}  \displaystyle  \int_{a^{(b)}_{i-1}}^{a^{(b)}_{i}} a^3 n(a)~ {\rm d}a ,
\end{equation}
with the normalisation factor
\begin{equation}
\chi = \displaystyle \int_{a_{\rm min}}^{a_{\rm cutoff}} a^3 n(a)~ {\rm d}a,
\end{equation}
see \cite{Bate2022} for more details. 

\subsection{The {\sc multigrain} method}
\label{sec:multigrain}

The {\sc multigrain} method for evolving the dynamical evolution of a population of dust grains that are strongly coupled to gas within the SPH method was developed by \cite{HutPriLai2018}.  We refer the interested reader to that paper for a full derivation of the method.  Here we simply discuss the main points of the method and present the main equations.  

\cite{LaiPri2014c} derived the continuum fluid equations for a mixture of gas and $N$ coupled dust species moving in a barycentric frame of reference.  They went on to show that in regimes in which the dust species are strongly coupled to the gas (commonly referred to as the terminal velocity approximation, e.g., \citealt{YouGoo2005}; \citealt{Chiang2008}; \citealt{Barranco2009};  \citealt*{JacBalLat2011}), the fluid equations can be written as
\begin{equation}
\frac{\rd \rho}{\rd t} = - \rho ( \nabla \cdot \mbox{\boldmath$v$} ),
\label{eq:mass1}
\end{equation}
\begin{equation}
\frac{\rd \mbox{\boldmath$v$}}{\rd t} = (1- \varepsilon) \mbox{\boldmath$f$}_{\rm g} + \sum_j \varepsilon_j \mbox{\boldmath$f$}_{{\rm d}j} +  \mbox{\boldmath$f$},
\end{equation}
\begin{equation}
\frac{\rd \varepsilon_j}{\rd t} = - \frac{1}{\rho} \nabla \cdot \left[ \rho \varepsilon_j ( \Delta \mbox{\boldmath$v$}_j - \varepsilon \Delta \mbox{\boldmath$v$}) \right],
\end{equation}
\begin{equation}
\frac{\rd u}{\rd t} = - \frac{P}{\rho_{\rm g}} \nabla \cdot \mbox{\boldmath$v$}  + \varepsilon \Delta \mbox{\boldmath$v$} \cdot \nabla u,
\label{eq:energy1}
\end{equation}
\begin{equation}
\Delta \mbox{\boldmath$v$}_j = \left[  \Delta \mbox{\boldmath$f$}_j - \sum_k \varepsilon_k \Delta \mbox{\boldmath$f$}_{{\rm d}k}  \right] \varepsilon_j t_j,
\end{equation}
where $\rd / \rd t$ is the convective derivative using the barycentric velocity
\begin{equation}
\mbox{\boldmath$v$} \equiv \frac{ \rho_{\rm g} \mbox{\boldmath$v$}_{\rm g} +  \displaystyle \sum_j \rho_{{\rm d}j} \mbox{\boldmath$v$}_{{\rm d}j} }{\rho}  = \frac{ \rho_{\rm g} \mbox{\boldmath$v$}_{\rm g} +  \rho_{\rm d} \mbox{\boldmath$v$}_{\rm d}}{\rho},
\end{equation}
and the subscripts g and d denote gas and dust properties, respectively, such that 
\begin{equation}
\rho \equiv \rho_{\rm g}+\rho_{\rm d} = \rho_{\rm g}+ \sum_j \rho_{{\rm d}j},
\end{equation}
and $\varepsilon_j = \rho_{{\rm d}j}/\rho$.  This also means that in Section \ref{sec:coagulation}, $\widetilde{\rho}_j \equiv {\rho}_{{\rm d}j}$.
The quantity $\Delta \mbox{\boldmath$v$}$ is the weighted sum of the differential velocities $\Delta \mbox{\boldmath$v$}_j \equiv \mbox{\boldmath$v$}_{{\rm d}j} - \mbox{\boldmath$v$}_{\rm g}$, so that
\begin{equation}
\Delta \mbox{\boldmath$v$} \equiv \frac{1}{\varepsilon} \sum_j \varepsilon_j \Delta \mbox{\boldmath$v$}_j.
\end{equation}
The specific internal thermal energy of the gas is represented by $u$, and $P$ is the gas pressure.
Accelerations acting on both gas and dust components are represented by \mbox{\boldmath$f$}, while $\mbox{\boldmath$f$}_{\rm g}$ and $\mbox{\boldmath$f$}_{{\rm d}j}$ represent the accelerations acting on the gas and dust components, respectively.  Thus, $\Delta \mbox{\boldmath$f$}_j \equiv \mbox{\boldmath$f$}_{{\rm d}j} - \mbox{\boldmath$f$}_{\rm g}$ is the differential acceleration between the gas and each dust component.

For the case of hydrodynamics and gravity, the only acceleration acting on the gas and not the dust is
\begin{equation}
\mbox{\boldmath$f$}_{\rm g} = - \frac{\nabla P}{\rho_{\rm g}}
\end{equation}
and $\mbox{\boldmath$f$}_{{\rm d}j} = 0$, so that
\begin{equation}
\Delta \mbox{\boldmath$f$}_{j} = \frac{\nabla P}{\rho_{\rm g}},
\label{eq:dustv}
\end{equation}
and 
\begin{equation}
\Delta \mbox{\boldmath$v$}_{j} = \frac{\varepsilon_j t_j \nabla P}{\rho} = \frac{T_{{\rm s}j} \nabla P}{\rho_{\rm g}},
\end{equation}
where \cite{HutPriLai2018} define (there is an erroneous factor of two in their equations 14 and 15) the quantities
\begin{equation}
T_{{\rm s}j}  = \varepsilon_j (1-\varepsilon)t_j,
\end{equation}
and
\begin{equation}
\widetilde{T}_{{\rm s}j}  = \varepsilon_j t_j - \sum_k \varepsilon_k^2 t_k,
\end{equation}
where the $t_j = \rho/K_j$ is related to the stopping time and $K_j$ is the drag coefficient for each dust species. For the Epstein regime \citep[suitable for strongly-coupled dust grains with low Mach numbers][]{Epstein1924}
\begin{equation}
K_j = \frac{\rho_{\rm g} \rho_{\rm d}}{\rho_{\rm gr} a_j} c_{\rm s} \sqrt{\frac{8}{\pi \gamma}},
\label{eq:Epstein}
\end{equation}
where $\gamma$ is the adiabatic constant of the gas and $c_{\rm s}$ is the sound speed. 
For a single dust species,  the stopping time $T_{{\rm s}j}  = \widetilde{T}_{{\rm s}j} = \varepsilon_j(1-\varepsilon)t_j$ with $j=1$.

In this case, equations \ref{eq:mass1} to \ref{eq:energy1} can be expressed as
\begin{equation}
\frac{\rd \rho}{\rd t} = - \rho ( \nabla \cdot \mbox{\boldmath$v$} ),
\label{eq:mass2}
\end{equation}
\begin{equation}
\frac{\rd \mbox{\boldmath$v$}}{\rd t} = - \frac{\nabla P}{\rho} +  \mbox{\boldmath$f$},
\label{eq:mom2}
\end{equation}
\begin{equation}
\frac{\rd \varepsilon_j}{\rd t} = - \frac{1}{\rho} \nabla \cdot \left( \varepsilon_j \widetilde{T}_{{\rm s}j}  \nabla P \right),
\label{eq:dedt2}
\end{equation}
\begin{equation}
\frac{\rd u}{\rd t} = - \frac{P}{\rho_{\rm g}} \nabla \cdot \mbox{\boldmath$v$} + \frac{\varepsilon T_{\rm s}}{\rho_{\rm g}} \nabla P \cdot \nabla u,, 
\label{eq:energy2}
\end{equation}
where
\begin{equation}
T_{\rm s} \equiv \frac{1}{\varepsilon} \sum_j \varepsilon_j T_{{\rm s}j},
\label{eq:Ts}
\end{equation}
acts like an effective stopping time for the mixture.
Equation \ref{eq:energy2} can also be written as
\begin{equation}
\frac{\rd \widetilde{u}}{\rd t} = - \frac{P}{\rho} \nabla \cdot \mbox{\boldmath$v$},
\label{eq:energy3}
\end{equation}
where $\widetilde{u} \equiv (1-\varepsilon)u$.
In the limit of a single dust species, equations \ref{eq:mass2} -- \ref{eq:energy2} reduce to equations 10 -- 13 in \cite{PriLai2015}.
The adiabatic equation of state is
\begin{equation}
P= (\gamma -1)(1-\varepsilon) \rho u = (\gamma-1)\rho \widetilde{u}.
\end{equation}

The derivation of the {\sc multigrain} method given by \cite{HutPriLai2018} includes sections including subtleties regarding the definition of drag time-scales when there are multiple dust components, explicit time-stepping considerations, and energy conservation.  We note that because we evolve the dust fractions using an implicit method (see Section \ref{sec:integration}), we do not need to concern ourselves with the explicit dust time-step constraints, although the {\sc sphNG} code includes both explicit and implicit methods for evolving the dust fractions and in some cases we compare the results of our implicit calculations with explicit versions.

\subsection{SPH forms of the dust-as-mixture equations}
\label{sec:SPHequations}

In the above barycentric dust-as-mixture formulation, the equations that express mass and momentum conservation (\ref{eq:mass2} and \ref{eq:mom2}) are identical to those in standard SPH (only the equation for $P$ changes due to the dust fraction), so the conservation of mass and momentum are retained.  Therefore, the standard {\sc sphNG} implementation is used, in which
\begin{equation}
\rho_a= \sum_b m_b W_{ab}(h_a),
\end{equation}
\begin{equation}
\begin{split}
\displaystyle  \frac{ \rd \mbox{\boldmath$v$}_a}{\rd t} = & \displaystyle - \sum_b m_b \left[ \frac{P_a}{\Omega_a \rho_a^2} \nabla_a W_{ab}(h_a) + \frac{P_b }{\Omega_b \rho_b^2} \nabla_a W_{ab}(h_b) \right] \\
 & \displaystyle  +  \mbox{\boldmath$f$}_a + \Pi_{\rm AV},
\end{split}
\end{equation}
where we have used the subscripts $a$,$b$ to denote SPH particles, $W_{ab}$ is the SPH kernel (for which we use the M$_6$ quintic spline kernel), $h$ is the smoothing length, and $\Omega$ is the term that accounts for the gradients of smoothing lengths \citep[see][]{PriMon2004b,PriMon2007}.  The  $\Pi_{\rm AV}$ term is the artificial viscosity; we use a form of the \cite{MorMon1997} artificial viscosity in which the parameter controlling the linear term, $\alpha^{\rm AV}_a \in [0.1, 1]$, but the parameter controlling the quadratic term is kept constant at $\beta^{\rm AV}=2$. See \cite{ElsBat2024} for the full details.

SPH dust-as-mixture methods do not usually evolve the dust fractions directly.  Rather, the dust fractions tend to be evolved using quantities that guarantee that the dust fraction remains positive.  Several different dust variables that have been used in the literature.  For a single dust species, \cite{PriLai2015} evolved the dust fractions using a variable, $s$, for which $\varepsilon \equiv s^2/\rho$.  \cite{Ballabio_etal2018} proposed using 
\begin{equation}
\varepsilon \equiv s^2/(1+ s^2), 
\label{eq:epsilon}
\end{equation}
which has the benefits that it enforces $\varepsilon \in [0,1]$ (for a single dust species) and it does not require the density to be evaluated in order to evaluate $s$ initially, or to obtain $\varepsilon$ from $s$.  \cite{Hutchison_etal2022} proposed $\varepsilon \equiv \sin^2(s)$, which also enforces $\varepsilon \in [0,1]$ and was found to produce more accurate solutions in their tests, but it is more computationally expensive due to the need to evaluate trigonometric functions.  Following \cite{ElsBat2024}, whose implicit method we used to evolve the dust fractions, we choose a formulation based on \cite{Ballabio_etal2018}.

\subsubsection{The dust equations for a single dust species}

Following \cite{Ballabio_etal2018} or \cite{ElsBat2024} for a {\it single dust species}, we have that
\begin{equation}
s = \sqrt{ \frac{\varepsilon}{1 - \varepsilon} },
\label{eq:s}
\end{equation}
so the time evolution of the dust variable can be written as
\begin{equation}
 \frac{\rd s}{\rd t}  =   \frac{1}{2 s (1 - \varepsilon)^2} \frac{\rd \varepsilon}{\rd t}.
\label{eq:dsdtdedt}
\end{equation}
After substituting in the single dust species equivalent of equation \ref{eq:dedt2} (in which $\varepsilon_j \rightarrow \varepsilon$ and $\widetilde{T}_{{\rm s}j}  \rightarrow t_{\rm s}$ for a single dust species), this can be manipulated into the form
\begin{equation}
 \frac{\rd s}{\rd t}  =  - \frac{1}{2 \rho (1 - \varepsilon)^2} \left[ \nabla \cdot \left[ s (1 - \varepsilon) t_{\rm s} \nabla P \right] + (1 - \varepsilon) t_{\rm s} \nabla P \cdot \nabla s \right],
\label{eq:single:dsdt}
\end{equation}
by writing $\varepsilon$ in terms of $s$ using equation \ref{eq:epsilon}, noting that $1/(1+s^2) = (1-\varepsilon) $, and using the vector identity $\nabla \cdot [\psi \mbox{\boldmath$A$}] = \psi \nabla \cdot \mbox{\boldmath$A$} + \mbox{\boldmath$A$} \cdot \nabla{\psi}$ in which $\psi \equiv s$ and $\mbox{\boldmath$A$} \equiv s(1-\varepsilon) t_{\rm s} \nabla P$.

Using an SPH form of a direct second derivative \citep[e.g.,][]{PriLai2015, Ballabio_etal2018}, the SPH discretisation of equation \ref{eq:single:dsdt} can be written
\begin{equation}
 \frac{\rd s_a}{\rd t}  =   - \frac{1}{2 \rho_a (1 - \varepsilon_a)^2} \sum_b \frac{m_b s_b}{\rho_b}(\mathcal{D}_a+ \mathcal{D}_b)(P_a - P_b) \frac{\bar{F}_{ab}}{r_{ab}},
\label{eq:single:dsdt:SPH}
\end{equation}
where $\mathcal{D}_a = t_{{\rm s},a}(1-\varepsilon_a)$, $\bar{F}_{ab}= \left[ F_{ab}(h_a) + F_{ab}(h_b) \right]/2$, $r_{ab} = | \mbox{\boldmath$r$}_{ab} |$, and $\hat{\mbox{\boldmath$r$}}_{ab} F_{ab}(h_a) = \nabla_a W_{ab}(h_a)$.
 \cite{ElsBat2024} also derive the SPH energy equation for a single dust species, which can be written as
 \begin{equation}
 \begin{split}
\displaystyle \frac{\rd u_a}{\rd t} & \displaystyle =  \frac{P_a}{(1 - \varepsilon_a)\Omega_a \rho_a^2} \sum_b m_b (\mbox{\boldmath$v$}_a - \mbox{\boldmath$v$}_b) \cdot \nabla_a W_{ab}(h_a) \\
&  \displaystyle  + \left( \frac{\rd u_a}{\rd t} \right)^{\rm AV}  - \\
& \displaystyle \frac{1}{2(1 - \varepsilon_a)} \sum_b m_b \frac{s_a s_b}{\rho_a \rho_b} (\mathcal{D}_a + \mathcal{D}_b) ( P_a - P_b) (u_a - u_b) \frac{\bar{F}_{ab}}{r_{ab}}.
 \end{split}
 \label{eq:energytot}
 \end{equation}
The first term in this equation is the standard work ($P \rd V$) term (adjusted due to the dust fraction), the second term is the contribution from the artificial viscosity (again, see \citealt{ElsBat2024} for full details).  The third term gives the contribution from dust drag.

\subsubsection{The dust equations for a multiple dust species}

When considering multiple dust species, denoted by the subscript, $i$, we have
\begin{equation}
\varepsilon_i \equiv s_i^2/(1+ s_i^2),
\end{equation}
and, equivalently,
\begin{equation}
s_i = \sqrt{ \frac{\varepsilon_i}{1 - \varepsilon_i} }.
\end{equation}
Following the same steps as above, the evolution equation for the dust variables then becomes
\begin{equation}
 \frac{\rd s_{i,a}}{\rd t}  =   - \frac{1}{2 \rho_a (1 - \varepsilon_{i,a})^2} \sum_b \frac{m_b s_{i,b}}{\rho_b}(\mathcal{D}_{i,a}+ \mathcal{D}_{i,b})(P_a - P_b) \frac{\bar{F}_{ab}}{r_{ab}},
 \label{eq:dsdt:SPHmuti}
\end{equation}
which is identical to the single dust species equation \ref{eq:single:dsdt:SPH}, except that each dust species has its own rate of change, and $\mathcal{D}_{i,a}=\widetilde{T}_{{\rm s}i,a}(1-\varepsilon_{i,a})$.  Essentially, a vector of dust variable values is evolved for each SPH particle, rather than just a single value.

For the energy evolution, the first two terms in equation \ref{eq:energytot} are unchanged, except to note that $\varepsilon_a = \sum_i \varepsilon_{i,a}$.  The third term becomes
 \begin{equation}
 \begin{split}
\displaystyle \left. \frac{\rd u_a}{\rd t} \right|_{\rm dust} & = - \displaystyle  \sum_i\ \left[ \frac{1}{2(1 - \varepsilon_{i,a})} \times  \right. \\
&  \displaystyle  \left.  \sum_b m_b \frac{s_{i,a} s_{i,b}}{\rho_a \rho_b} (\mathcal{D}_{i,a} + \mathcal{D}_{i,b}) ( P_a - P_b) (u_a - u_b) \frac{\bar{F}_{ab}}{r_{ab}} \right]. 
 \end{split}
 \label{eq:dudt_drag}
 \end{equation}

\subsection{Time integration of the equations}
\label{sec:integration}

In the {\sc sphNG} code, two explicit integration methods are available to evolve the hydrodynamical equations in time: a second-order Runge--Kutta--Fehlberg integrator \citep{Fehlberg1969}, and a second-order leap-frog integrator \cite[as used in the {\sc phantom} SPH code;][]{Price_etal2018}.  Both integration methods use individual particle time-steps \cite[see ][for further details]{BatBonPri1995}.  In the calculations presented in this paper we use the former integrator.  

In most previous papers, dust-as-mixture SPH calculations have evolved the dust variable, $s$, explicitly using the same integrator as for the hydrodynamics.  Recently, however, \cite{ElsBat2024} developed an implicit method to evolve the dust variable.  This has two main advantages over explicit integration.  First, the explicit time-step criterion gives shorter time-steps for longer stopping times (i.e., larger grains, or low gas densities); using an implicit method can be much quicker for evolving dust with comparatively long stopping times (although the dust must still be well coupled for the terminal velocity approximation to be appropriate).  Second, using a dust variable such as equation \ref{eq:s} avoids the possibility of negative dust fractions, but it is still possible for the dust variable itself to become negative.  This typically occurs in low-density regions when there are strong dust density gradients, and it can result in the dust fraction becoming finite in regions where it should be vanishingly small.  \cite{Ballabio_etal2018} proposed using a stopping time limiter to prevent dust conservation issues in low-density regions.  However, this can result in unphysical behaviour; for example, grains with comparatively long stopping times may be artificially given shorter stopping times and, therefore, act like smaller grains.  Since the stopping time limiter is automatically applied during a calculation, it is essentially impossible to know whether the dust dynamics are realistic for their actual size or not. Using the implicit integration method means that it is not necessary to use a stopping time limiter \citep[see][for further details]{ElsBat2024}.  Thus, although we use the dust variable of \cite{Ballabio_etal2018}, we do not use the stopping time limiter in any of our calculations.

\cite{ElsBat2024} use a backwards Euler method with Gauss-Seidel iterations to solve the dust diffusion (equation \ref{eq:single:dsdt:SPH}) implicitly.  The basic method was originally used in SPH by \cite*{WhiBatMon2005} to solve the the diffusion equation for radiative transfer in the flux-limited diffusion approximation.  Following \cite{ElsBat2024}, the backwards Euler method to advance a time dependent variable, $A$, associated with particle $a$ from time $t=n$ to $t=n+1$ can be stated as
\begin{equation}
A_a^{(n+1)} = A_a^{(n)} + \rd t \left(  \frac{\rd A_a}{\rd t}  \right)^{(n+1)}.
\label{eq:BE}
\end{equation}
For interactions between particles $a$ and $b$, $A_a^{(n+1)}$ can be solved for using Gauss-Seidel iterations by rearranging the above equation into the form
\begin{equation}
A_a^{(n+1)} = \frac{\displaystyle  A_a^{(n)} - \rd t \sum_b \sigma_{ab} \left(  A_b^{(n+1)}  \right)}{ \displaystyle  1 - \rd t \sum_b \sigma_{ab}},
\label{eq:GS}
\end{equation}
where $\sigma_{ab}$ contains quantities other than the variable $A$.  At the beginning of each series of iterations, $A_b^{(n+1)} = A_b^{(n)}$ and is updated to $A_b^{(n+1)}$ as soon as this value becomes available, and the iterations continue until a specified convergence criterion is met.

In the implicit solution of the {\sc multigrain} dust variable equation \ref{eq:dsdt:SPHmuti} then becomes 
\begin{equation}
\begin{split}
s_{i,a}^{(n+1)} = & s_{i,a}^{(n)} - \rd t \left[ \displaystyle  \frac{1+ \left(s_{i,a}^{(n+1)} \right)^2}{2 \rho_a} \widetilde{T}_{{\rm s}i,a} \sum_{b} s_{i,b}^{(n+1)} L_{ab} + \right. \\
& \left.  \displaystyle  \frac{(1+ \left(s_{i,a}^{(n+1)} \right)^2)^2}{2 \rho_a} \sum_{b} \frac{\widetilde{T}_{{\rm s}i,b} }{1+ \left(s_{i,b}^{(n+1)} \right)^2}  s_{i,b}^{(n+1)} L_{ab} \right],
\label{eq:s_implicit}
\end{split}
\end{equation}
for the $i$th dust species of SPH particle, $a$, where 
\begin{equation}
L_{ab} \equiv \displaystyle \frac{m_b}{\rho_b} ( P_a - P_b) \frac{\bar{F}_{ab}}{r_{ab}}.
\end{equation}
Again, this is essentially identical to the single dust diffusion equation given by \cite{ElsBat2024}, except that a vector of dust variables (one component for each dust species) is evolved.  Equation \ref{eq:s_implicit} can be rearranged into a quartic equation for $s_{i,a}^{(n+1)} $.
For the details how we solve the quartic equation and the convergence criterion for the Gauss-Seidel iterations, see \cite{ElsBat2024}.

\subsection{Dust grain relative velocities}
\label{sec:relative}

Dust grains move relative to the gas, and they collide and stick together due to relative motions of the dust grains ($v_{kj}$ in equation \ref{eq:alpha}).  In \cite{Bate2022}, the diffusion of dust between fluid elements (SPH particles) was neglected, meaning that the distance moved by dust particles over the duration of the simulation had to be smaller than the spatial resolution of the simulations.  This limited the duration of the simulations (dust growth was only studied to the end of the first hydrostatic core phase of star formation).  Furthermore, because the relative motions of dust and gas, and of different dust grains was not explicitly solved for, analytic prescriptions were used to determine the relative velocities between grains.

\cite{Bate2022} considered Brownian motion, terminal velocities in the envelopes of collapsing molecular cloud cores, turbulence, and the effects of pressure gradients in protostellar discs that lead to radial and azimuthal velocity differences and vertical settling of the dust towards the mid-plane.  These effects all vary with the sizes of the dust particles, and thereby produce differential velocities between dust particles.  The relative dust velocities from each of these terms were then added in quadrature to estimate $v_{kj}$ (see \citealt{Bate2022} for details).

In this paper, we perform some calculations using the same prescriptions for the relative dust velocities as \cite{Bate2022}, but because the {\sc multigrain} dust-as-mixture method allows the bulk velocities of different dust sizes relative to the gas to be computed directly  (equation \ref{eq:dustv}) we usually perform calculations that use these velocities to replace the bulk relative grain velocities (i.e., those that \citealt{Bate2022} referred to as terminal velocities in the envelopes, and radial, azimuthal and vertical settling velocities in discs).  Thus, using equation \ref{eq:dustv} we have 
\begin{equation}
[v_{kj}]_{\rm bulk} = | \Delta \mbox{\boldmath$v$}_{k} - \Delta \mbox{\boldmath$v$}_{j} |.
\label{eq:vbulk}
\end{equation}

In addition to these bulk relative grain velocities, we retain the same formulations as those used by \cite{Bate2022} to account for Brownian motion, and turbulence.  For Brownian motions we use 
\begin{equation}
\left[ v_{kj} \right]_{\rm Brownian}= \sqrt{\frac{8 k_{\rm B}T}{\pi \mu_{kj}}}.
\label{eq:brownian_reduced}
\end{equation}
where $\mu_{ki}$ is the reduced mass of the two grains \citep*[e.g.,][]{BirFanJoh2016}. 

For the relative grain velocities due to the grains being coupled to gas turbulence via drag \citep{Ossenkopf1993, WeiRuz1994, Ormel_etal2009}, we use the parameterisation of \cite{OrmCuz2007}.  For protoplanetary discs the Stokes number, ${\rm St}= t_{\rm s}\Omega$, where  $\Omega$ is the orbital angular frequency.   \citeauthor{OrmCuz2007} consider both class I (slow) eddies and class II (fast) eddies.  Since we are only concerned with relatively small dust particles that are relatively well coupled to the gas, we implement equations 26 and 28 from \cite{OrmCuz2007} which cover the `tightly coupled' (class I eddies) and the `fully intermediate' (class II eddies) regimes.  For the former:
\begin{equation}
\left[ v_{12} \right]^2_{\rm turb,I} = V_{\rm g}^2  \frac{\St_1 - \St_2}{\St_1 + \St_2} \left( \frac{\St_1^2}{\St_{1} + {\rm Re}^{-1/2}}  - \frac{\St_2^2}{\St_{2} + {\rm Re}^{-1/2}} \right) 
\label{eq:class1}
\end{equation}
where the two dust grains are ordered such that ${\rm St}_1 \geq {\rm St}_2$, and Re is the Reynolds number.
For the latter:
\begin{equation}
\left[ v_{12} \right]^2_{\rm turb,II} = V_{\rm g}^2 {\rm St}_1 \left[ 2 y_{\rm a} - (1+\varepsilon) + \frac{2}{1+\varepsilon} \left( \frac{1}{1+y_{\rm a}} + \frac{\varepsilon^3}{y_{\rm a} + \varepsilon} \right) \right],
\end{equation}
where $\varepsilon=\St_2/\St_1 \leq 1$ and we take $y_{\rm a}=1.6$.  This latter contribution from class II eddies is only applied when  $\St_1>{\rm Re}^{-1/2}/y_{\rm a}$.  We limit St$_2$ to be no greater than $0.9~{\rm St}_1$ so that two similarly-sized dust particles have some relative motion (otherwise equation \ref{eq:class1} gives zero relative velocity for two identical dust grains).  This is justified by the fact that, in reality, two equal-mass dust grains will not have identical shapes (see also \citealt{Okuzumi_etal2012,Krijt_etal2016}; \citealt*{SatOkuIda2016}).

For turbulence in a disc, we assume an $\alpha_{\rm SS}$-disc model \citep{ShaSun1973}, for which the normalisation can be expressed as $V_{\rm g}=\alpha_{\rm SS}^{1/2} c_{\rm s}$ \citep*[e.g.,][]{NakSekHay1986,CuzWei2006}, and the Reynolds number as ${\rm Re} = \alpha_{\rm SS} c_{\rm s} H/\nu = \alpha_{\rm SS} c_{\rm s}^2/(\nu \Omega)$ where $c_{\rm s}$ is the isothermal gas sound speed, and $H$ is the gas disc scale height.  For the calculations reported in this paper, we set $\alpha_{\rm SS}=10^{-3}$.  \cite{Bate2022} took a constant value ($10^8$) for the Reynolds number in their calculations, but here we compute it locally using
\begin{equation}
{\rm Re} = 6.2 \times 10^7 \left(\frac{n_H}{10^5~{\rm cm}^{-3}}\right)^{1/2} \left(\frac{T_{\rm g}}{10~{\rm K}}\right)^{1/2},
\label{eq:Re}
\end{equation}
as in \cite{Ormel_etal2009} and \cite{Lebreuilly_etal2023}.

\cite{Bate2022} set the turbulence in the envelope of a collapsing molecular cloud core using $V_{\rm g}=(3/2)^{1/2} c_{\rm s}$ (e.g., \citealt{Ormel_etal2009,Guillet_etal2020}), and defining the Stokes number in the envelope as ${\rm St} = t_{\rm s} c_{\rm s}/\lambda_{\rm Jeans} = 2 t_{\rm s} \sqrt{G \rho_{\rm G}/\pi}$ (e.g., \citealt{Guillet_etal2020}), where the Jeans length $\lambda_{\rm Jeans}$ has been defined by relating the sound crossing time to the free-fall time.  We retain the same turbulence model for the envelope in the star formation calculations presented here.  As shown by \cite{Bate2022}, there is negligible grain growth in the envelope so this is of little importance.

For the star formation calculations, as in \cite{Bate2022} we define an SPH particle to lie in a disc if the magnitude of its radial velocity is less than half the magnitude of its tangential velocity, computing these velocities relative to the direction of the gravitational acceleration experienced by the SPH particle (i.e., the radial velocity is $v_{\rm rad} = \mbox{\boldmath{$v$}} \cdot \hat{\mbox{\boldmath{$a$}}}_{\rm grav}$ and $v_{\rm tan}^2 = v^2 - v_{\rm rad}^2$).  
 \cite{Bate2022} found that this prescription was sufficient to identify first hydrostatic 
cores as being part of a `disc' even though they are more pressure supported than
rotationally supported; see \cite{Bate2022} for further discussion.
Note that this prescription assumes that the centre of mass of the star/disc system is stationary, otherwise the centre of mass velocity must be subtracted first.  When using the relative grain velocity prescriptions of \cite{Bate2022}, whether a particle lies in a disc or envelope affects which bulk relative velocity prescriptions are included and the definition of $V_{\rm g}$ for the turbulent contributions (see \citealt{Bate2022} for further details).  When using the bulk relative velocities provided by the {\sc multigrain} method, whether a particle is determined to be in the disc or envelope only determines the contribution of the turbulence to the relative grain velocities (i.e., the magnitude of $V_{\rm g}$ and the definition of the Stokes number).

Finally, the above sources of relative dust velocity are combined in quadrature to give the overall relative grain velocity that appears in equation \ref{eq:alpha}, i.e.,
\begin{equation}
v_{kj}^2 =  \left[ v_{kj} \right]_{\rm Brownian}^2 +  \left[ v_{kj} \right]_{\rm turb,I}^2 +  \left[ v_{kj} \right]_{\rm turb,II}^2 + \left[ v_{kj} \right]_{\rm bulk}^2
\label{eq:quadrature}
\end{equation}

\subsection{Turbulent stirring of dust}
\label{sec:stirring}

As discussed in the previous section, grain growth depends on dust grains having relative motions so that they can collide.  For the smallest grains, the dominant source is Brownian motion, but for other small grains turbulence is likely to provide the dominant source.  However, the dust dynamical equations of Sections \ref{sec:multigrain} to \ref{sec:SPHequations} only describe bulk motions of grains, e.g., due to pressure gradients.  For discs, in particular, this leads to an inconsistency -- we are assuming some $\alpha_{\rm SS}$ type turbulence is responsible for providing relative grain velocities that allow grain growth, but we are not including the effects of such turbulent stirring on the bulk dynamics of the dust grains.  For example, as it stands, the dust dynamical equations given above will allow weakly-coupled dust grains to settle towards the mid-plane of a laminar protoplanetary disc and the dust layer will continually get thinner and thinner with time.  In reality, however, if the disc is turbulent the dust grains should be stirred up and a dust layer should maintain a finite thickness in which settling is balanced by turbulent stirring (with more strongly-coupled grains being confined to thicker layers).

The bulk effects of turbulent stirring can be described using a conventional diffusion equation for the dust fraction
\begin{equation}
\frac{\rd \varepsilon_i}{\rd t} = \frac{1}{\rho} \nabla \cdot \left( D_i \rho  \nabla \varepsilon_i \right),
\label{eq:dedtdiff}
\end{equation}
where $D_i$ is the diffusion constant for dust species, $i$.
This has a similar form to a heat conduction equation or a radiative transfer equation in the diffusion limit, both of which have been modelled in the past using SPH and a direct second derivative \citep{CleMon1999,WhiBat2004,WhiBatMon2005}.  Using the SPH form of a direct second derivative (similar to equation \ref{eq:single:dsdt:SPH} above), equation \ref{eq:dedtdiff} can be written
\begin{equation}
 \frac{\rd \varepsilon_{i,a}}{\rd t}  =   \sum_b \frac{m_b}{\rho_a \rho_b}(D_{i,a}+ D_{i,b})\left(\frac{\rho_a + \rho_b}{2}\right) (\varepsilon_{i,a} - \varepsilon_{i,b}) \frac{\bar{F}_{ab}}{r_{ab}},
\label{eq:diffdustfrac}
\end{equation}
which, due to the symmetry of SPH particles $a,b$, inherently conserves the dust mass (at least for global time steps).  Using equation \ref{eq:dsdtdedt} this can be written as
\begin{equation}
\begin{split}
 \frac{\rd s_a}{\rd t}  =  & \frac{1}{2 s_a (1 - \varepsilon_a)^2} \sum_b \frac{m_b}{\rho_a \rho_b}(D_{i,a}+ D_{i,b})\left(\frac{\rho_a + \rho_b}{2}\right) \times \\
& (\varepsilon_{i,a} - \varepsilon_{i,b}) \frac{\bar{F}_{ab}}{r_{ab}},
\end{split}
\label{eq:s_diffdustfrac}
\end{equation}
for the evolution of the \cite{Ballabio_etal2018} dust variable, $s$.  This equation can be used to evolve the dust variable explicitly and, when combined with equation \ref{eq:dsdt:SPHmuti}, the resulting equations allow the dust distributions to evolve both due to bulk motions and turbulent diffusion.  Again, as with the addition of the dust drag equations, the equations that express mass and momentum conservation are unchanged so the conservation of mass and momentum are retained, but an additional term needs to be added to the energy if energy is to be conserved.  This has a similar form to equation \ref{eq:dudt_drag}
 \begin{equation}
 \begin{split}
& \displaystyle \left. \frac{\rd u_a}{\rd t} \right|_{\rm diffusion} =  \displaystyle  \sum_i\ \left[ \frac{1}{2(1 - \varepsilon_{i,a})} \times  \right. \\
&  \displaystyle  \left.  \sum_b  \frac{m_b}{\rho_a \rho_b} (D_{i,a} + D_{i,b}) \left(\frac{\rho_a + \rho_b}{2}\right)(\varepsilon_{i,a} - \varepsilon_{i,b}) (u_a - u_b) \frac{\bar{F}_{ab}}{r_{ab}} \right]. 
 \end{split}
 \label{eq:dudt_diffusion}
 \end{equation}
 The physical interpretation is that if some dust is diffused from particle $a$ to particle $b$, some gas (which has its own specific internal energy) is moved in the opposite direction to maintain the total mass of the SPH particle.

A model is, of course, required to specify the magnitude of the turbulent diffusion, $D_i$.  Here we again assume an $\alpha_{\rm SS}$-type fluid dynamical disc turbulence for which we can write $D_i = \alpha_{\rm SS} c_{\rm s} H$.  This is appropriate for grains that are strongly coupled to the gas, but not for grains that are weakly coupled (e.g., larger grains).  Following \cite{YouLit2007} we therefore divide this diffusion parameter by the Schmidt number (the ratio of gas to particle diffusivity) which we express as $Sc = 1 + (\Omega t_{\rm s})^2$, and writing $H=c_{\rm s}/\Omega$, we have that the diffusion parameter for dust species, $i$, of SPH particle, $a$, is given by
\begin{equation}
D_{i,a} = \frac{\alpha_{\rm SS} c_{{\rm s},a}^2}{\Omega_a(1 + (\Omega_a t_{{\rm s}i,a})^2)},
\label{eq:diffconst}
\end{equation}
where $\Omega_a$ is the orbital frequency experienced by SPH particle, $a$.

If explicit integration is used, an additional time-step constraint is required, based on the usual time-step constraint for a diffusion equation, i.e., $\delta t < h^2/D$.  This can potentially become restrictive at high resolution.  To avoid such time-step constraints we can again develop an implicit scheme using a similar Gauss-Seidel method to that described above.  In equation \ref{eq:single:dsdt:SPH}, we can replace $1/(1-\varepsilon_a)^2$ with $(1+s_a^2)$, and $\varepsilon_i,a = s_{i,a}^2/(1+s_{i,a}^2)$ and a similar substitution for $\varepsilon_i,b$.  This allows an implicit equation to be expressed, similar to equation \ref{eq:s_implicit}.  However, the presence of the quantity $s_a$ in the denominator of the first term in equation \ref{eq:single:dsdt:SPH} means that we cannot obtain a pure quartic equation as we did for equation \ref{eq:s_implicit}.  We have implemented and tested an iterative Gauss-Seidel implicit method based on the solution of a quartic in which we simply replace the offending $1/s_a$ with the value of $s_a$ from the previous iteration.  This does work when the dust variable (equivalently, the dust fraction) is substantial.  However, if the dust fraction tends towards zero we have found that the iterations can struggle to converge (i.e., the iterative solution becomes very slow and/or requires sub-cycling with smaller time-steps).  This makes such a method essentially unusable (in that it can become much slower than an explicit method).

An alternative is to treat the evolution of the dust fraction, $\varepsilon_{i,a}$, using an implicit Gauss-Seidel method based on equation \ref{eq:diffdustfrac} rather than using the dust variable, $s_{i,a}$.  This avoids the offending $1/s_a$ term, and has the added benefit that the backwards Euler form of equation \ref{eq:diffdustfrac} is a simple linear equation (i.e., there is no need to solve a quartic equation).  Thus, we solve
\begin{equation}
\varepsilon_{i,a}^{(n+1)} =  \left( \varepsilon_{i,a}^{(n)} - \rd t \displaystyle   \sum_{b} M_{ab} ~ \varepsilon_{i,b}^{(n+1)}  \right) / \left(  1 - \rd t \displaystyle  \sum_{b}  M_{ab}  \right),
\label{eq:epsilon_implicit}
\end{equation}
for the $i$th dust species of SPH particle, $a$, where 
\begin{equation}
M_{ab} \equiv \displaystyle \frac{m_b}{\rho_a \rho_b} (D_{i,a}+ D_{i,b}) \left(\frac{\rho_a + \rho_b}{2} \right) \frac{\bar{F}_{ab}}{r_{ab}}.
\end{equation}
Thus, we use an `operator splitting' approach to evolving the dust variable / dust fraction, first computing the effect of the bulk dust dynamics using Gauss-Seidel iteration of equation \ref{eq:s_implicit}, and then solving for the effect of the dust turbulent diffusion using separate Gauss-Seidel iteration of equation \ref{eq:epsilon_implicit}.  Computationally, it is a little inconvenient to evolve both the dust variable, $s$, and the dust fraction, $\varepsilon$, but this combination does result in a purely implicit scheme to evolve the dust fractions that converges well. 

\subsection{The SPH code}
\label{sec:sph}

The smoothed particle hydrodynamics method was original devised by 
\cite{Lucy1977} and \cite{GinMon1977}.  The {\sc sphNG} code used here originated 
from that of \citeauthor{Benz1990} 
(\citeyear{Benz1990}; \citealt{Benzetal1990}), but has been substantially
modified as described in \citet*{BatBonPri1995}, \cite{PriBat2007}, and 
parallelised using both OpenMP and MPI.

A particle's nearest neighbours and the gravitational forces between particles
are calculated using a binary tree.  The smoothing lengths of particles are 
variable in space and time, set iteratively such that the smoothing
length of each particle 
$h = 1.2 (m/\rho)^{1/3}$ where $m$ and $\rho$ are the 
SPH particle's mass and density, respectively
\citep{PriMon2007}.  In the calculations presented here the
\cite{MorMon1997} artificial viscosity is used
with $\alpha_{\rm_v}$ varying between 0.1 and 1 and with $\beta_{\rm v}=2$
\citep[see also][]{PriMon2005}.  The SPH equations are 
integrated using a second-order Runge-Kutta-Fehlberg 
integrator \citep{Fehlberg1969} with individual particle 
time steps (see \citealt{BatBonPri1995} for further details). 
Following \cite{Bate2022}, in addition to the usual SPH time-step criteria (e.g., the 
Courant condition), we add a time-step constraint for the dust
fractions such that the change in $\varepsilon_i$ for any 
dust bin $i$ should not exceed 0.3 of the value of $\varepsilon_i$.

For the calculations of star formation that are reported towards the
end of this paper, we use the combined radiative transfer and diffuse 
ISM model that was developed by \cite{BatKet2015} to model the thermal 
evolution of the gas.  For the details of this method, the reader
is directed to that paper, or the brief summary provided by \cite{Bate2019}. 
We assume solar metallicity, and we note that the dust opacities for the
radiative transfer and diffuse ISM model are constant (i.e., grain growth
and migration are not taken into account for the radiative transfer).  Linking
the radiative transfer to the dust evolution is an obvious next step, but is
beyond the scope of the current paper.

Finally, we use sink particles \citep{BatBonPri1995} to follow the calculations 
beyond the point of stellar core formation.   \cite{Bate2022} only followed 
the collapse of the molecular cloud cores up to the onset of
the second collapse phase due to the dissociation of molecular hydrogen that results in
the formation of the stellar core \citep{Larson1969}.  This end point was chosen since
up until this point any drift of the dust relative to the gas is negligible when beginning
with a population of strongly-coupled grains.  
In this paper, because we treat both the growth 
and dynamical evolution of the dust, we are able to continue the calculations 
much further. But to allow us evolve the calculations over tens of thousands of
years beyond stellar core formation, we need to use sink particles to avoid the 
high densities and temperatures, and thus short time steps, in the vicinity of the 
stellar core.

\section{Test calculations}
\label{sec:tests}

As in \cite{HutPriLai2018}, basic tests of the {\sc multigrain} implementation into the {\sc sphNG} code  were performed using the {\sc dustywave} and {\sc dustydiffuse} test cases to confirm that using $N$ identical dust bins with dust fractions $\varepsilon_i=\varepsilon/N$ gave the same result as using a single dust phase.  Similarly, some of the calculations of dust growth from \cite{Bate2022} were repeated with the new version of the code, without allowing dust diffusion, to confirm that the results were unchanged.  In the following subsections we describe the outcome of tests that combine the dust growth and the dust diffusion.

\subsection{The dust settling test}
\label{sec:settle0}

The first substantial test performed by \cite{HutPriLai2018}, was a version of the dust settling test from \cite{PriLai2015} that mimics the settling of strongly-coupled dust grains in protoplanetary discs.  We use the same basic set up as that described by \cite{HutPriLai2018}.  The test case consists of a Cartesian column of dusty gas in vertical hydrostatic equilibrium that is subject to an external acceleration
\begin{equation}
\mbox{\boldmath$a$}_{\rm ext} = - \frac{GMz}{(r^2+z^2)^{3/2}} {\mbox{\boldmath$\hat{z}$}},
\end{equation}
where $G$ is the gravitational constant and $z$ is the vertical coordinate.  This acceleration mimics the vertical acceleration that would be experienced by the gas in a disc orbiting a star of mass, $M$, at a distance $r$ from the star, except that $r=50$ au is set to a constant value (i.e., it does not vary with the $x$ and $y$ coordinates within the column).  There are no velocities in the $x$,$y$ directions.  The gas density of the column is given by 
\begin{equation}
\rho_{\rm G} = \rho_{\rm g,0} \exp \left( - \frac{z^2}{2H^2} \right),
\label{eq:vert}
\end{equation}
where $H/r = 0.05$ (i.e., a disc scale height of 2.5 au).  We use an isothermal equation of state ($\gamma=1$) with $P=c_{\rm s}^2\rho_{\rm g}$, where $c_{\rm s} \equiv H\Omega$ and $\Omega \equiv \sqrt{GM/r^3}$. This corresponds to an orbital period of $t_{\rm orb} = 2\pi/\Omega \approx 353$ yr.  The code units are set to a distance unit of 10 au, mass in solar masses, and $G=1$, giving an orbital time of $\approx 70.2$ in code units.

\begin{figure*}
\centering \vspace{-0.25cm} \vspace{-0.0cm}
    \includegraphics[height=6.5cm]{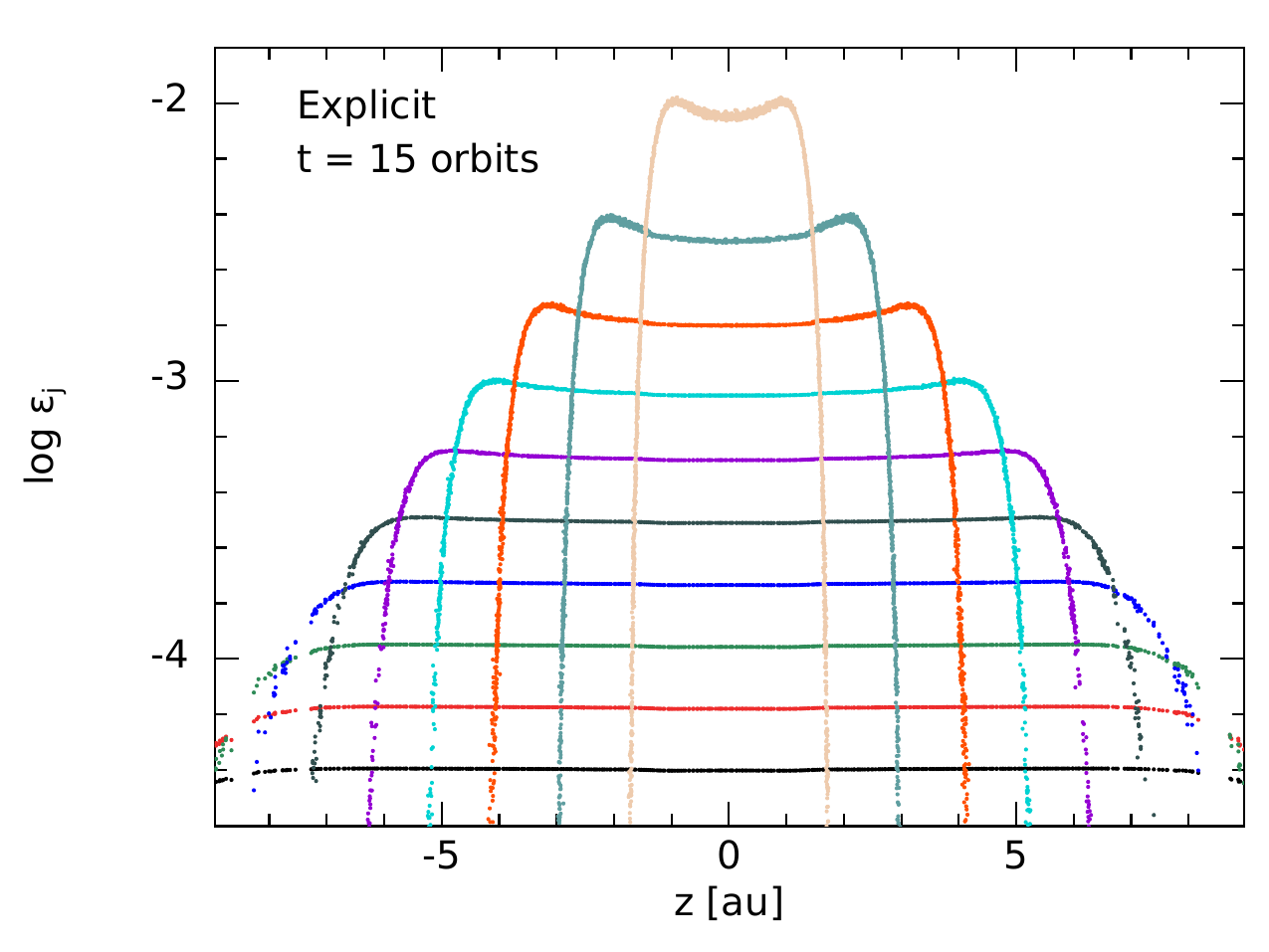}
    \includegraphics[height=6.5cm]{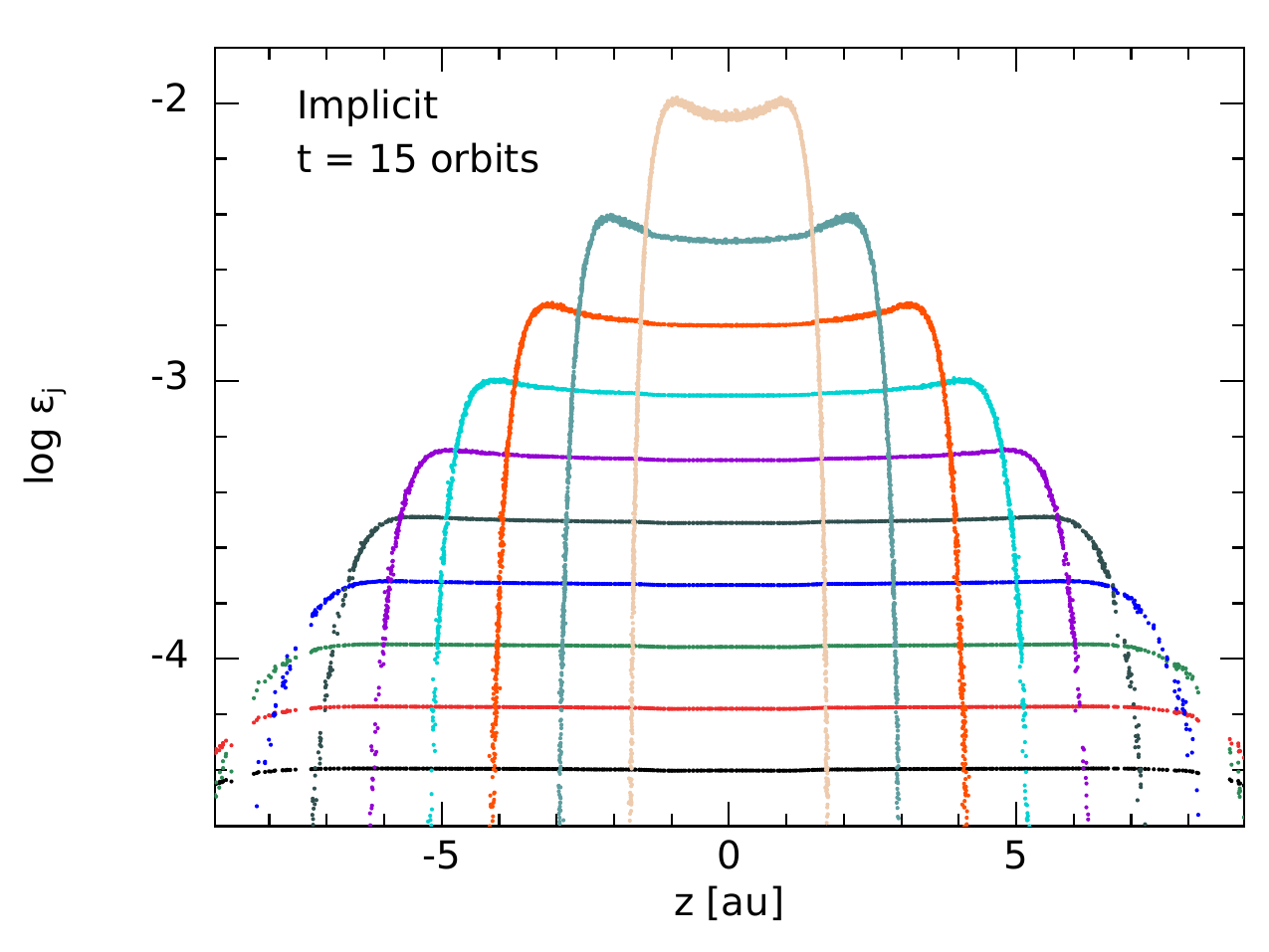}
\caption{The vertical distributions of the dust fractions of the 10 dust size bins after 15 orbits of dust settling when calculated using the explicit (left) and implicit (right) {\sc multigrain} methods.  The results are almost identical.  The different grain size bins are ordered from 1~mm to 0.1~$\mu$m  from top to bottom at $z=0$ (the mid-plane).  The largest grains quickly settle towards the mid-plane, while the smallest grains maintain a near uniform dust fraction distribution.}
\label{fig:settle1}
\end{figure*}

We use essentially the same particle spatial resolution as that of \cite{HutPriLai2018}, but we use a column with smaller $x$,$y$ dimensions (using larger dimensions in the `plane' of the disc simply requires more particles).  We first set up 25942 particles on a close-packed lattice in the domain $[x,y,z]=[\pm 0.2, \pm 0.15, \pm 1]$ in code units (\citeauthor{HutPriLai2018} used 670\,800 particles in a column that had 5 times larger dimensions in the $x-y$ plane).  We then stretch the particle positions along the vertical direction using the method described in \cite{Price2004} to give the Gaussian vertical density profile (equation \ref{eq:vert}).  To allow the column of gas to settle into and maintain a stable hydrostatic equilibrium with small particle velocities during the dust settling test we included velocity damping.  An extra term was added to the momentum equation of the form ${\rm d}\mbox{\boldmath{$v_a$}}/{\rm d}t = - \mbox{\boldmath{$v$}}_a/\tau$, for each SPH particle $a$ (we used $\tau=2$).  For each of the calculations discussed below, we evolved the SPH calculation for 5 orbital periods with the dust growth and dust diffusion equations disabled (i.e., the dust fractions remained constant) to achieve a hydrostatic gas distribution, after which the dust settling and/or dust growth were enabled.  A total mass of $7.5\times 10^{-5}$~M$_\odot$ gives $\rho_{\rm g,0} \approx 10^{-3}$ in code units, corresponding to $\approx 6 \times 10^{-13}$~g~cm$^{-3}$ in physical units.  This gives a mass surface density of approximately 56 g~cm$^{-2}$ at 50 au for the column of disc material, which is  on the high side compared to realistic discs (if the disc had an overall surface density profile $\Sigma \propto r^{-1/2}$, this would give a total disc mass of $0.065~{\rm M}_\odot$ within 50 au).  We used periodic boundary conditions in all three directions, but set the vertical boundaries at $z = \pm 40 H$ to avoid periodicity in the vertical direction.

Following \cite{HutPriLai2018}, for the first tests we used 10 dust bins to model grain sizes ranging from $0.1~\mu$m to 1~mm.  The total dust fraction was set to be $\varepsilon=1/101$, corresponding to a dust-to-gas ratio of 0.01, and the magnitudes of the dust fractions of each bin were set according to a power-law distribution
\begin{equation}
{\rm d}\varepsilon = \varepsilon_0 a^{3-p} {\rm d}a,  ~~~~~~ {\rm for} ~ a_{\rm min} \leq a \leq a_{\rm max},
\label{eq:powerlaw}
\end{equation}
where ${\rm d}\varepsilon$ is the differential dust fraction with respect to grain size, $p$ is the power-law index for number density as a function of grain size \citep[e.g.,][]{MatRumNor1977}, and $\varepsilon_0$ is a normalisation factor.  Taking $p=3.5$ and setting $a_{\rm min} = 0.0599~\mu$m and $a_{\rm max} = 1.67$~mm, the initial dust fractions in each bin are set by integrating equation \ref{eq:powerlaw} across each grain-size bin and normalising the values.  For the dust settling tests with 10 bins we use geometric means describe the mean dust grain size in a bin for consistency with \cite{HutPriLai2018}.  This gives mean dust sizes of 0.10, 0.28, 0.77, 2.2, 6.0, 17, 46 $\mu$m and 0.13, 0.36, 1.0 mm for the 10 dust bins. 

\subsection{A dust settling test, with and without grain growth}
\label{sec:settle1}

\begin{figure}
\centering \vspace{-0.25cm} \vspace{-0.0cm}
    \includegraphics[height=6.5cm]{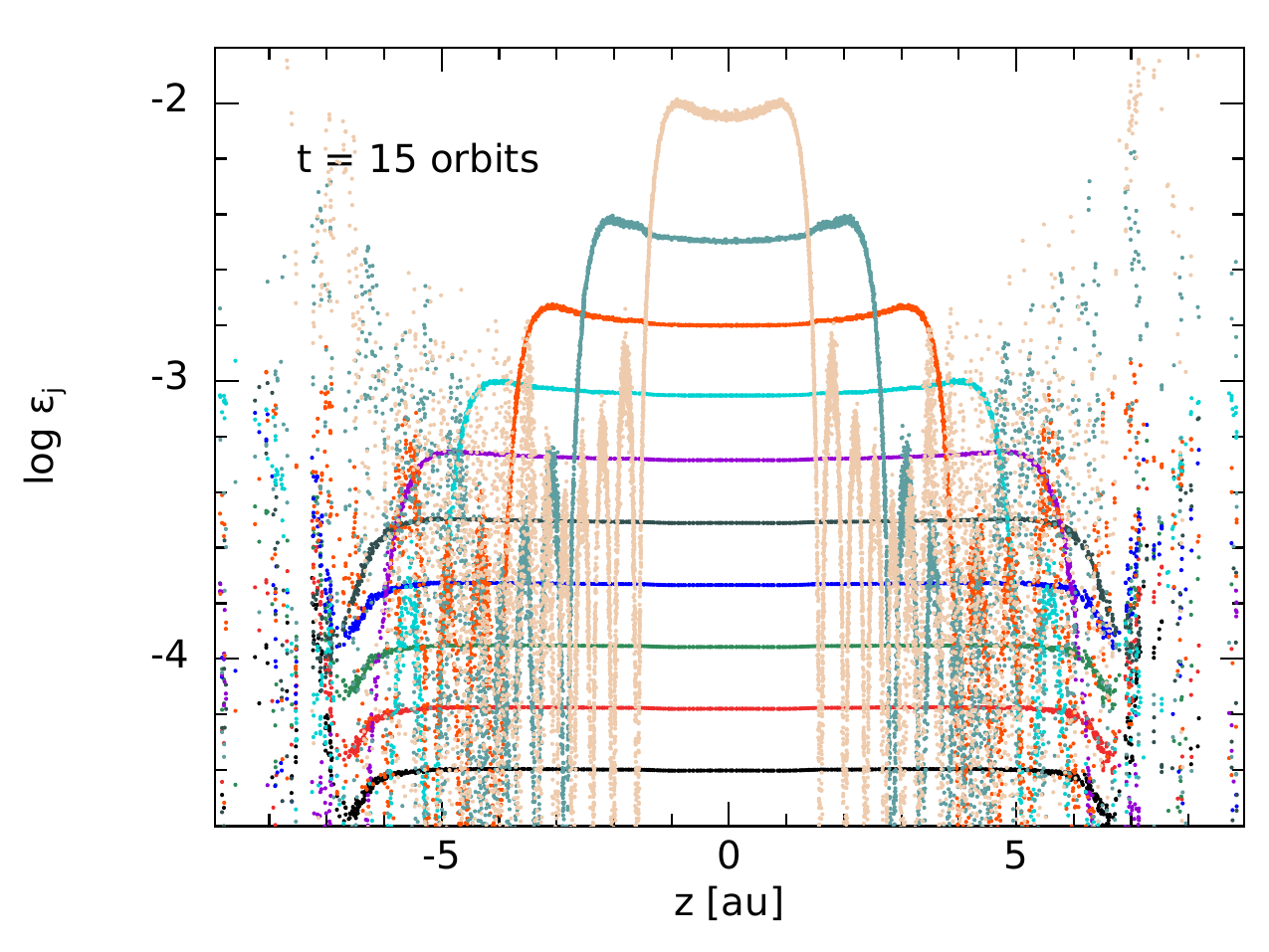}
\caption{The vertical distributions of the dust fractions of the 10 dust size bins after 15 orbits of dust settling when calculated using the explicit {\sc multigrain} method {\it without enforcing that the dust variable should never become negative}.  Although the solution is correct in regions with a substantial dust fraction, spurious dust appears above and below the dust layers where, in reality, all the dust in that size bin should have settled out.}
\label{fig:settle2}
\end{figure}

In Fig.~\ref{fig:settle1}, we show the vertical profiles of the dust fractions for each grain size bin after 15 orbits of dust settling, without any dust growth.  The left panel shows the results using explicit integration of the dust evolution equation formulated with the \cite{Ballabio_etal2018} dust variable (equation \ref{eq:epsilon}).  The left panel is the equivalent of Figure 4 in \cite{HutPriLai2018}, although they used the \cite{PriLai2015} dust variable formulation.   The results are very similar to those obtained by \cite{HutPriLai2018} except that there is less noise in the dust fractions here.  \cite{HutPriLai2018} settled a gas-only disc before adding the dust to it which likely introduced small waves, whereas here we used velocity damping to maintain an almost static gas distribution.  The small dust remains well mixed with the gas (i.e., the dust fraction remains almost constant) while large dust grains settle towards the mid-plane as expected \citep{HutPriLai2018}.  The right panel of Fig.~\ref{fig:settle1} shows the results obtained using the implicit method \citep{ElsBat2024} generalised to the {\sc multigrain} formulation, as described in Section \ref{sec:integration}.  The results obtained by the explicit and implicit formulations are almost identical.   We note that to obtain this result using explicit method requires that the dust variable is not allowed to go negative, i.e., any explicit update of the dust variable is of the form $s_{j}(t+\delta t) = {\rm MAX}(0, s_{j}(t) + ({\rm d}s_{j} /{\rm d}t) \delta t)$.  Without enforcing positivity of the dust variable the solution to the dust settling test is as seen in Figure ~\ref{fig:settle2}.  In this case it is guaranteed that the dust fraction always lies between 0 and 1 due to the formulation of the dust variable (equation \ref{eq:epsilon}), but when the dust variable goes negative this shows up in the solution as spurious dust in regions that should be devoid of dust.  This was pointed out by \cite{ElsBat2024}.  The implicit method does not suffer from this issue -- the dust variable always stays positive without any enforcement being necessary.

\begin{figure*}
\centering \vspace{-0.25cm} \vspace{-0.0cm}
    \includegraphics[height=6.5cm]{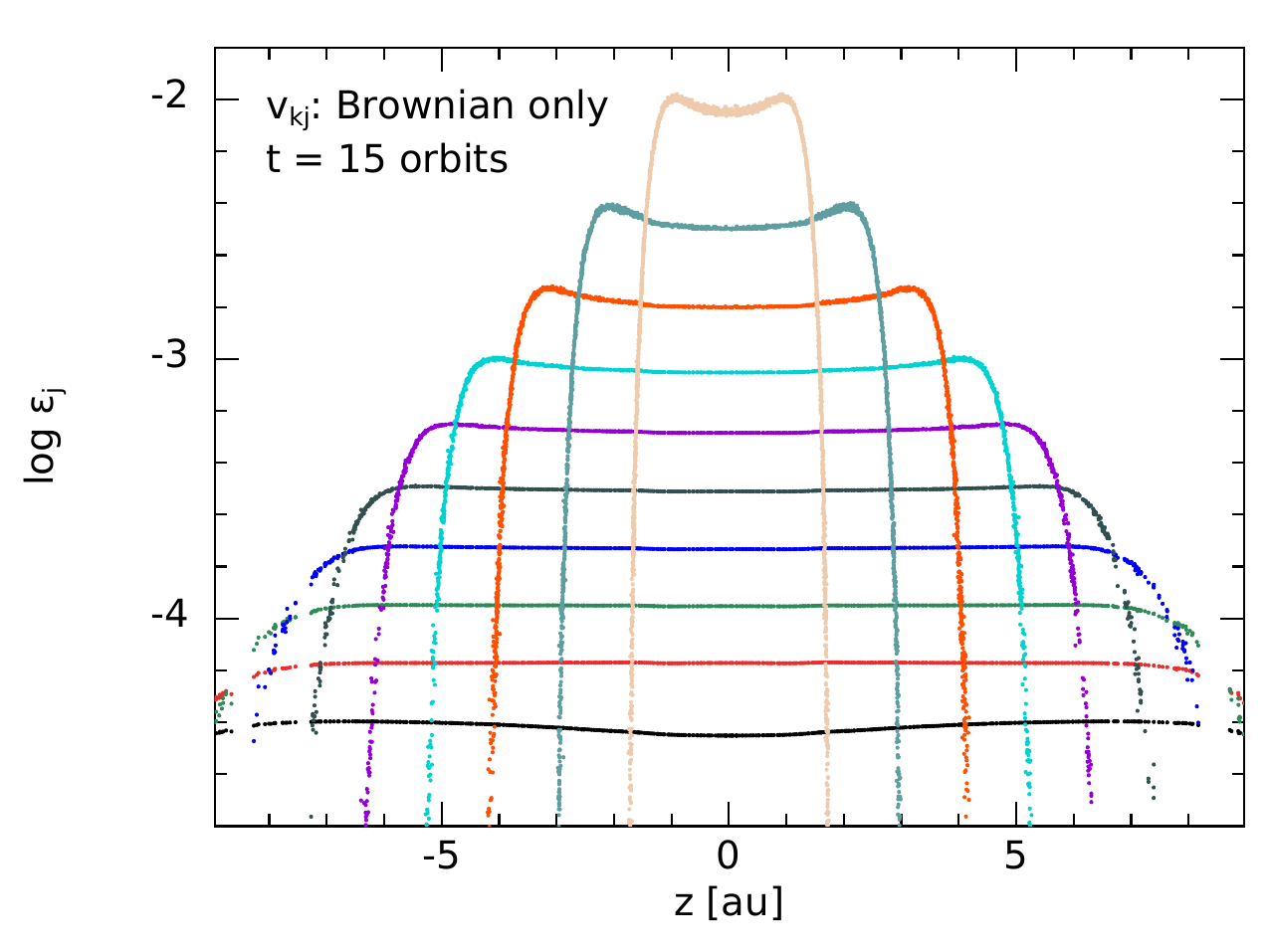}
    \includegraphics[height=6.5cm]{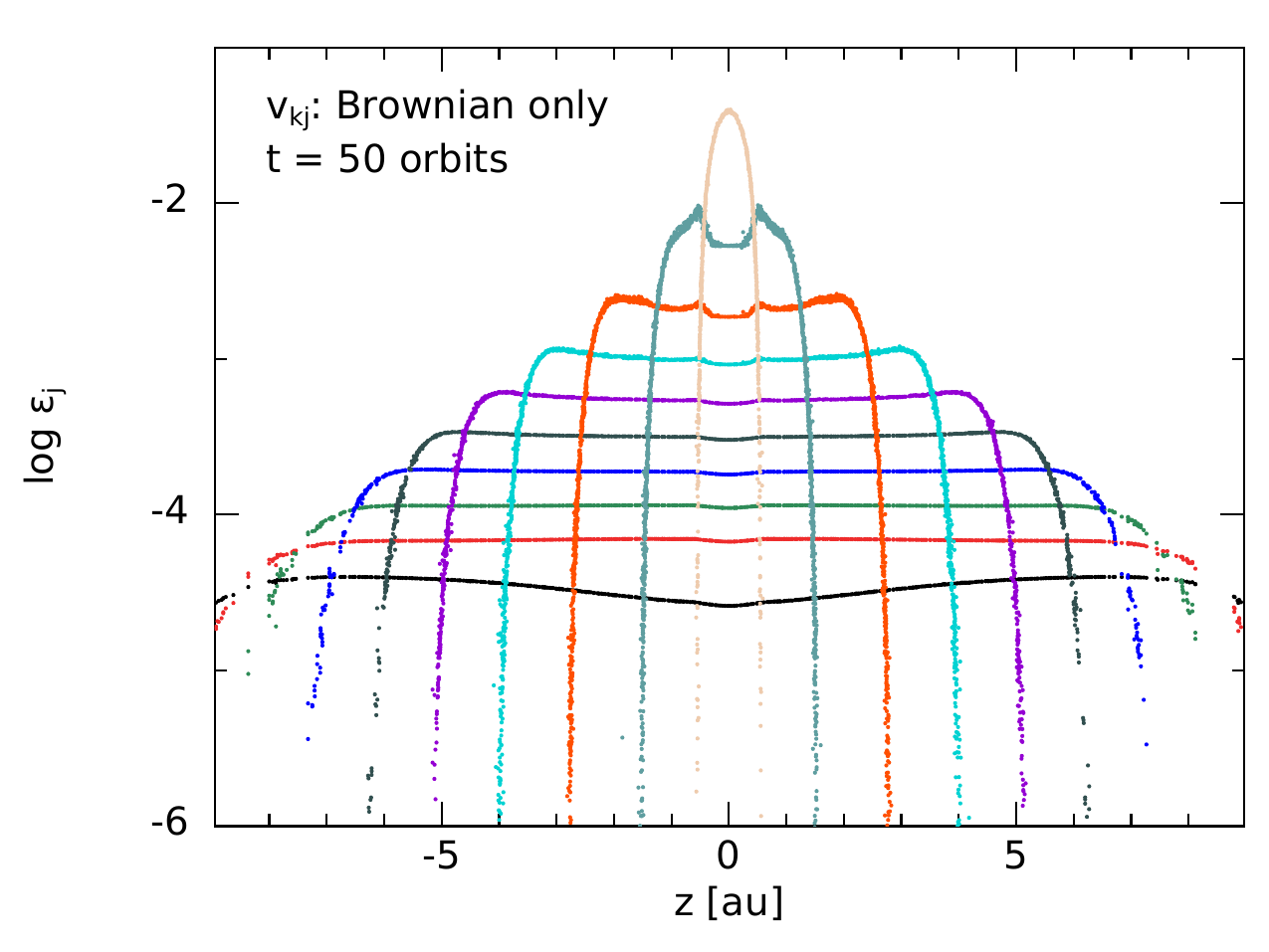}
    \includegraphics[height=6.5cm]{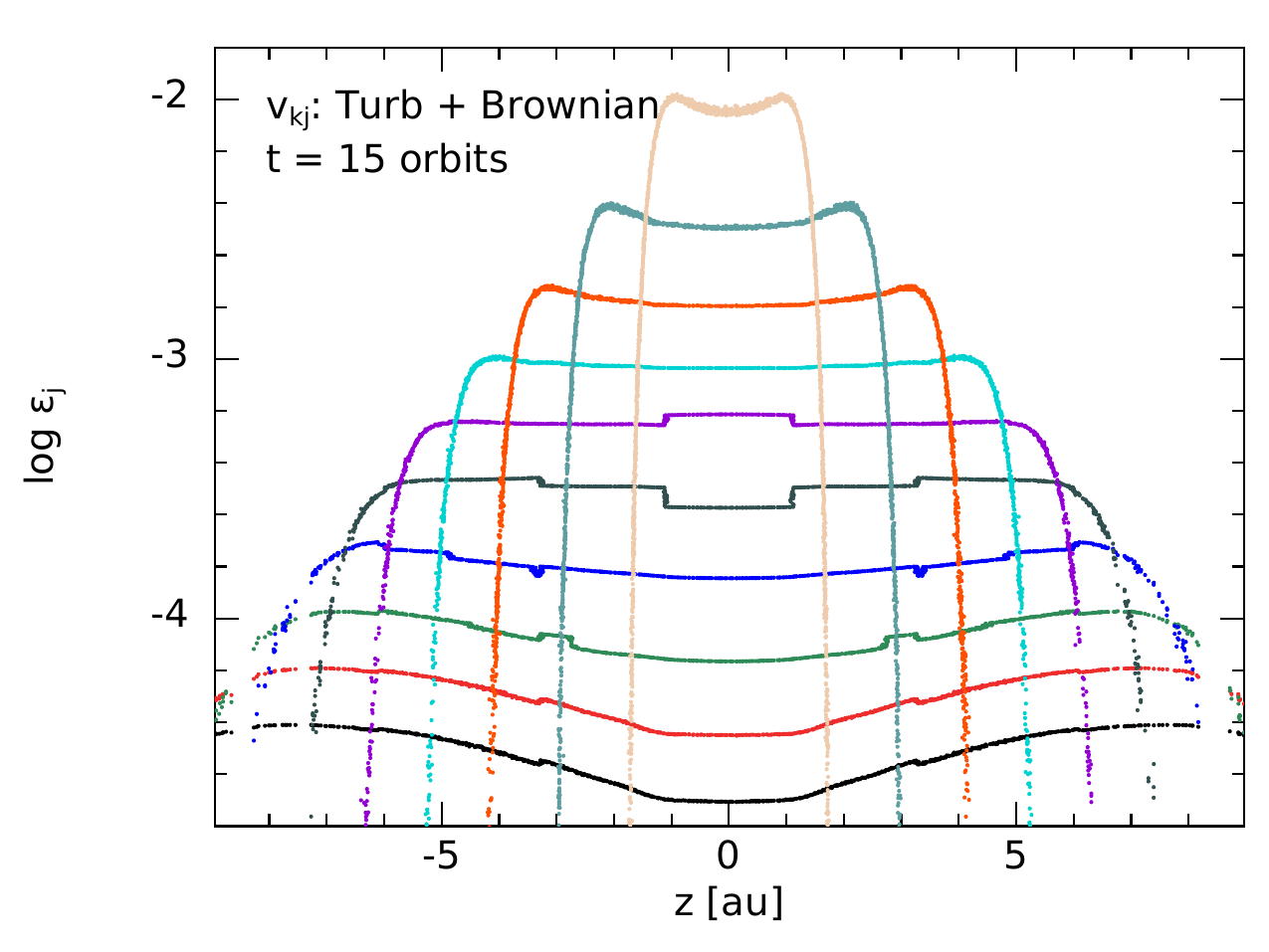}
    \includegraphics[height=6.5cm]{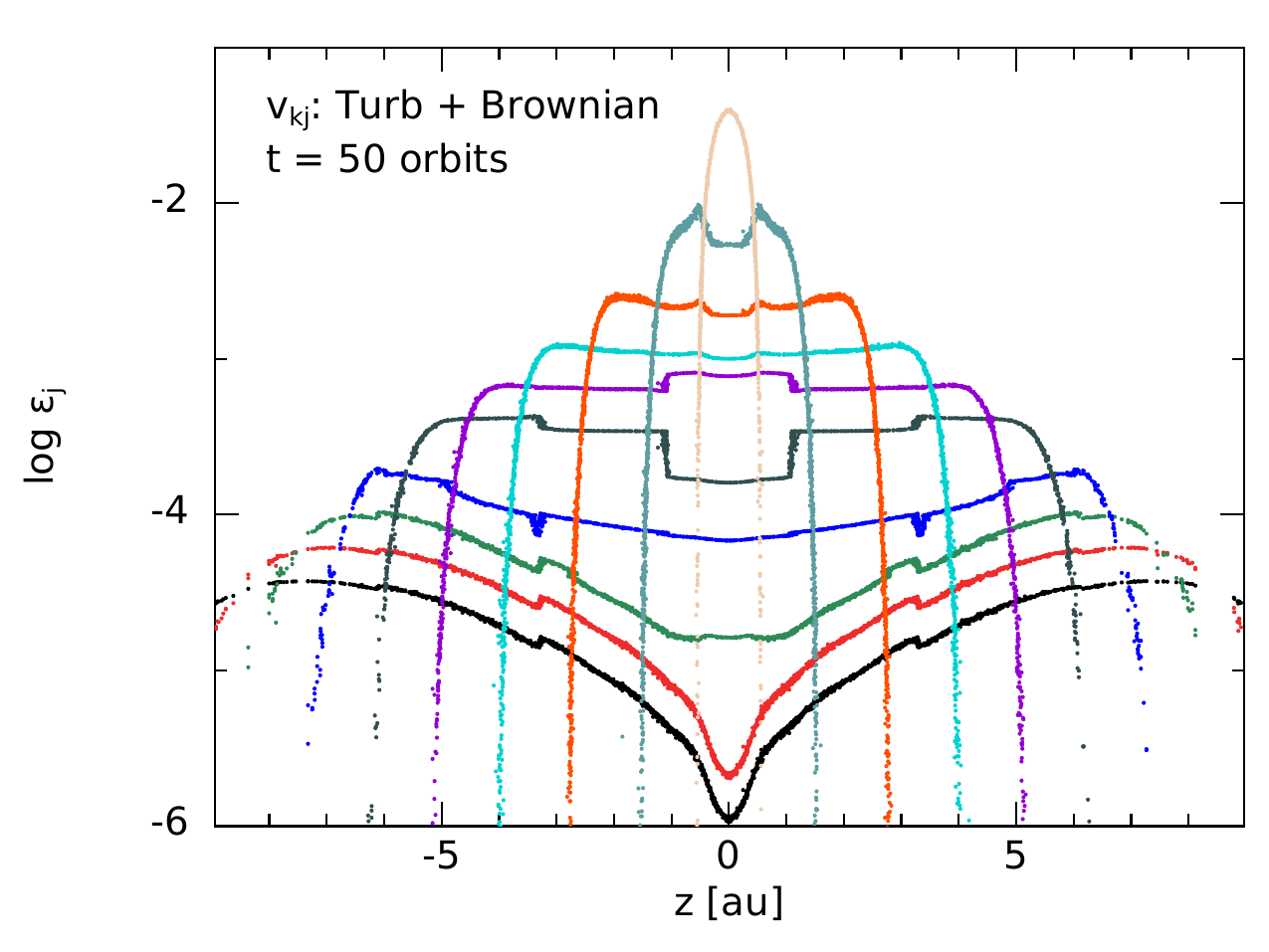}
    \includegraphics[height=6.5cm]{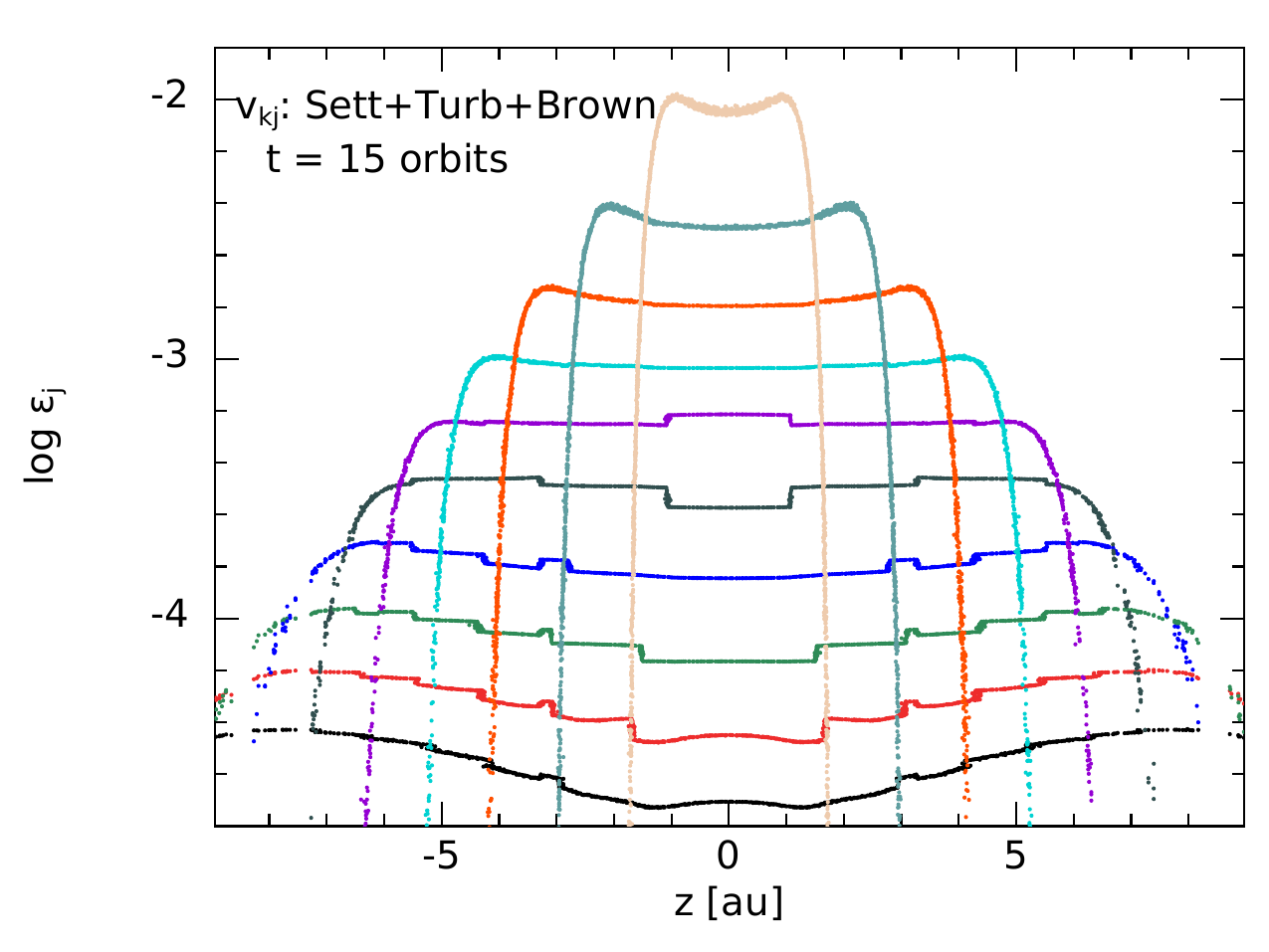}
    \includegraphics[height=6.5cm]{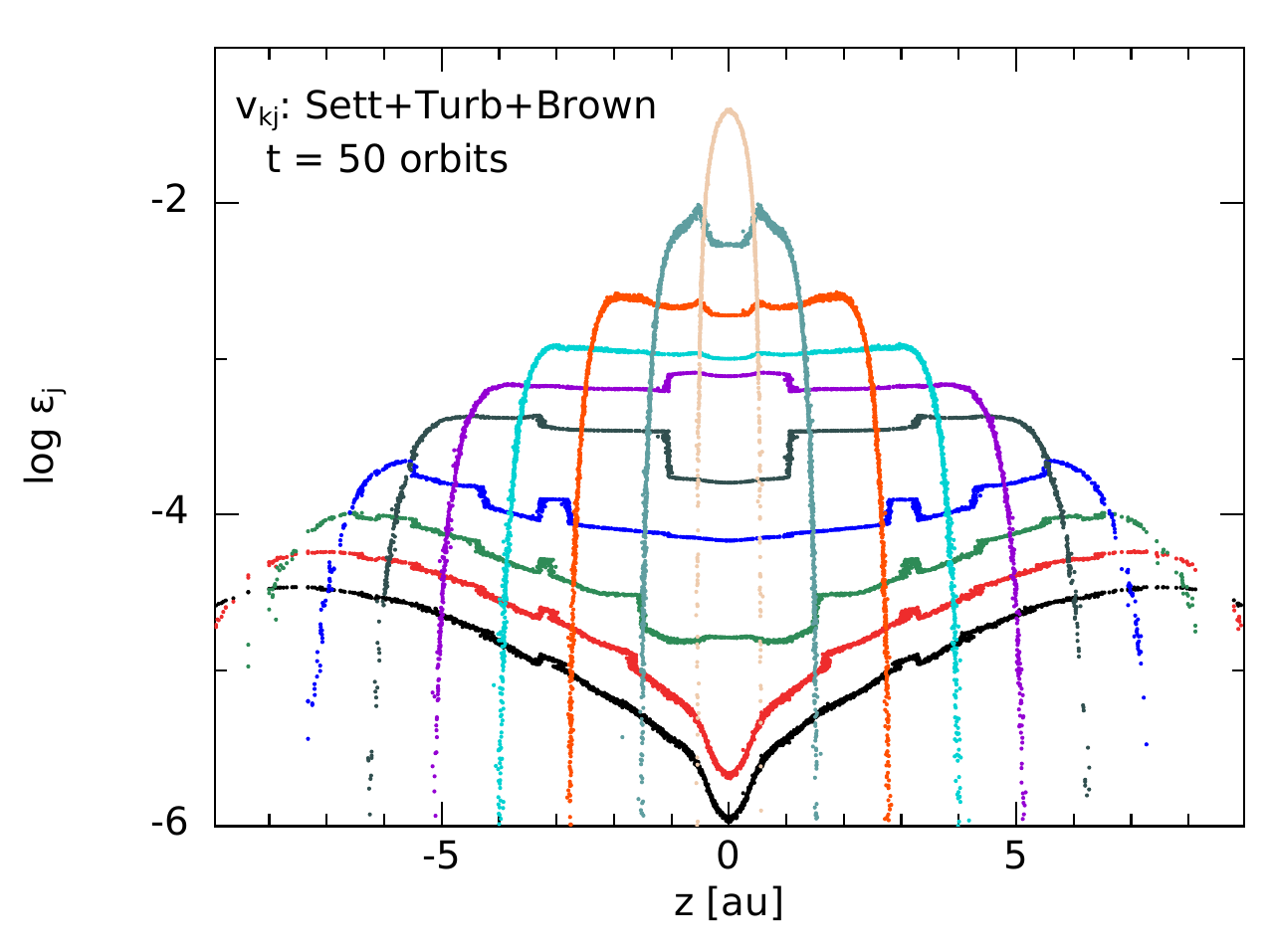}
\caption{The vertical distributions of the dust fractions of the 10 dust size bins after 15 orbits (left column) or 50 orbits (right column; note the change of vertical scale) of the dust settling test but also including grain growth.  The different grain size bins are ordered from 1~mm to 0.1~$\mu$m  from top to bottom at $z=0$ (the mid-plane).  The top panels only include Brownian motions when calculating the relative grain velocities, $v_{kj}$, that are required for grain growth.  The centre panels include both Brownian and turbulent motions (with $\alpha_{\rm SS}=10^{-3}$).  The lower panels include Brownian and turbulent motions, and the vertical (settling) velocity differences between grains of different sizes, as computed by the {\sc multigrain} method (i.e., in the terminal velocity approximation).  As is the case without grain growth the largest grains quickly settle towards the mid-plane, but with grain growth the smallest grains are depleted.  The steps in the abundances of small grains in the centre and lower panels are due to the coarse discretisation with the small number of dust size bins.  This means there are large jumps in both the Stokes numbers and vertical velocities between different bins.  Using more bins results in smoother distributions. }
\label{fig:settle3}
\end{figure*}

From this point on we only use the implicit method for modelling the dust drag.  Fig.~\ref{fig:settle3} shows the results of the dust settling test when we allow dust grain growth.  We present results from three separate calculations (top row to bottom row) and at two times (left: after 15 orbits and right: after 50 orbits).  In the top row, the relative grain velocities, $v_{kj}$, that are important for grain growth only include the contribution from Brownian motion.  Apart from the smallest grains (0.1~$\mu$m; black points) the top left panel of Fig.~\ref{fig:settle3} is almost identical to the panels in Fig.~\ref{fig:settle1}.  Over this very short timescale (15 orbits, $\approx 1050$~yrs), only the dust fractions of the smallest grains are significantly affected by coagulation because Brownian motion is only important for very small dust grains.  Even then, significant coagulation only occurs where the dust grain number density is the highest (i.e., near the mid-plane).  In the right panel, after 50 orbits, the large dust grains have settled into thin layers near the mid-plane of the disc.  Without anything to stir the dust, this settling would continue indefinitely.  

In the second row of Fig.~\ref{fig:settle3}, the relative grain velocities include contributions from Brownian motion and turbulence, taking $\alpha_{\rm SS}=0.001$ as described in Section \ref{sec:relative}.  The turbulent contributions to the relative grain velocities are much more important than Brownian motion for driving grain coagulation and the small grains become highly depleted, particularly near the mid-plane.  Nevertheless, the large grains (sizes $\gsim 46~\mu$m) do not undergo significant coagulation, at least for this choice of parameters.  The `steps' in the vertical distributions of the small and intermediate grains that are visible in the second and third rows of Fig.~\ref{fig:settle3} are due to the coarse dust size discretisation.
Only 10 bins are used to model grains ranging in size from 0.1~$\mu$m to 1~mm (there is a factor of $\approx 2.78$ increase in grain size from one bin to the next).  This results in large jumps in both the Stokes numbers (used when computing the turbulent contribution to the relative grain velocities) and vertical settling velocities between different bins.  The solutions become smoother if more bins are used.

In the bottom row of Fig.~\ref{fig:settle3}, the relative grain velocities also include the bulk relative velocities provided by the {\sc multigrain} formation, which in this case are the terminal velocities of the grains settling towards the mid-plane.  Comparing the second and third rows, there are small additional effects visible from the settling velocities, particularly for the small grains at 15 orbits.  For example, there is less depletion of the small grains right at the mid-plane ($z=0$) compared to $\approx 1-2$ au away from the mid-plane.  This is because the vertical component of the gravity tends to zero at the mid-plane and so the settling terminal velocities also tend to zero. Nevertheless, it is clear that the dominant contribution to the relative grain velocities that drives the coagulation of the small and intermediate sized dust grains comes from the turbulence, at least for this choice of $\alpha_{\rm SS}$.  The small grains have relatively large settling velocities far from the mid-plane where the gas density is low, but the rate of coagulation is low there due to the low grain number densities.  The larger grains rapidly settle toward the mid-plane, but they do not grow significantly. Presumably the settling velocities would play a much greater role if much lower values of $\alpha_{\rm SS}$ were used.

\subsection{Adding dust diffusion due to turbulence}
\label{sec:diffusion}

As discussed in Section \ref{sec:stirring}, the dust dynamics present in the tests presented in Section \ref{sec:settle1} result in dust layers that become thinner and thinner with time as the dust settles toward the disc mid-plane.  This is clear in Fig.~\ref{fig:settle3}.  However, such evolution is inconsistent with our assumption that gas turbulence provides a source of relative grain velocities that aid in grain growth -- such turbulence should also mix the dust and prevent it from forming very thin dust layers.

In this section we present that results of tests that include the effects of dust diffusion due to gas turbulence (i.e., using the method described in Section \ref{sec:stirring}).  Neglecting radial evolution and dust growth, a vertical column of disc material in which the dust is subject both to dust settling due to gravity and dust diffusion due to turbulence will eventually settle to a steady state vertical dust distribution.  The vertical evolution of single species of dust in such a vertical column of disc can be expressed \citep*{DubMorSte1995,SchHen2004,FroNel2009} as
\begin{equation}
\frac{\partial \rho_{\rm d}}{\partial t} - \frac{\partial}{\partial z} \left( z \Omega^2 T_{\rm s} \rho_{\rm d} \right) = \frac{\partial}{\partial z}  \left[ D \rho \frac{\partial}{\partial z} \left( \frac{\rho_{\rm d}}{\rho} \right) \right],
\end{equation}
where $\rho_{d} = \varepsilon \rho$ is the dust particle density.  For the steady state vertical profile, the time derivative vanishes and the left and right sides can be integrated to give
\begin{equation}
\frac{\partial}{\partial z} \left( \ln \varepsilon \right)= - \frac{ \Omega^2 T_{\rm s}}{ D} z.
\label{eq:steadystate}
\end{equation}
To integrate further requires knowledge of how the stopping time, $T_{\rm s}$, and $D$ depend on $z$.  However, if we take the vertical gas density profile to be a Gaussian (i.e., a vertically-isothermal disc), then $T_{\rm s} = T_{{\rm s},0} \exp[z^2/(2H^2)]$ where $T_{{\rm s},0}$ is the dust stopping time at the mid-plane, since $T_{\rm s} \propto 1/\rho_{\rm g}$ in the Epstein regime (equation \ref{eq:Epstein}).  Then, taking $D$ to be a constant, we can solve equation \ref{eq:steadystate} analytically to give
\begin{equation}
\varepsilon = \varepsilon_0 \exp \left( - \frac{ \Omega^2 H^2 T_{{\rm s},0}}{ D} \left[ \exp\left( \frac{z^2}{2H^2} \right) - 1 \right] \right),
\label{eq:analytic1}
\end{equation}
where $\varepsilon_0$ is the dust fraction at the mid-plane.  Noting that $\Omega=c_{\rm s}/H$ and expressing $D= \widetilde{D} c_{\rm s}H$, this can be further simplified to
\begin{equation}
\varepsilon = \varepsilon_0 \exp \left( - \frac{ \Omega T_{{\rm s},0}}{ \widetilde{D}} \left[ \exp\left( \frac{z^2}{2H^2} \right) - 1 \right] \right).
\label{eq:analytic2}
\end{equation}
In taking $\widetilde{D}$ to be a constant we ignore the dependence on the Schmidt number (Section \ref{sec:stirring}).  We can further assume $\widetilde{D}=\alpha_{\rm SS}$, which we do in most of the tests that follow.  However, we do allow the quantities $\widetilde{D}$ and $\alpha_{\rm SS}$ to differ in the code because it is not necessarily the case that there is a one-to-one correspondence between the turbulent diffusion of dust and other effects of hydrodynamic turbulence.  For example, \cite{FroNel2009} studied the turbulent diffusion of dust particles in magnetohydrodynamic (MHD) turbulent discs and obtained vertical dust distributions that were not well modelled by a constant vertical diffusion coefficient (see the end of Section \ref{sec:roles}).

This steady-state analytic solution provides an ideal test case for the combination of dust settling and dust turbulent diffusion with which to test the SPH code.  Using the same initial conditions that were specified in Section \ref{sec:settle0} (i.e., modelling a vertical column of an isothermal disc) we run several calculations until they reach steady-state vertical dust profiles.  For this set of initial conditions
\begin{equation}
\frac{\Omega T_{{\rm s},0}}{\widetilde{D}} \approx \frac{8.294\times 10^{-2}}{\alpha_{\rm SS}} \left( \frac{a}{{\rm 1~cm}} \right) 
\label{eq:dimensionless}
\end{equation}
where the dust grain size is specified in cgs units.  In what follows, we compare the steady-state vertical dust fraction profiles with the analytic solution given by equation \ref{eq:analytic2}.  The unknown quantity $\varepsilon_0$ in equation \ref{eq:analytic2} can be determined by noting that the total dust mass in the column (with $x,y$ dimensions of $\Delta x$ and $\Delta y$, respectively) is given by 
\begin{equation}
M_{\rm d, col}  = \bar{\varepsilon} M_{\rm col} =  \Delta x \Delta y \int_{-\infty}^{\infty} \rho \varepsilon ~ {\rm d}z,
\end{equation}
where the initial dust fraction, $\bar{\varepsilon}$, and the total mass in the column, $M_{\rm col}$, are set by the initial conditions of the SPH simulation.
Making the approximation that $\rho \approx \rho_{\rm g}/(1-\bar{\varepsilon})$ where we again assume that the vertical gas density profile is Gaussian, $\rho_{\rm g}=\rho_{\rm g,0} \exp(-z^2/(2H^2))$ (which are good approximations for sufficiently low dust fractions), this allows us to set $\varepsilon_0$ by numerically integrating
\begin{equation}
\frac{1}{\varepsilon_0} \approx \frac{ \Delta x \Delta y}{M_{\rm d, col} } \int_{-\infty}^{\infty} \rho_{0} \exp \left( - \frac{ \Omega T_{{\rm s},0}}{ \widetilde{D}} \left[ \exp\left( \frac{z^2}{2H^2} \right) - 1 \right] - \frac{z^2}{2H^2}  \right) {\rm d}z,
\label{eq:epsilon0}
\end{equation}
where $\rho_0=\rho_{g,0}/(1-\bar{\varepsilon})$ is the total mass density at the mid-plane.

\begin{figure}
\centering \vspace{-0.25cm} \vspace{-0.0cm}
    \includegraphics[height=6.5cm]{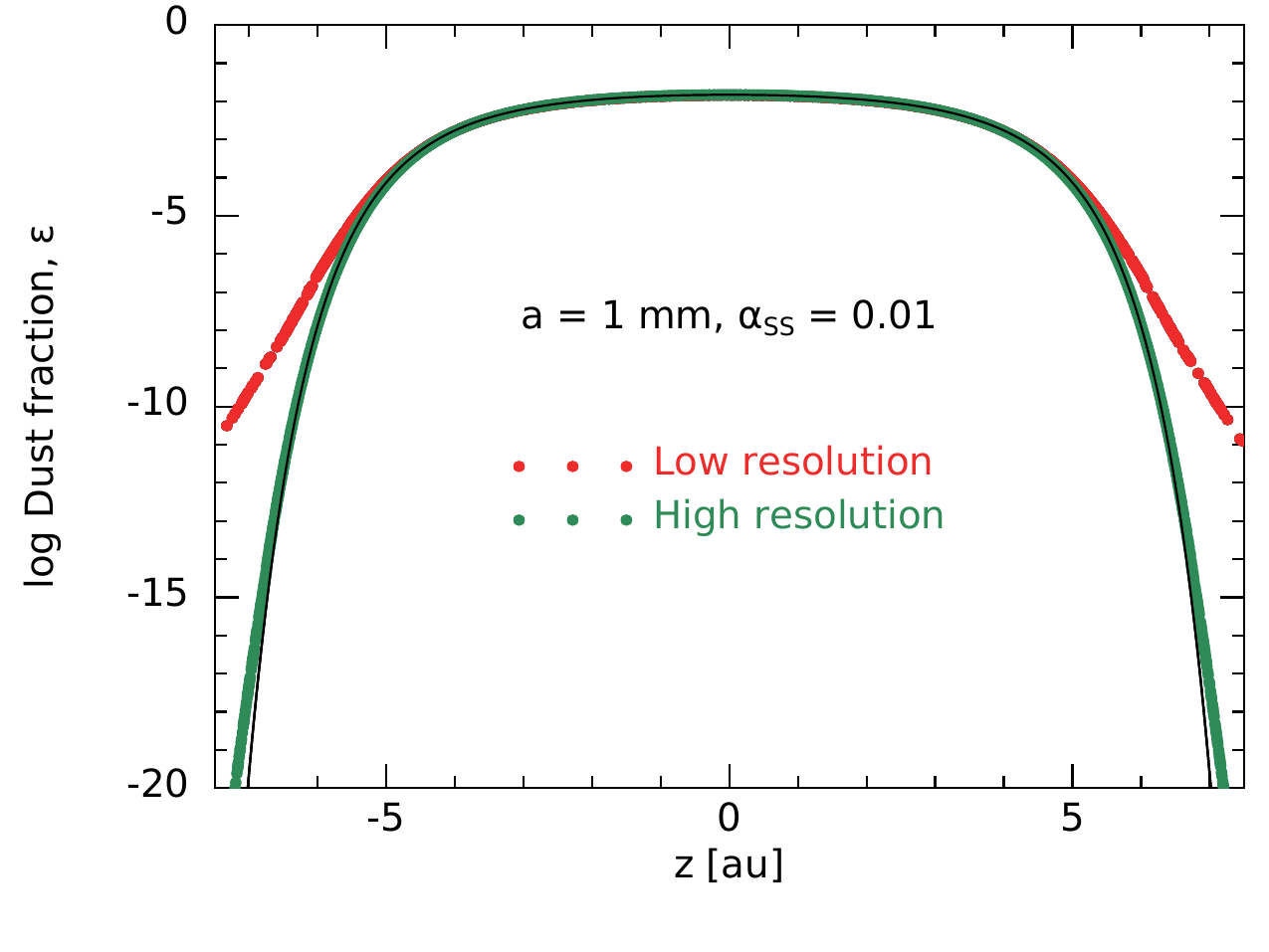}
    \vspace{-1cm}
\caption{The vertical distribution of the dust fraction of 1 mm dust grains from calculations with both dust settling and turbulent stirring after the equilibrium distribution has been attained.  The solid black line gives the analytic solution (equations \ref{eq:analytic2} and \ref{eq:epsilon0}).  The red points are from an implicit SPH calculation at our standard particle resolution (25942 particles in $[x,y]=[\pm 0.2,\pm 0.15]$; after 50 orbits).  The green points are from an equivalent calculation with 10 times the linear spatial resolution (68080 SPH particles in $[x,y]=[\pm 0.01,\pm 0.0075]$; after 50 orbits). Both simulations match the analytic solution well at high dust fractions near the mid-plane, but the rate at which the dust fraction drops off at large distances from the mid-plane is under-estimated at lower resolution. }
\label{fig:settleTurbEq}
\end{figure}

\begin{figure}
\centering \vspace{-0.25cm} \vspace{-0.0cm}
    \includegraphics[height=6.5cm]{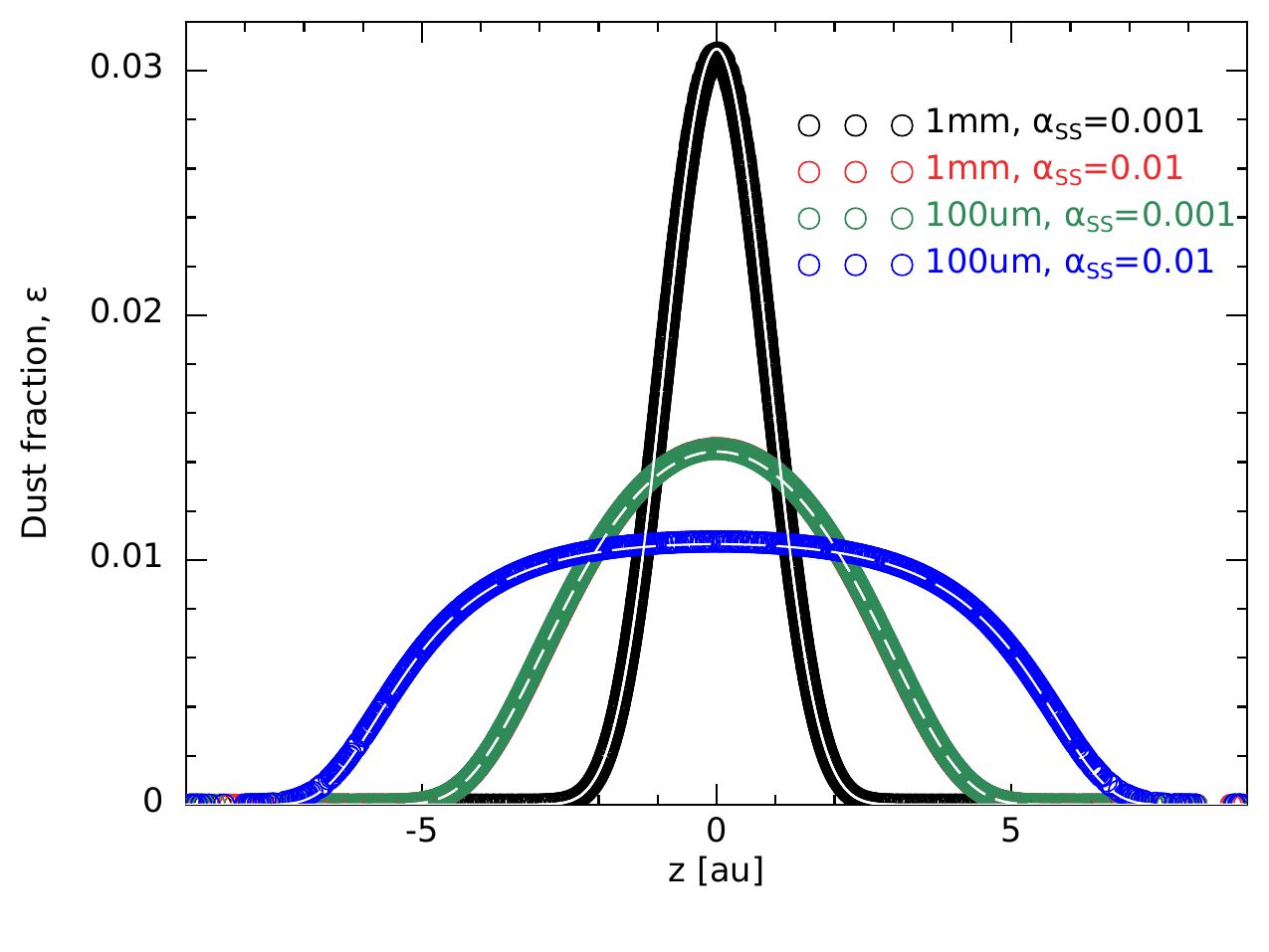}
    \vspace{-1cm}
\caption{The vertical distribution of the dust fraction from SPH calculations (standard resolution) with both dust settling and turbulent stirring after the equilibrium distribution has been attained.  Four calculations are performed, using 1 mm or 100 $\mu$m grains each with $\alpha_{\rm SS}=0.001$ or $0.01$ (see the legend).  The solutions for the 1 mm grains with $\alpha_{\rm SS}=0.01$ and 100 $\mu$m grains with $\alpha_{\rm SS}=0.001$ lie on top of each other, as they should since they have the same value of the dimensionless quantity in equation \ref{eq:dimensionless}.  The white lines give the analytic solutions (equations \ref{eq:analytic2} and \ref{eq:epsilon0}).  Each of the simulations match the analytic solutions well.  Cases with larger dust grains and/or less turbulence produce thinner dust layers.}
\label{fig:settleTurbEq2}
\end{figure}

\subsubsection{Testing the SPH equilibrium settling/diffusion solution}

To test that the SPH code correctly recovers the equilibrium dust settling / turbulent diffusion solution, we perform calculations with a single dust species and a uniform dust-to-gas ratio of 1/100.  We run the SPH calculations until the vertical dust density profile stops evolving with time.  The time this takes depends on the dust particle size (i.e., the stopping time) and the strength of the turbulence (i.e., the value of $\alpha_{\rm SS}$).  

In Fig.~\ref{fig:settleTurbEq} we present the results from two SPH calculations performed with $\widetilde{D}=\alpha_{\rm SS}=10^{-2}$ with a grain size $a=1$mm.  For these conditions the SPH solution reaches an equilibrium in less than 20 orbits and we show the solution at 50 orbits. The two calculations are identical except for the linear spatial resolution which is 10 times higher in one simulation than the other (this is achieved by altering the cross section of the fluid column (i.e., the $x,y$ dimensions) as well as increasing the number of SPH particles; see the figure caption for details).  The vertical distribution of the dust fractions are shown using red (low resolution) and green (high resolution) points.  The analytic solution (equations \ref{eq:analytic2} and \ref{eq:epsilon0}) is overplotted with the solid black line.  The SPH solutions match the analytic solution very well near the mid-plane (where the dust fraction is significant; note that the vertical scale plots $\log_{10}\varepsilon$).  A long way from the mid-plane, where the turbulent stirring is ineffective at maintaining a significant dust fraction due to the low gas density and the resulting low dust/gas drag ($ z \gsim 5$~au $\approx 2 H$) the SPH solutions tend to over-estimate the dust fraction, but the SPH solutions converge toward the analytic solution as the resolution is increased (compare the green points and the black line).  Note also that, even with the lower resolution, significant divergence only occurs at dust fractions less than $\varepsilon < 10^{-5}$.

In Fig.~\ref{fig:settleTurbEq2} we present the results from four SPH calculations performed with $\widetilde{D}=\alpha_{\rm SS}=10^{-2}$ or $10^{-3}$, each with grain sizes of $a=1$mm or 100~$\mu$m.  The analytic solutions (equations \ref{eq:analytic2} and \ref{eq:epsilon0}) are overplotted with the white lines (solid or dashed).  This figure uses a linear vertical scale to illustrate the good agreement of the dust fractions with the analytic solutions near the disc mid-plane.  Note that the profiles for 100~$\mu$m with $\alpha_{\rm SS}=10^{-3}$ and for 1~mm with $\alpha_{\rm SS}=10^{-2}$ coincide as expected: the stronger settling of the larger grains is exactly offset by the increased turbulent stirring from the higher value of $\alpha_{\rm SS}$.

\begin{figure}
\centering \vspace{-0.25cm} \vspace{-0.0cm}
    \includegraphics[height=6.5cm]{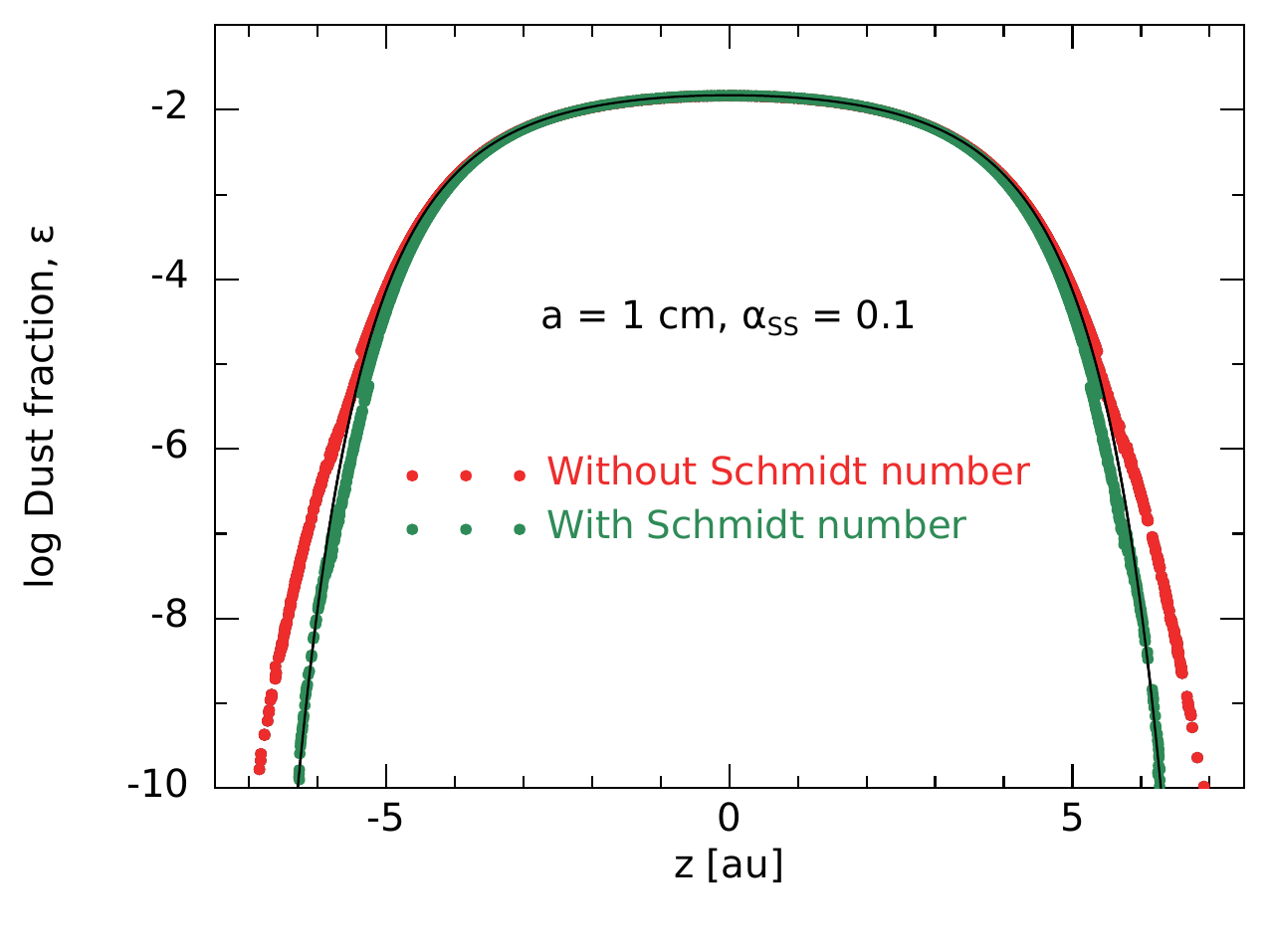}
    \vspace{-1cm}
\caption{The vertical distribution of the dust fraction of 1 cm dust grains from calculations with both dust settling and turbulent stirring after the equilibrium distribution has been attained (10 orbits).  The red points are from a calculation in which the turbulent diffusion, $D$, is constant.  The green points are from an identical calculation but with turbulent diffusion including the Schmidt number.  Both calculations were performed at the standard resolution.  The solid black line gives the analytic solution for a constant turbulent diffusion (equations \ref{eq:analytic2} and \ref{eq:epsilon0}).  Including the Schmidt number reduces the turbulent diffusion substantially when the Stokes number exceeds unity, resulting in less dust transport into the low density regions well away from the disc mid-plane, but relatively large grain sizes are necessary to observe a significant effect. }
\label{fig:settleTurbEqSchmidt}
\end{figure}

\subsubsection{Including the Schmidt number}

In reality, $D$ or $\widetilde{D}$ are not constants.  Even if the disc is vertically isothermal (i.e., $c_{\rm s}$ and $H$ are constant with height), the effectiveness of the turbulence in stirring dust particles depends on how well they are coupled to the gas.  The coupling between dust and gas weakens both with distance from the mid-plane, due to lower gas density, and with increasing grain size.  To account for this, rather than setting $D$ to be a constant, we use equation \ref{eq:diffconst} in which the $D$ depends on the Schmidt number (which, in turn, is a function of the orbital frequency and the dust stopping time).  Although analytical solutions are now no longer possible, including the Schmidt number for the test cases in Figs.~\ref{fig:settleTurbEq} and \ref{fig:settleTurbEq2} yields results nearly identical to the originals, since the Stokes numbers are lower than unity throughout the vertical extent that is shown.

For our chosen gas initial conditions, to observe a noticeable effect requires performing calculations with larger dust grains, and with strong turbulence so that a significant dust fraction is maintained far enough from the disc mid-plane that the Stokes number is much greater than unity.  In Fig.~\ref{fig:settleTurbEqSchmidt} we present the results from two calculations using 1~cm dust grains and $\widetilde{D}=\alpha_{\rm SS}=0.1$ after 10 orbits, when the equilibrium distribution has been attained (the equilibrium is quickly attained with large dust grains and high turbulent stirring).  The two calculations are identical, except that the calculation plotted using the red points has a constant value of $D$, while the calculation plotted using the green points includes the Schmidt number dependence of the diffusion (equation \ref{eq:diffconst}). The solid black line again shows the analytic solution {\it with the constant value of} $D$ (i.e., without the Schmidt number).  It can be seen that the inclusion of the Schmidt number dependence in the diffusion has the expected effect on the simulations; the dust with high Stokes numbers (i.e., long stopping times, at low gas densities) far from the disc mid-plane has a reduced dust fraction (i.e., the green points lie below the red points).  However, with this resolution both of the simulations suffer from enhanced dust fractions far from the mid-plane (the red points should lie on the black line).

\subsection{Multigrain dust settling and turbulent diffusion, with and without grain growth}

We now report results from simulations with multiple grain sizes that settle (similar to those presented in Section \ref{sec:settle1}) but that also include turbulent diffusion/stirring of the dust (including the Schmidt number).  We begin with calculations without dust growth, and then include dust growth as well.

\begin{figure*}
\centering \vspace{-0.25cm} \vspace{-0.0cm}
    \includegraphics[height=6.0cm]{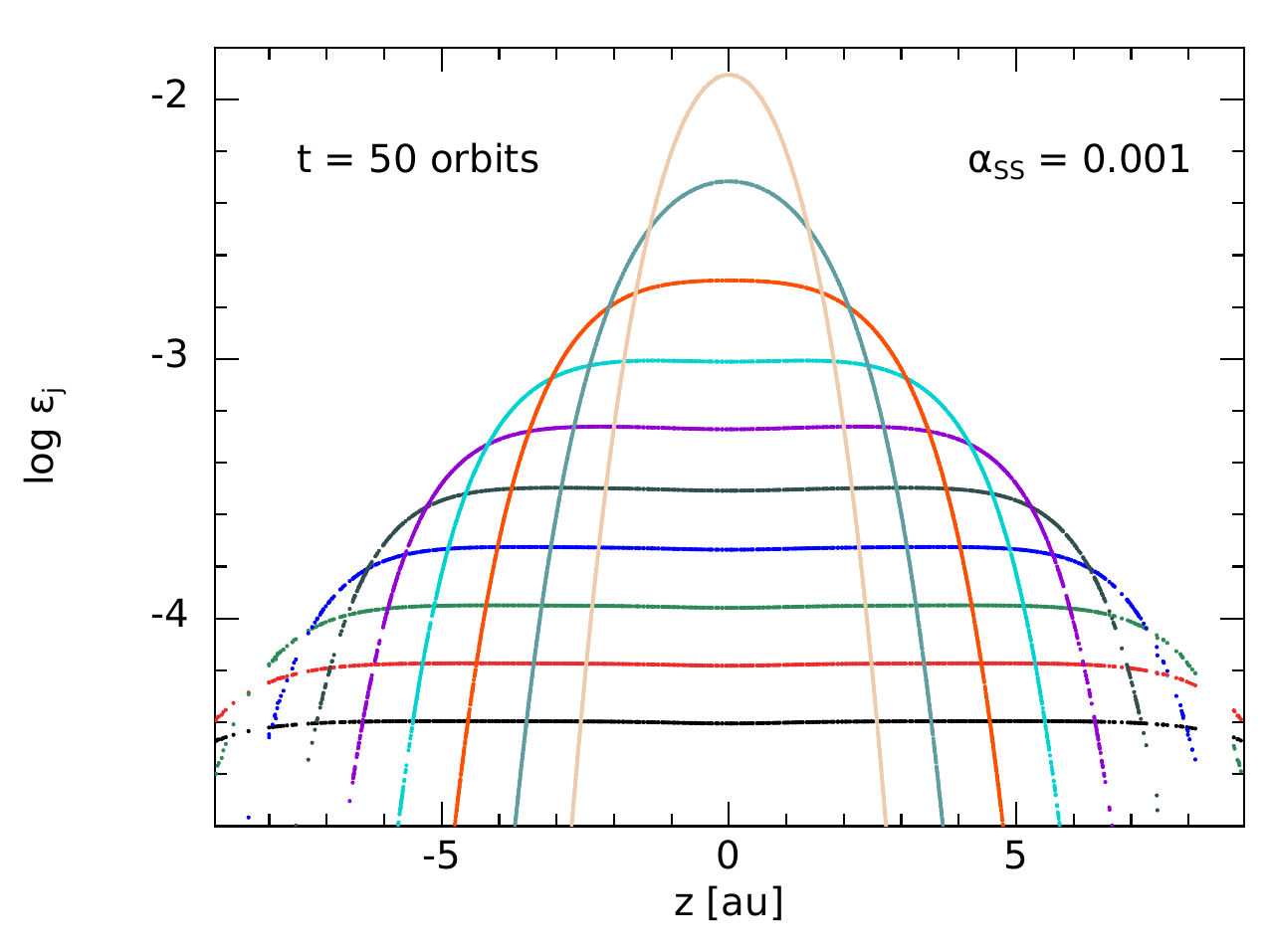}
    \includegraphics[height=6.0cm]{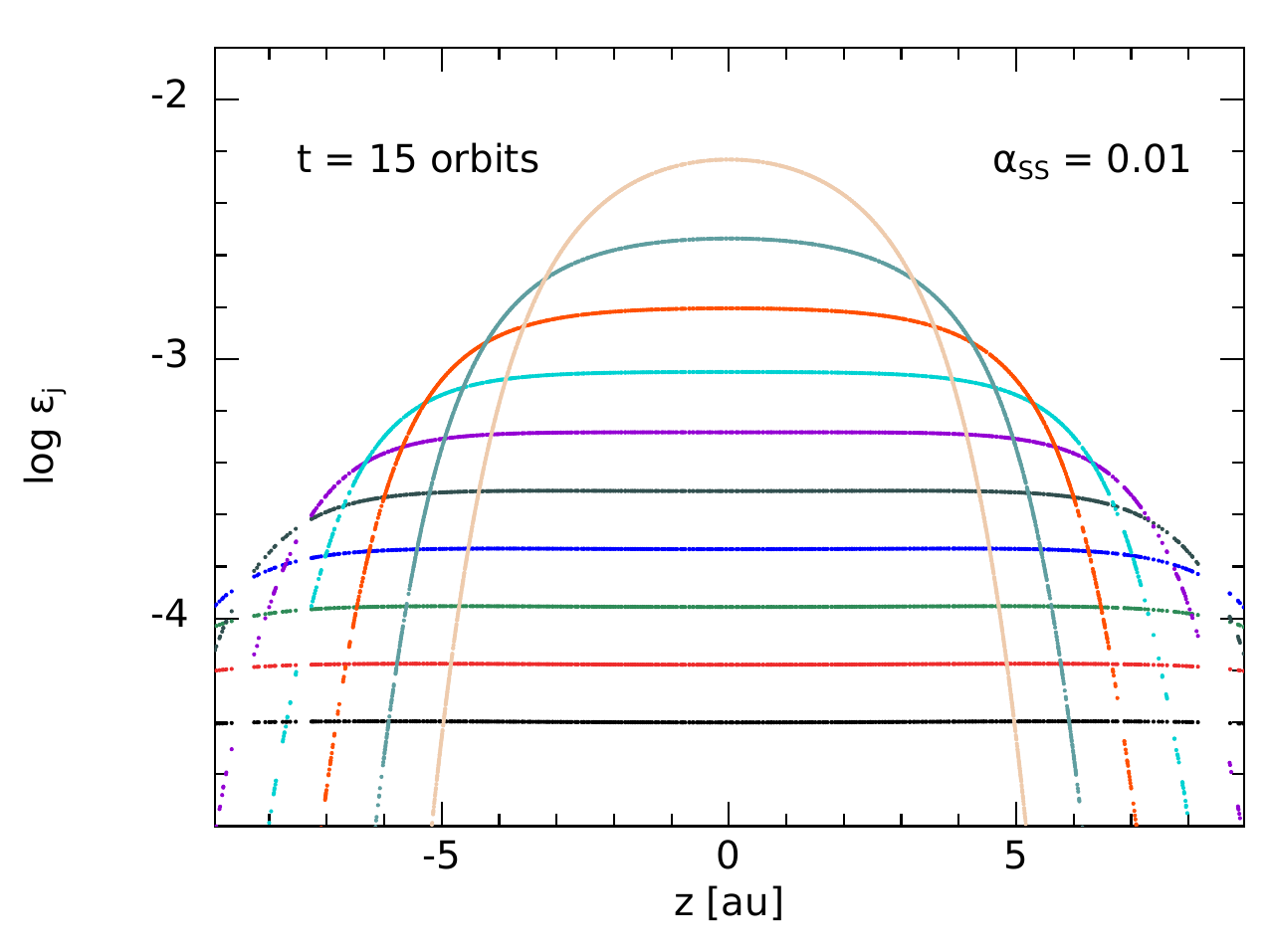}
\vspace{-0.25cm}
\caption{The vertical distributions of the dust fractions of the 10 dust size bins in the dust settling test including both dust dynamics and dust turbulent diffusion, but without any grain growth.  These simulations reach equilibrium states (shown in the two panels) in which the vertical settling of the dust towards the mid-plane is balance by vertical diffusion away from the mid-plane.  A shorter time is taken to reach the equilibrium state with a higher level of turbulence, parameterised using $\alpha_{\rm SS}$ (equilibrium distributions are established after $\gsim 30$ orbits with $\alpha_{\rm SS}=0.001$, but only $\approx 15$ orbits are required with $\alpha_{\rm SS}=0.01$).  When turbulent diffusion is included the dust grains do not settle as close to the mid-plane as they do in the absence of turbulent diffusion (compare with Fig. \ref{fig:settle1} and the top panels of Fig. \ref{fig:settle3}), and stronger turbulence results in thicker dust layers (compare the right and left panels).}
\label{fig:diff1}
\end{figure*}

Fig.~\ref{fig:diff1} shows the results from {\sc multigrain} calculations using the same 10 dust size bins as in Section \ref{sec:settle1}.  Rather than the grains settling closer and closer to the disc mid-plane with increasing time (Figs.~\ref{fig:settle1} and \ref{fig:settle3}), each grain size forms an equilibrium dust layer with thicker layers for smaller grains, or for stronger turbulence.  Fig.~\ref{fig:diff1} shows results for two levels of turbulence: $\alpha_{\rm SS} = 0.001$ (left panel) and $\alpha_{\rm SS} = 0.01$ (right panel).  The calculation with stronger turbulence only requires $\approx 15$ orbits to establish the equilibrium profiles, while the calculation with less turbulence takes a bit longer (at least $30$ orbits; we show the result at 50 orbits).

\begin{figure*}
\centering \vspace{-0.4cm} \vspace{-0.0cm}
    \includegraphics[height=6.0cm]{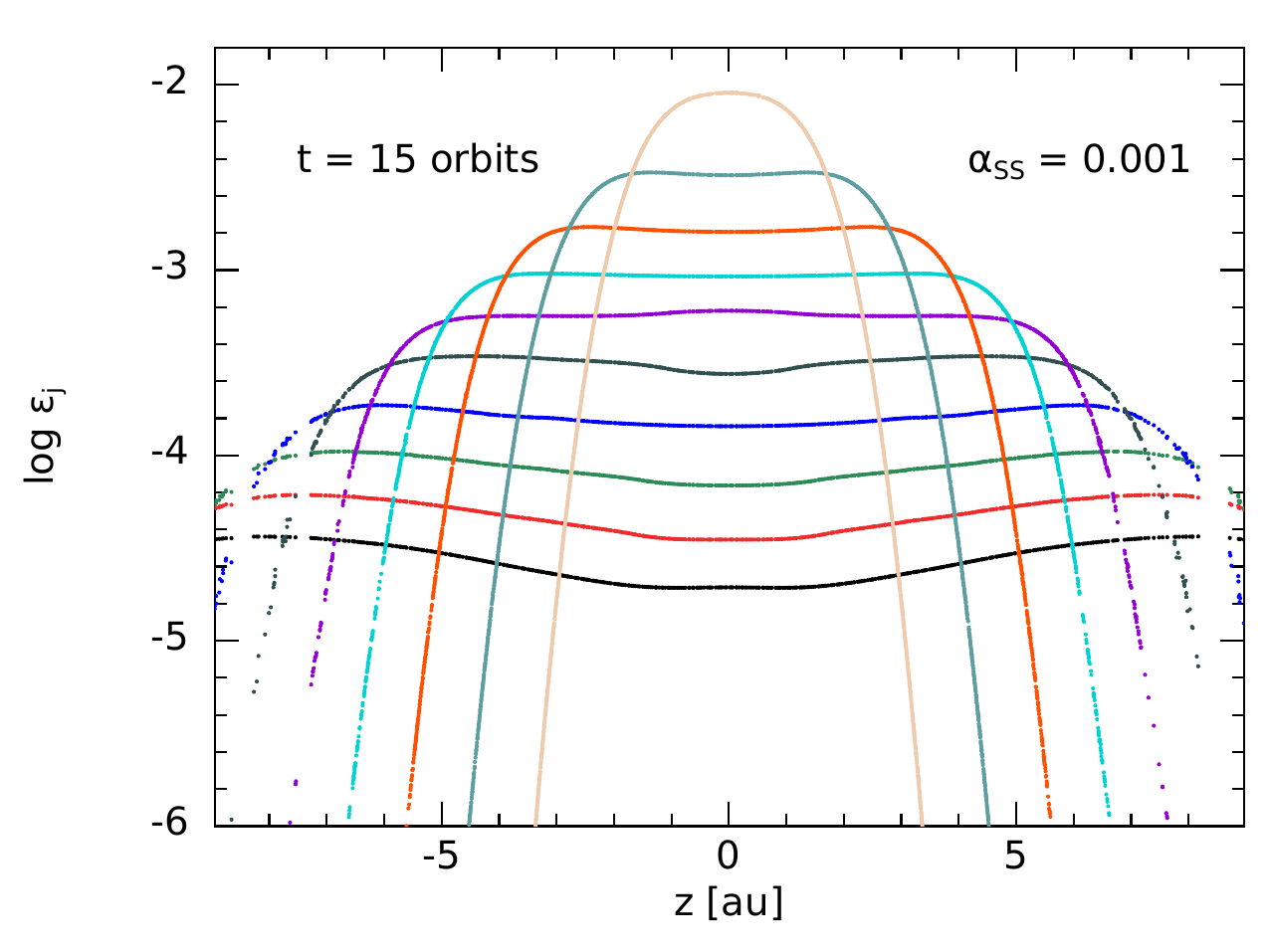}
    \includegraphics[height=6.0cm]{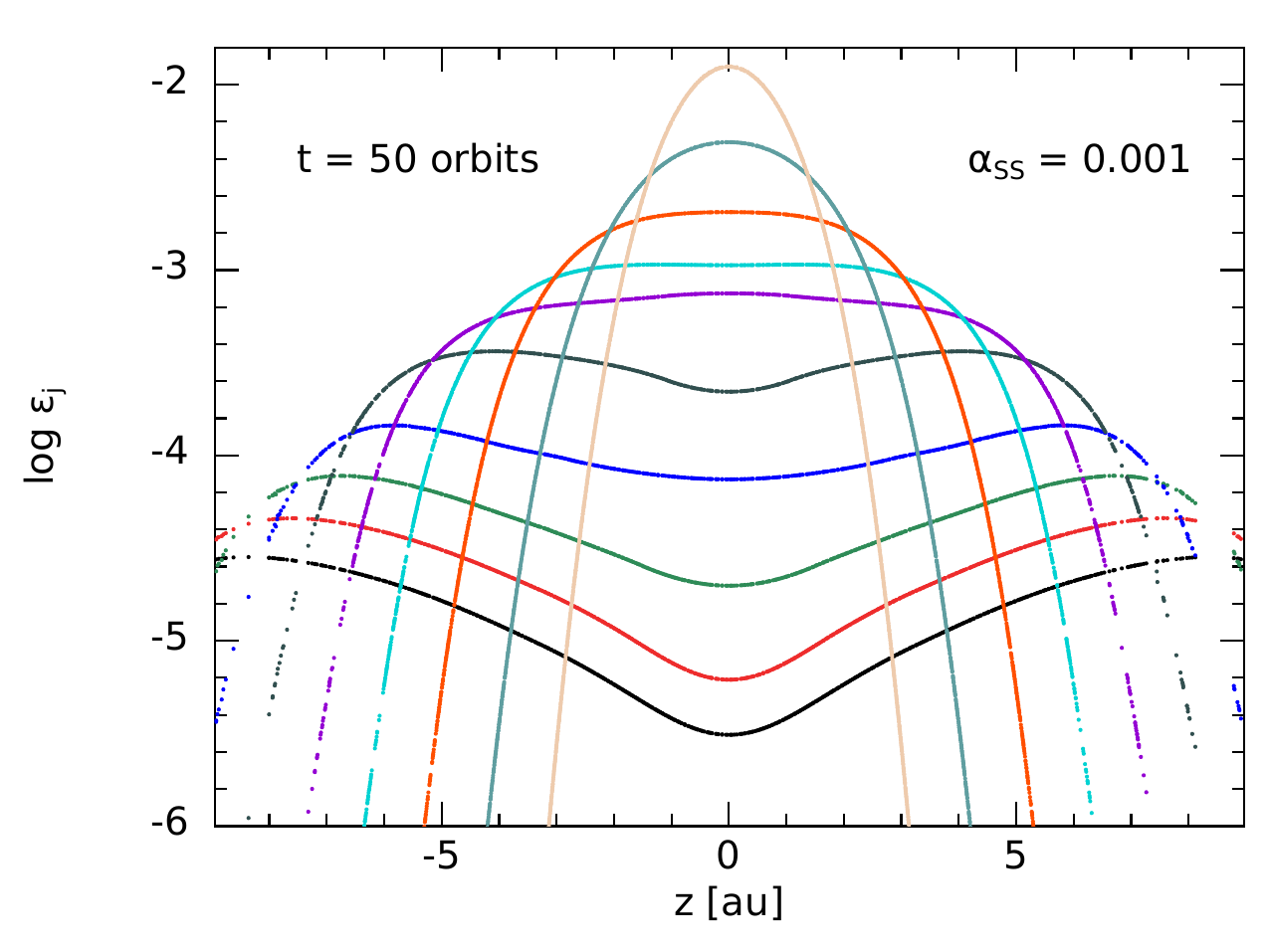}
    \includegraphics[height=6.0cm]{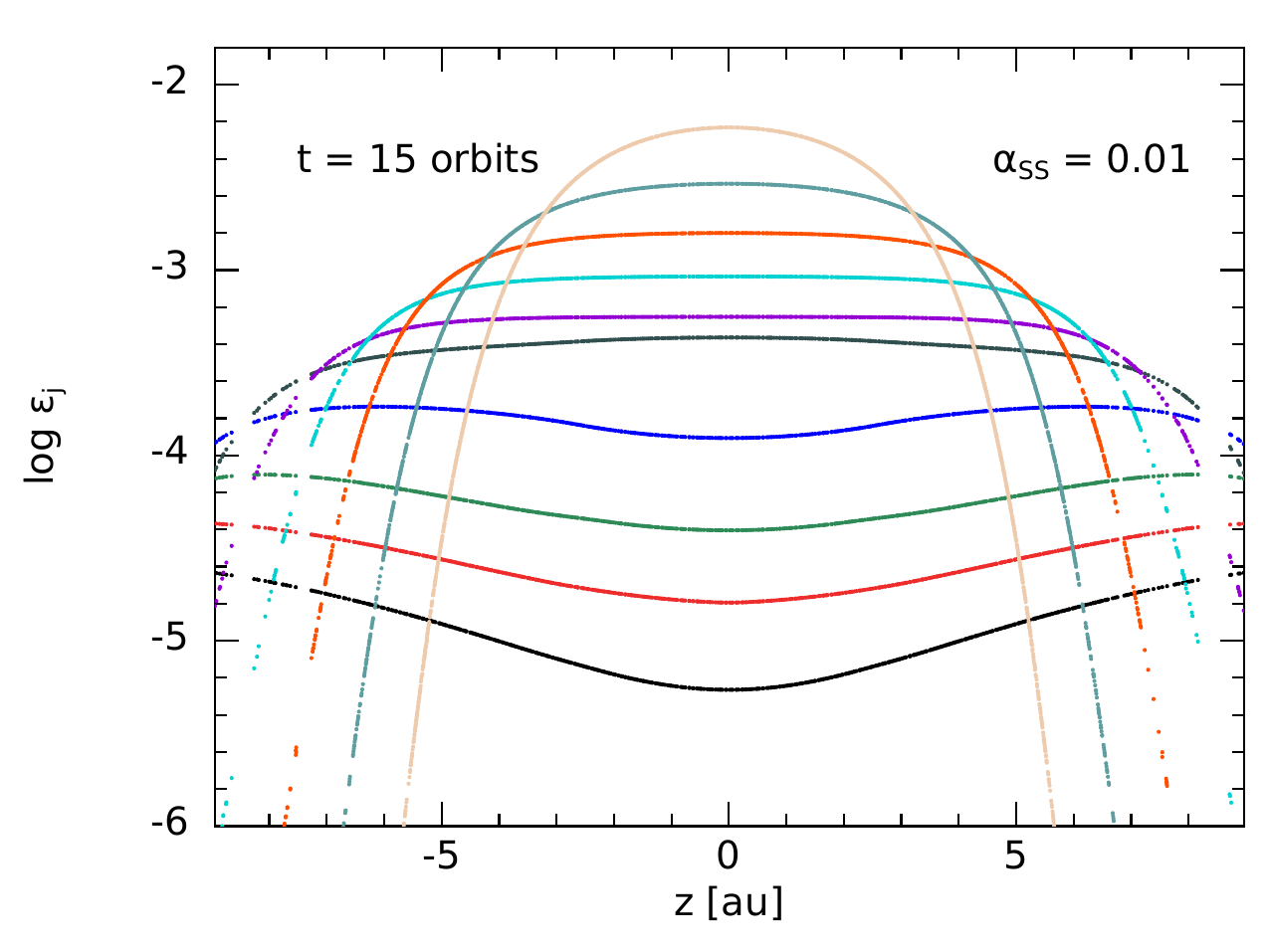}
    \includegraphics[height=6.0cm]{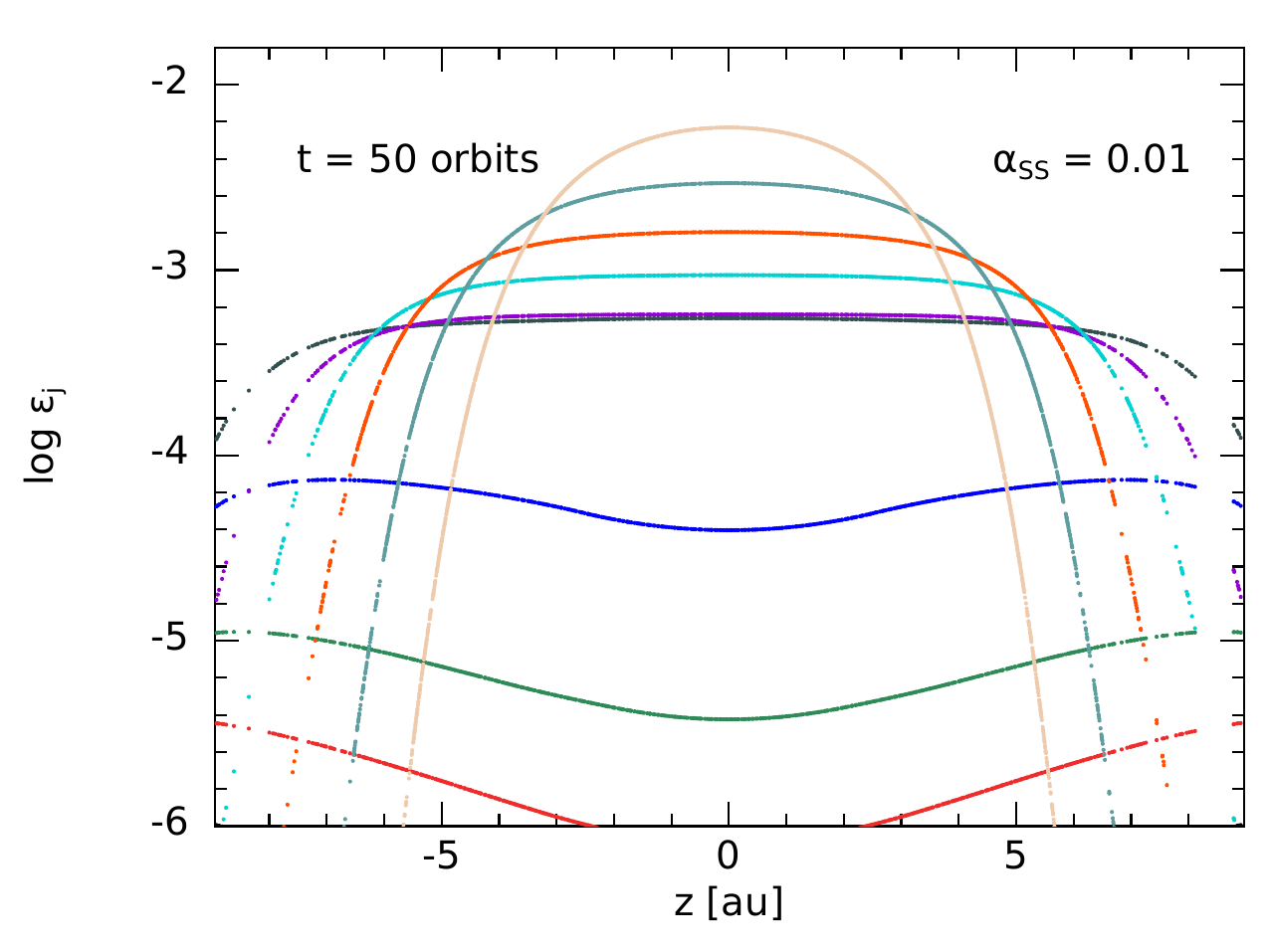}
\vspace{-0.25cm}
\caption{The vertical distributions of the dust fractions of the 10 dust size bins after 15 orbits (left column) or 50 orbits (right column) of the dust settling test but also including grain growth.  Results from two calculations, one with $\alpha_{\rm SS}=0.001$ (top row) and one with $\alpha_{\rm SS}=0.01$ (bottom row) are shown.  The same values of  $\alpha_{\rm SS}$ are used to compute both the relative grain velocities (for grain growth) and the turbulent diffusion (for vertical mixing).  The different grain size bins are ordered from 1~mm to 0.1~$\mu$m  from top to bottom at $z=0$ (the mid-plane).  The relative grain velocities on which the grain growth depends include Brownian and turbulent motions, and the bulk drift (in this case, vertical settling) velocity differences between grains of different sizes, as computed by the {\sc multigrain} method (i.e., in the terminal velocity approximation).  As is the case without grain growth the largest grains quickly settle towards the mid-plane (compare with Fig.~\ref{fig:diff1}), but with grain growth the smallest grains become progressively more depleted with time (i.e., there is no equilibrium state).  By the end of the simulation, the smallest grains have their greatest dust fractions far from the mid-plane.  With a lower level of turbulence (top row), the small grains coagulate more slowly than with greater turbulence (bottom row), but with less turbulence there also tends to be a greater difference in the small grain dust fractions with height.
}
\label{fig:diff2}
\end{figure*}

Next we performed simulations combining dust settling, turbulent stirring, and grain growth (Fig.~\ref{fig:diff2}).  These simulations never achieve an equilibrium because the grains keep growing.  We show two cases in Fig.~\ref{fig:diff2} -- the top row has the level of turbulence $\alpha_{\rm SS} = 0.001$, while the bottom row has stronger turbulent diffusion ($\alpha_{\rm SS} = 0.01$).  Note that the level of turbulence affects both the turbulent diffusion of dust grains and also the grain growth (because it affects the relative grain velocities; see Section \ref{sec:relative}).

In the top row of Fig.~\ref{fig:diff2}, the level of turbulence is set at $\alpha_{\rm SS} = 0.001$, which is the same as was assumed when calculating the relative grain velocities and dust growth in the centre and lower panels of Fig.~\ref{fig:settle3}.  Comparing the results from Fig.~\ref{fig:settle3} with the row panels of Fig.~\ref{fig:diff2}, we see that because of the turbulent diffusion of grains within the disc the small grains are depleted near the disc mid-plane more slowly than without turbulent diffusion and the large grains do not settle as quickly towards the mid-plane.  The vertical profiles are also much smoother with height in Fig.~\ref{fig:diff2} compared to Fig.~\ref{fig:settle3}, both because the edges of the dust layers are not as `sharp' when including dust stirring, and also because of the diffusion of the dust.

In the lower row of Fig.~\ref{fig:diff2}, the level of the turbulence is set at $\alpha_{\rm SS} = 0.01$, both for the turbulent diffusion and for the relative dust grain velocities that are important for dust growth.  Compared to the upper panels (with $\alpha_{\rm SS} = 0.001$), the dust layers are much thicker, but the higher relative grain velocities also lead to very different grain growth.  The small grains coagulate much more rapidly and are depleted much more quickly with the higher level of turbulence throughout the vertical extent of the disc.  The is due to both the higher rate of coagulation and because grains near the surface layers of the disc are being more rapidly diffused toward the mid-plane.  By contrast the larger grains do not grow effectively because with the higher level of turbulence the collisions result in bouncing rather than coagulating.  This leads to a 'pile up' with the dust fractions of bins 5 and 6 being almost identical to each other after 50 orbits (lower right panel of Fig.~\ref{fig:diff2}) because the number density of grains in bin 5 is increased by the rapid growth of the small grains, but the grains in bins 6 -- 10 barely grow at all.  Essentially the dust initially contained in size bins 1 -- 5 coagulates and would become mono-disperse (i.e., with most of the dust in a narrow range of grain sizes), except for the fact that due to our choose initial conditions there is a substantial population of larger grains that don't grow.  This demonstrates that the combined effects of grain settling, diffusion, and growth can lead to complex grain size evolution.

\subsection{Grain growth and settling of an MRN grain population}
\label{sec:settle2}

The dust settling test of \cite{PriLai2015} (with 10 bins covering a large range of dust sizes) is a good test of the numerical code.  However, having a large fraction of the dust in the form of large grains suspended high in the `atmosphere' of the disc (i.e., far from the mid-plane) is not a realistic initial condition (as is clearly demonstrated by the rapid settling of these grains).  A more realistic set of initial conditions is to begin with a typical interstellar size distribution (i.e., small grains, as would be expected for a disc that is just forming from a collapsing molecular cloud core) and to let them grow and settle towards the mid-plane as they grow.  For this test, we retain the same set up as that in the previous section, except that we begin with a typical MRN  \citep{MatRumNor1977} dust grain size distribution with power-law index $p=3.5$ and with $a_{\rm min}=5$~nm and $a_{\rm cutoff}=0.25$~$\mu$m.  We use 53 grain size bins with logarithmic spacing from $a_{\rm min}$ to $a_{\rm max}=1$~mm (i.e., 10 bins per decade in dust grain size), or 63 bins with $a_{\rm max}=10$~mm for some cases with higher sticking velocities (see below).

In the calculations for this section, we include dust grain growth with relative grain velocities arising from all three sources: Brownian motion, turbulence, and the bulk relative velocities between grain sizes (i.e., vertical settling).  We perform one calculation without turbulent dust diffusion (i.e., assuming $\alpha_{\rm SS}=0.001$ for the relative grain velocities, but not adding turbulent dust diffusion).  In all the other calculations, we use the same value of $\alpha_{\rm SS}$ for both the dust growth and the dust diffusion.

\begin{figure}
\centering \vspace{-0.05cm} \vspace{-0.0cm}
    \includegraphics[height=6.0cm]{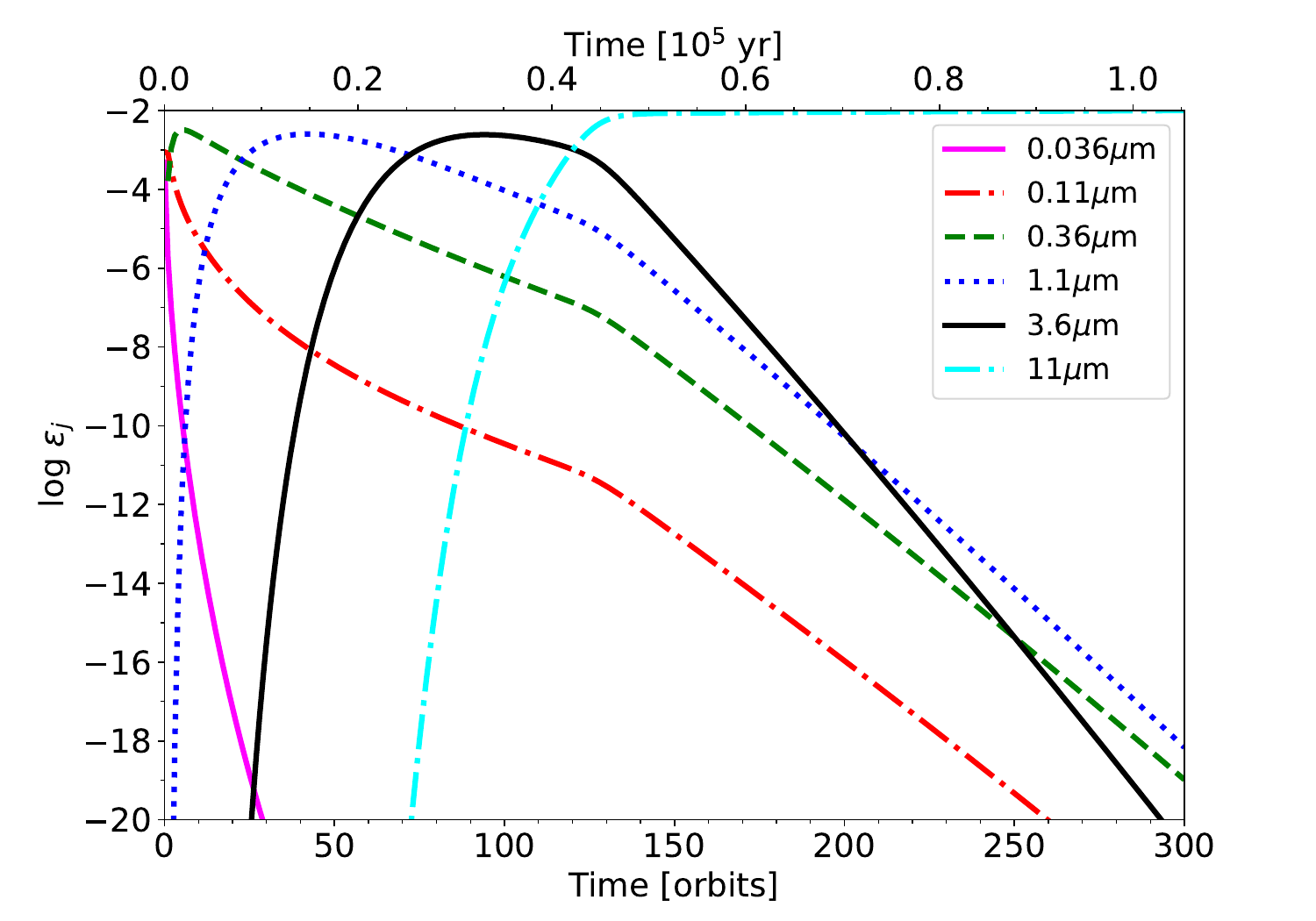}
    \vspace{-0.0cm}
\caption{The evolution with time of the dust fractions of grains with sizes 0.036, 0.11, 0.36, 1.1, 3.6, and 11 $\mu$m at the mid-plane of the disc when beginning with an MRN grain population (maximum initial size of 0.25 $\mu$m).  The smallest grains (e.g., 0.036, 0.11 $\mu$m) quickly coagulate with other grains.  Grains slightly larger than the initial maximum (0.36 $\mu$m) are produced rapidly initially, but later decline as they also evolve into even larger grains (e.g., 1.1, 3.6 $\mu$m).  The growth of large grains depends on the level of turbulence, and the sticking criteria.  Using our default coagulation threshold for grains to bounce rather than stick the growth stalls at a maximum grain size of $\approx 11$ $\mu$m. }
\label{fig:MRNgrow_and_settle_only_mid-plane}
\end{figure}

\begin{figure*}
\centering \vspace{0.0cm} 
    \includegraphics[height=4.4cm]{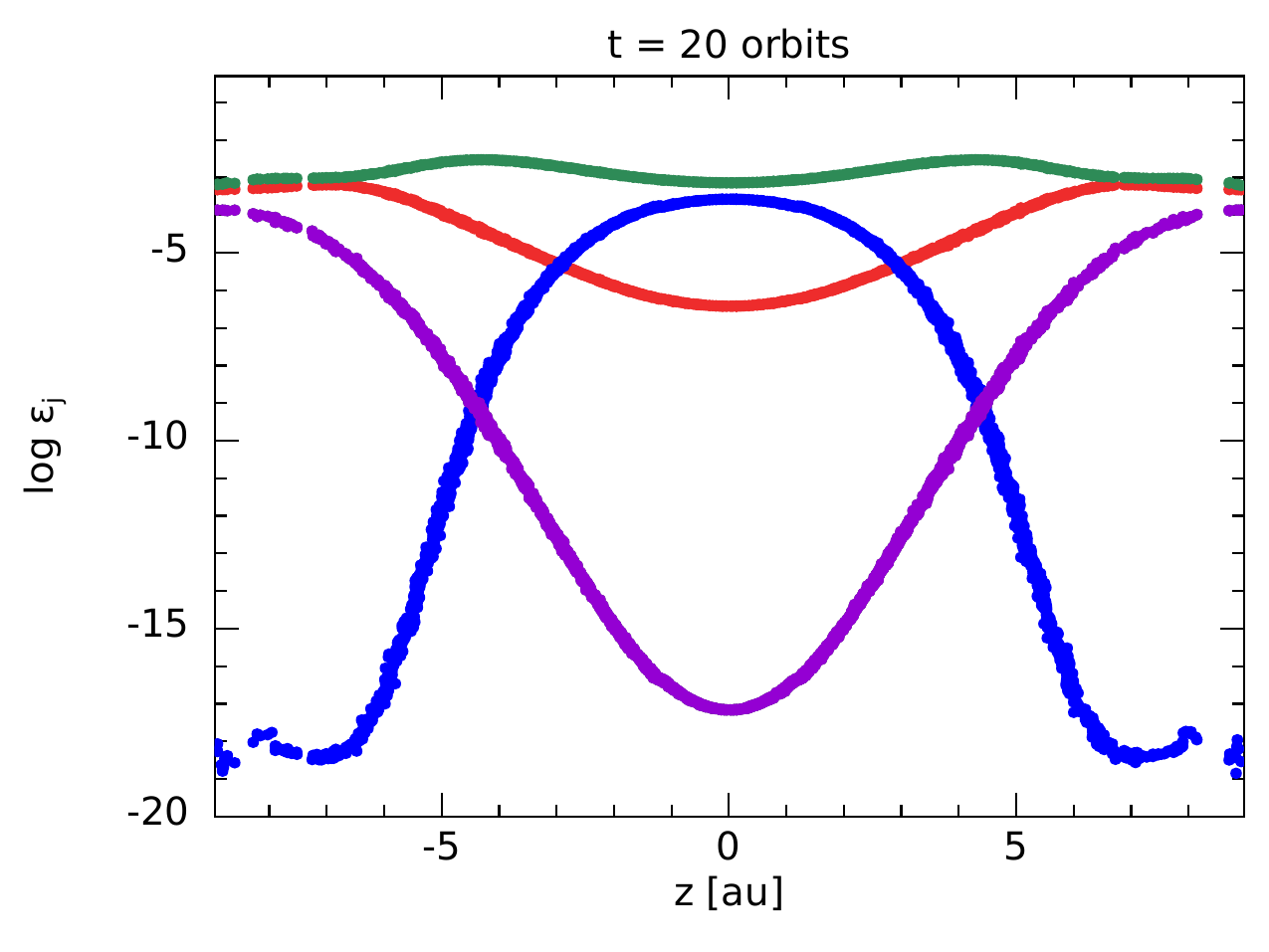}
    \includegraphics[height=4.4cm]{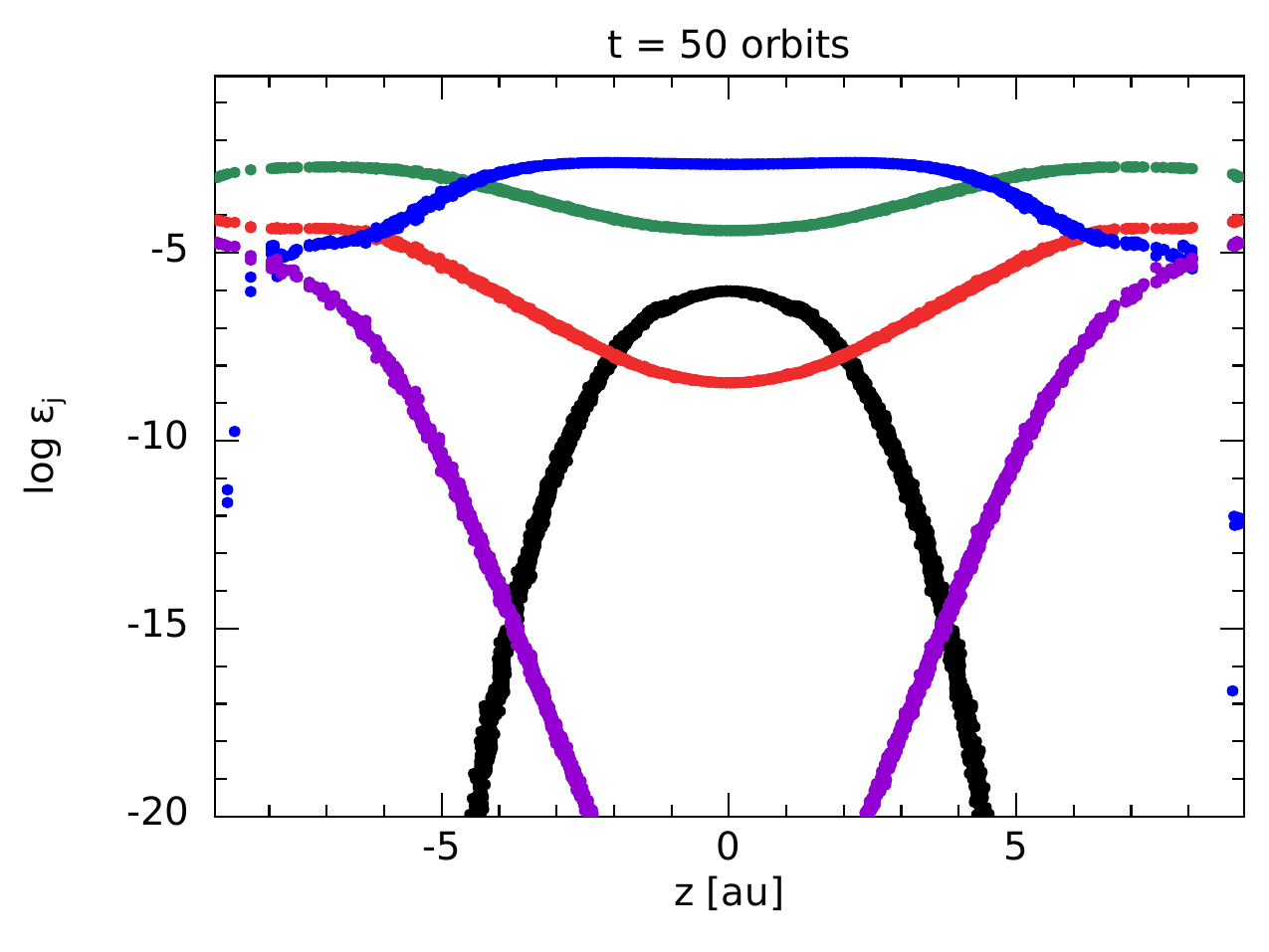} 
    \includegraphics[height=4.4cm]{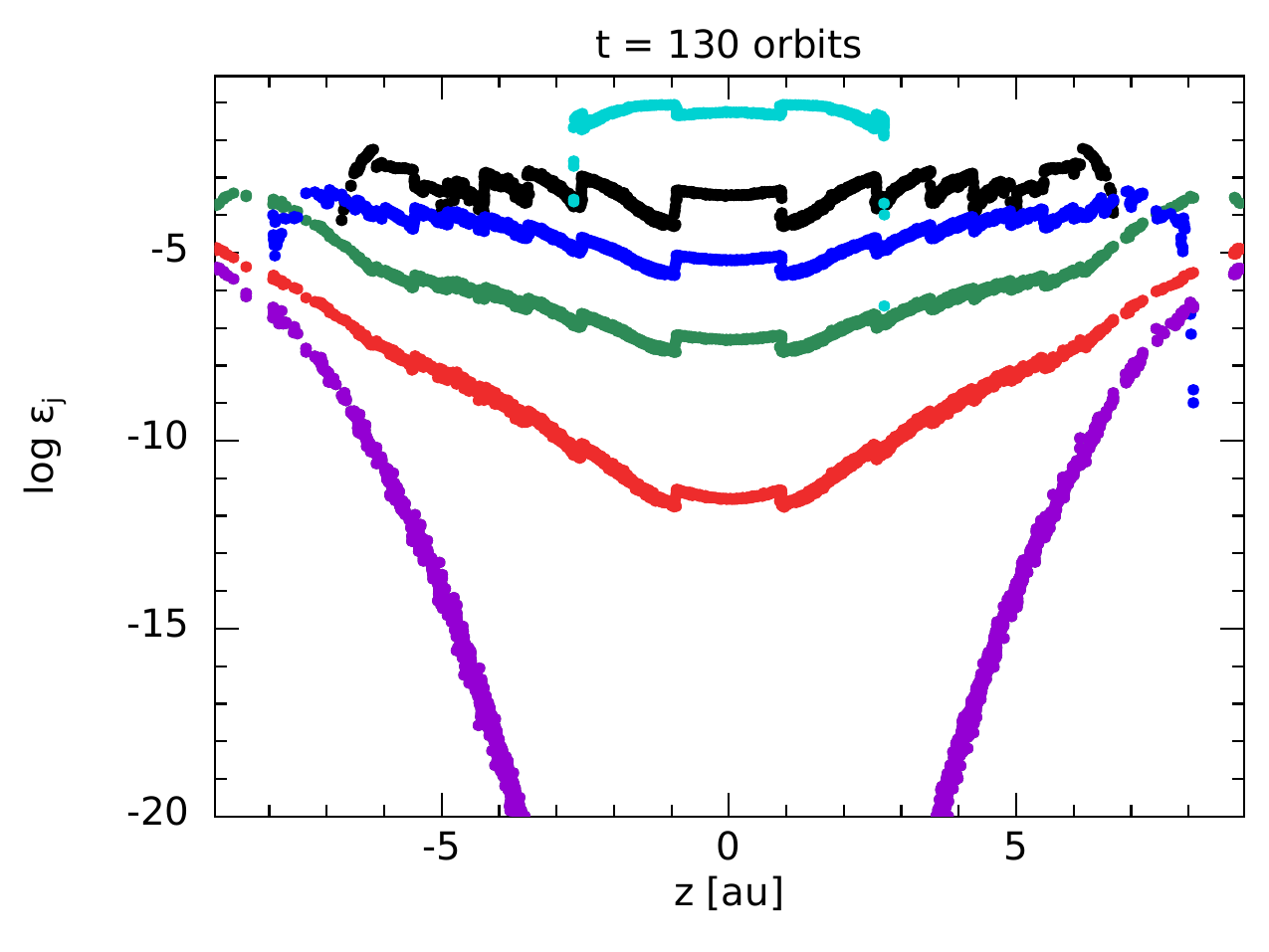}
     \includegraphics[height=4.4cm]{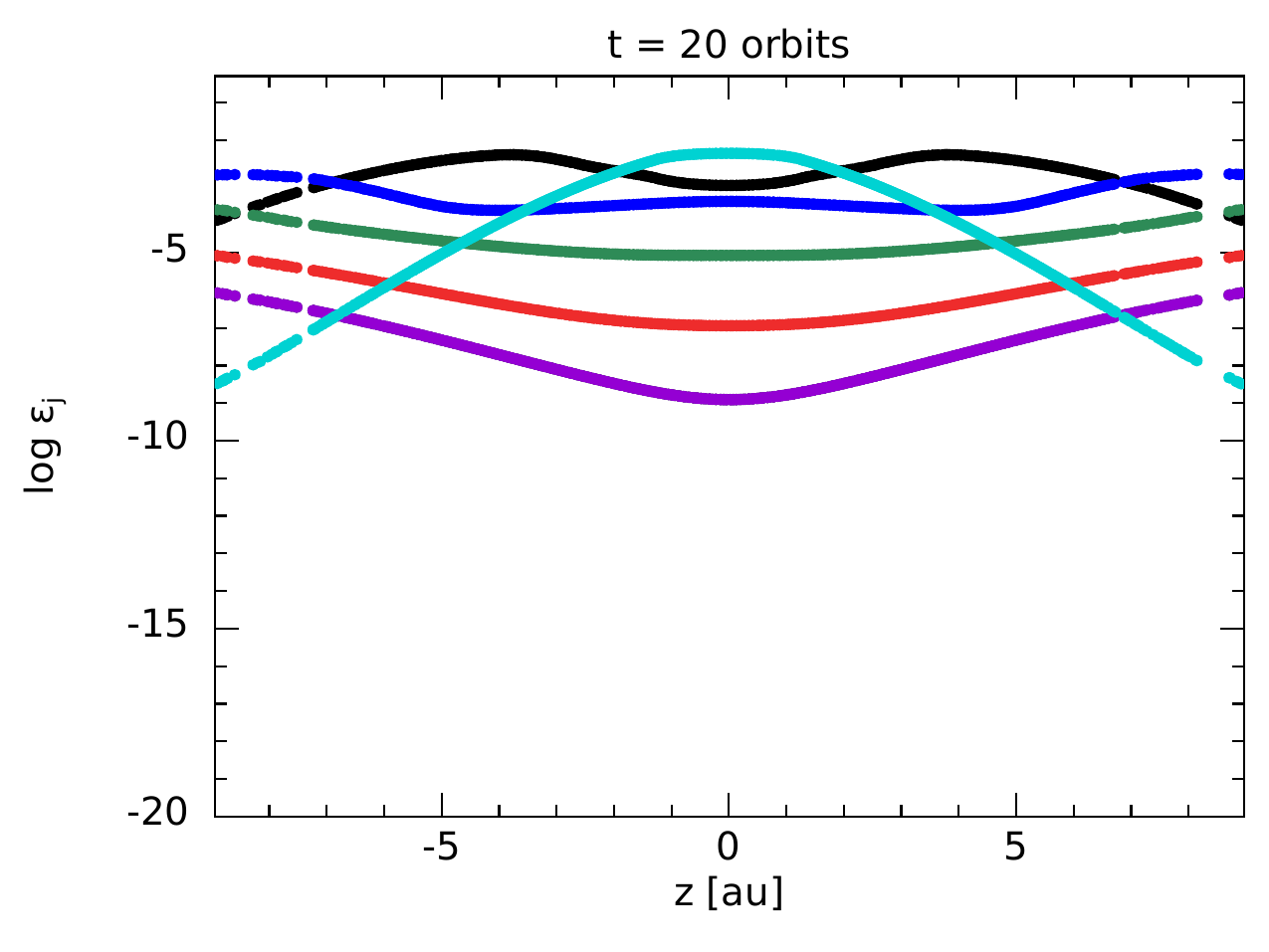}
    \includegraphics[height=4.4cm]{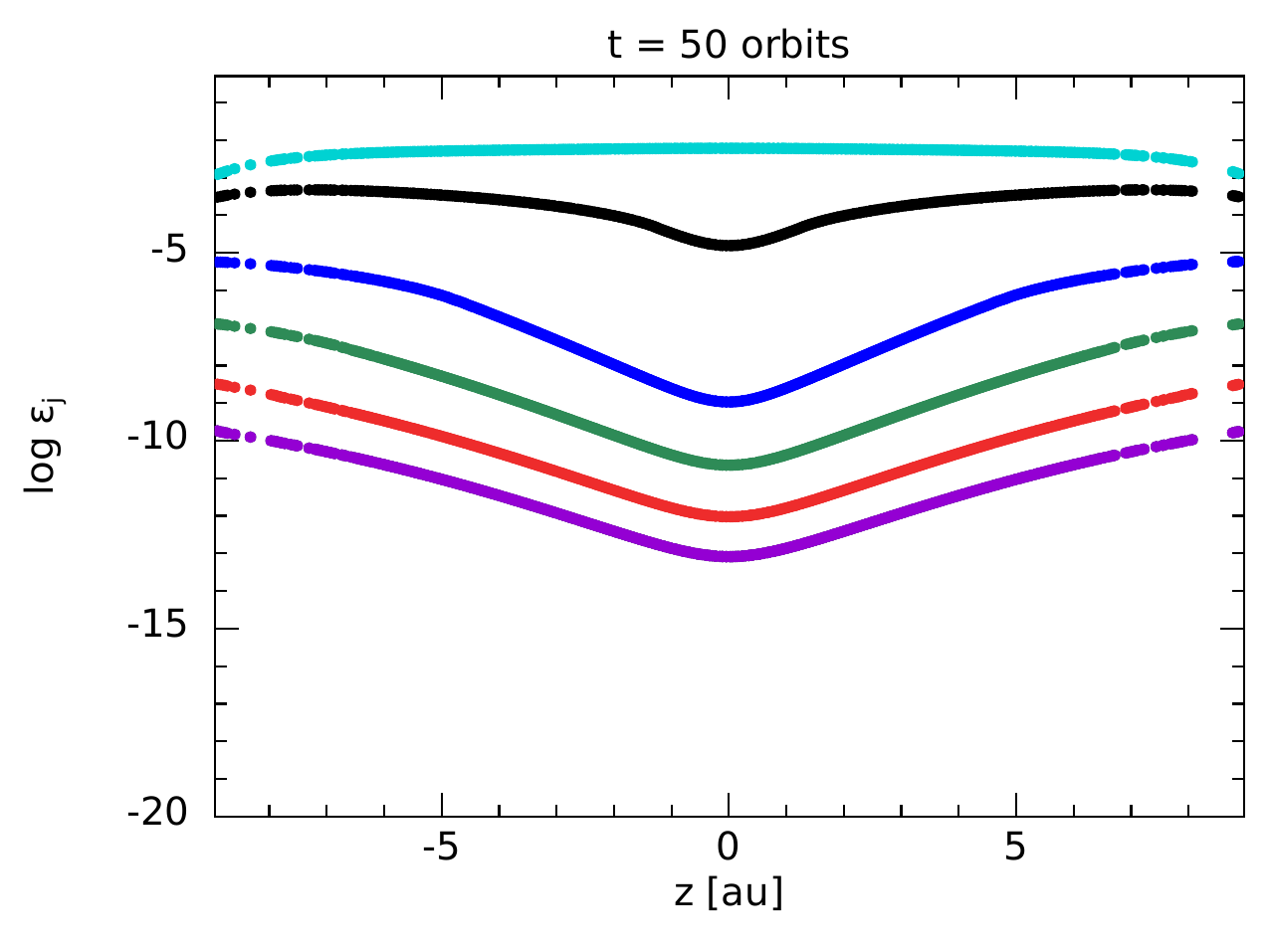} 
    \includegraphics[height=4.4cm]{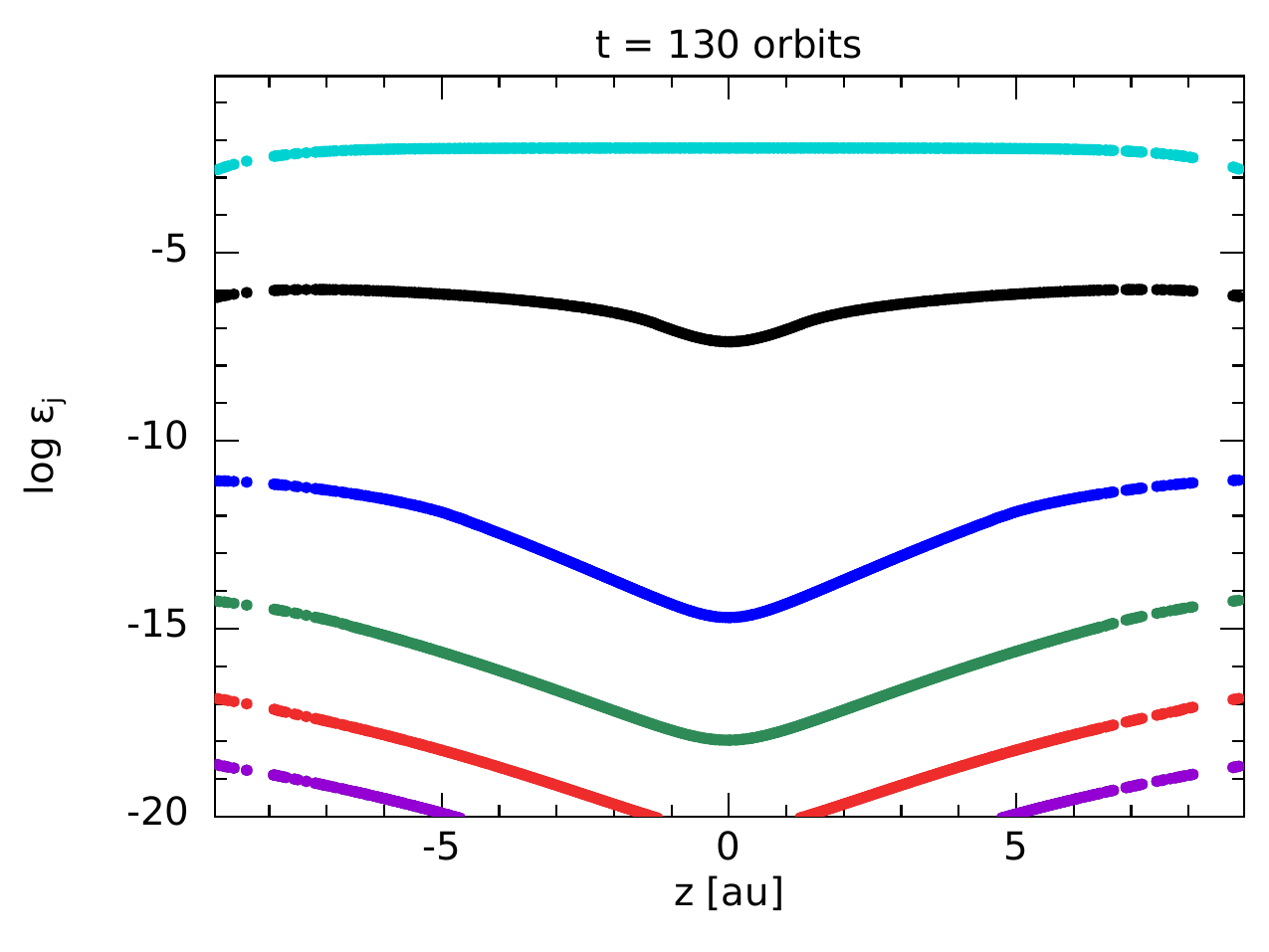}
    \includegraphics[height=4.4cm]{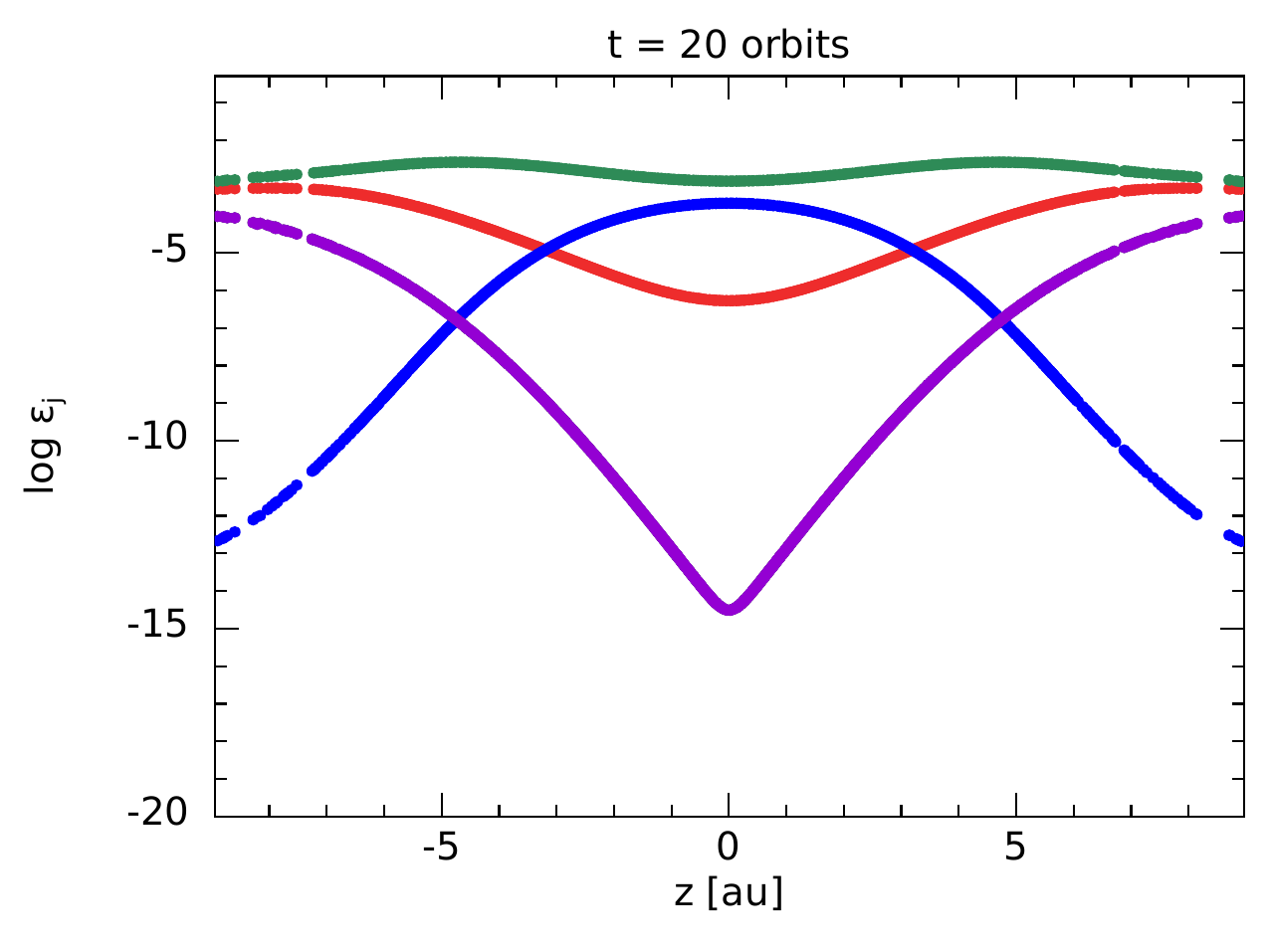}
    \includegraphics[height=4.4cm]{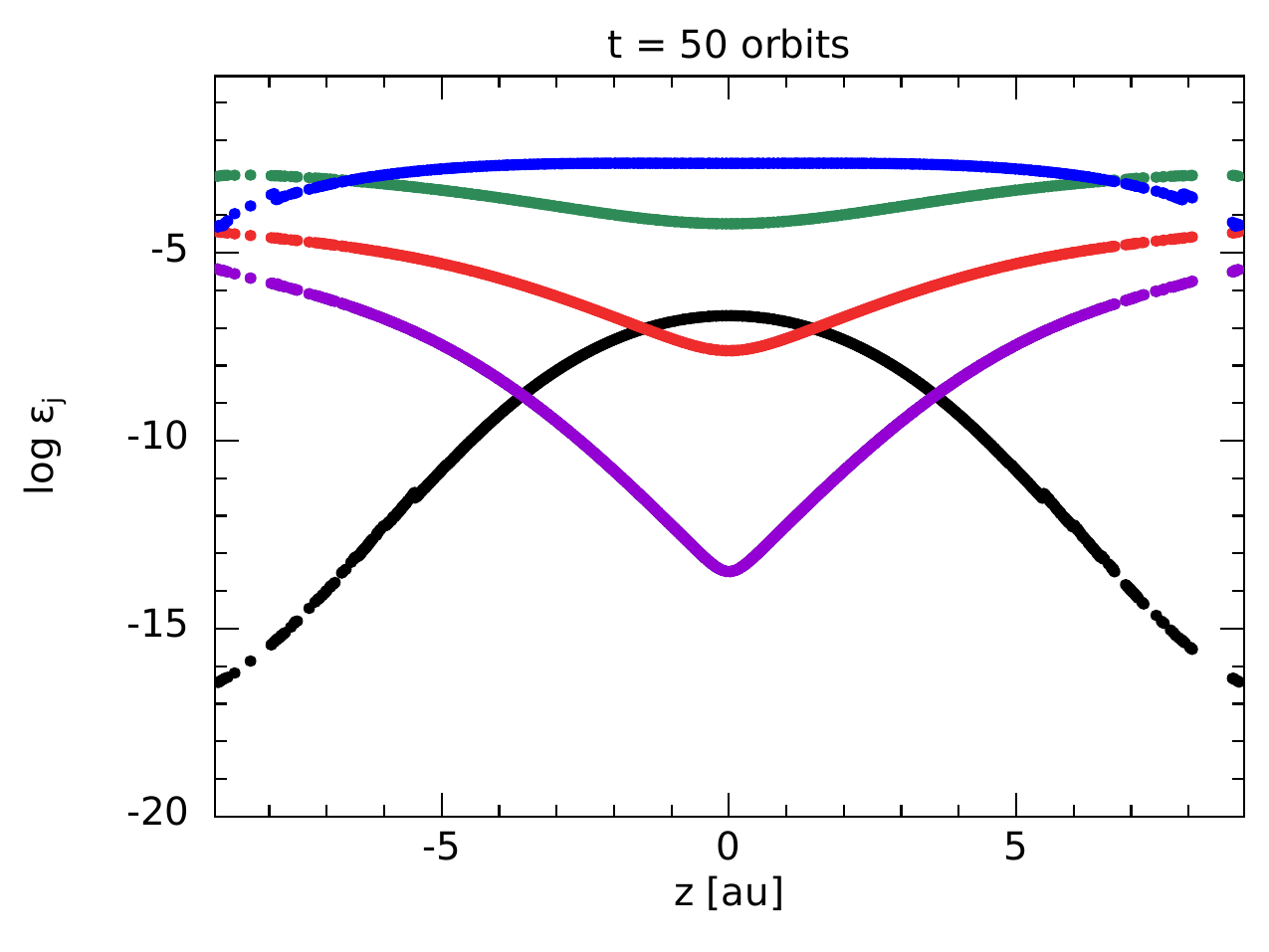} 
    \includegraphics[height=4.4cm]{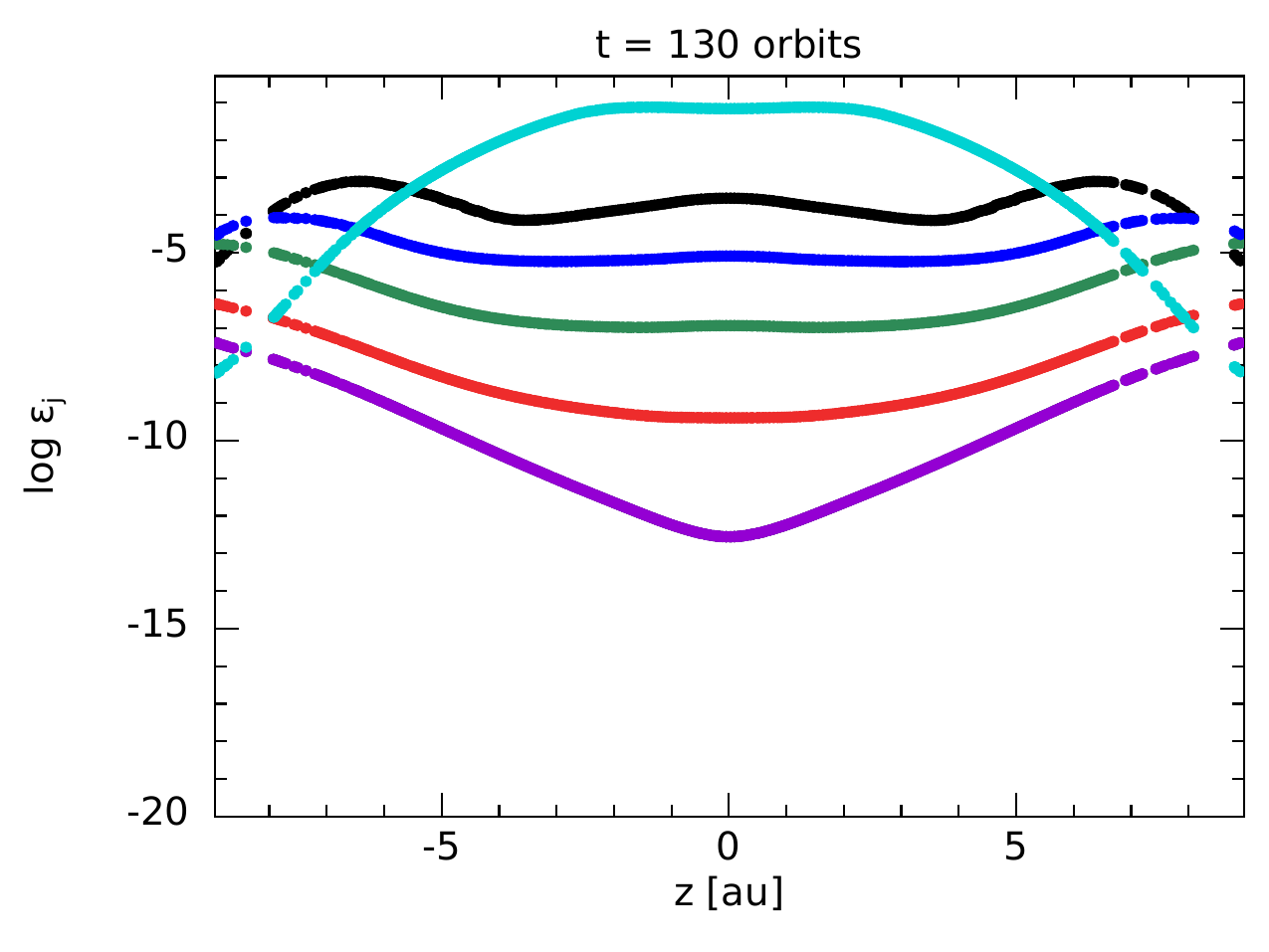}
     \includegraphics[height=4.4cm]{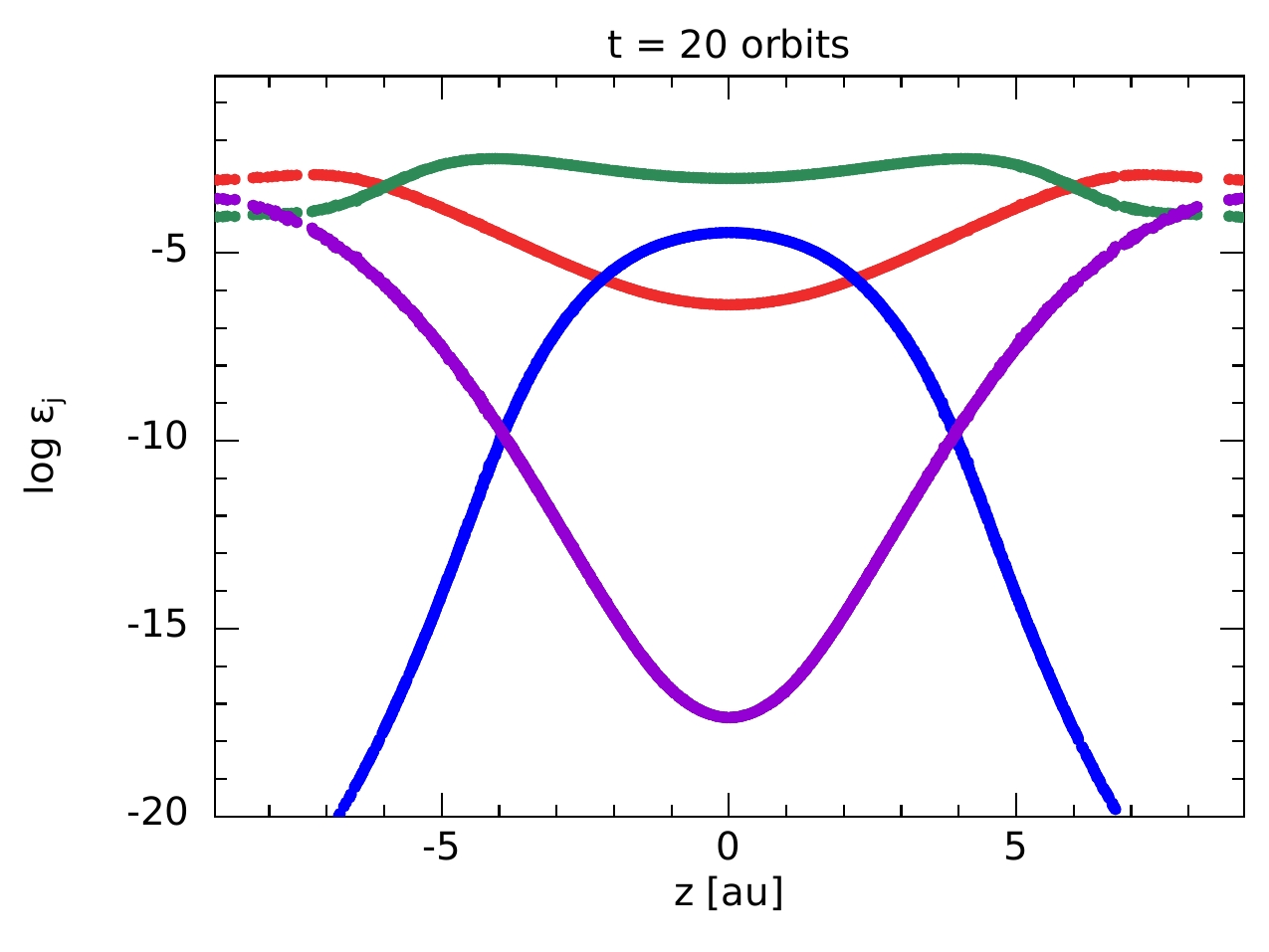}
    \includegraphics[height=4.4cm]{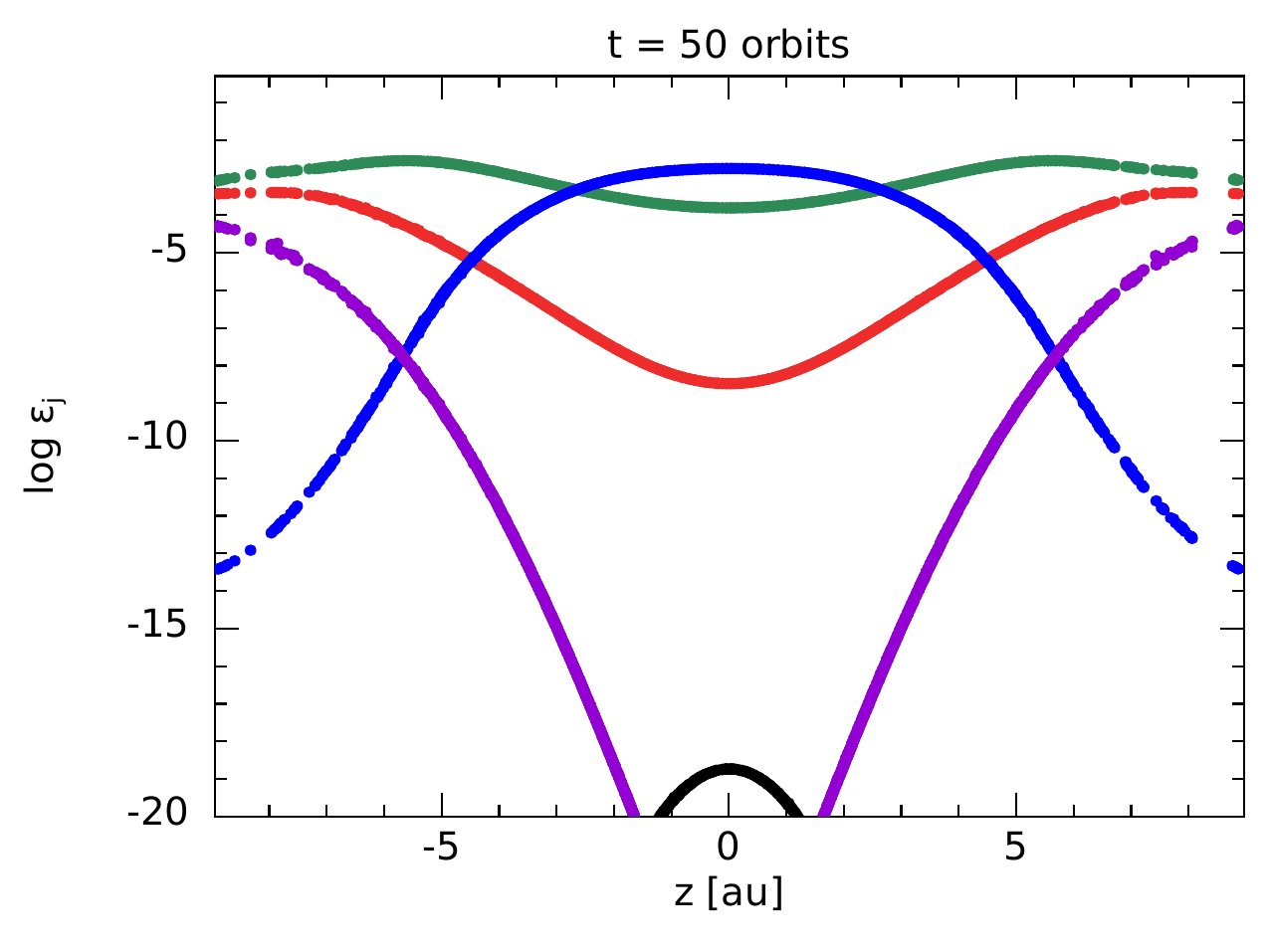} 
    \includegraphics[height=4.4cm]{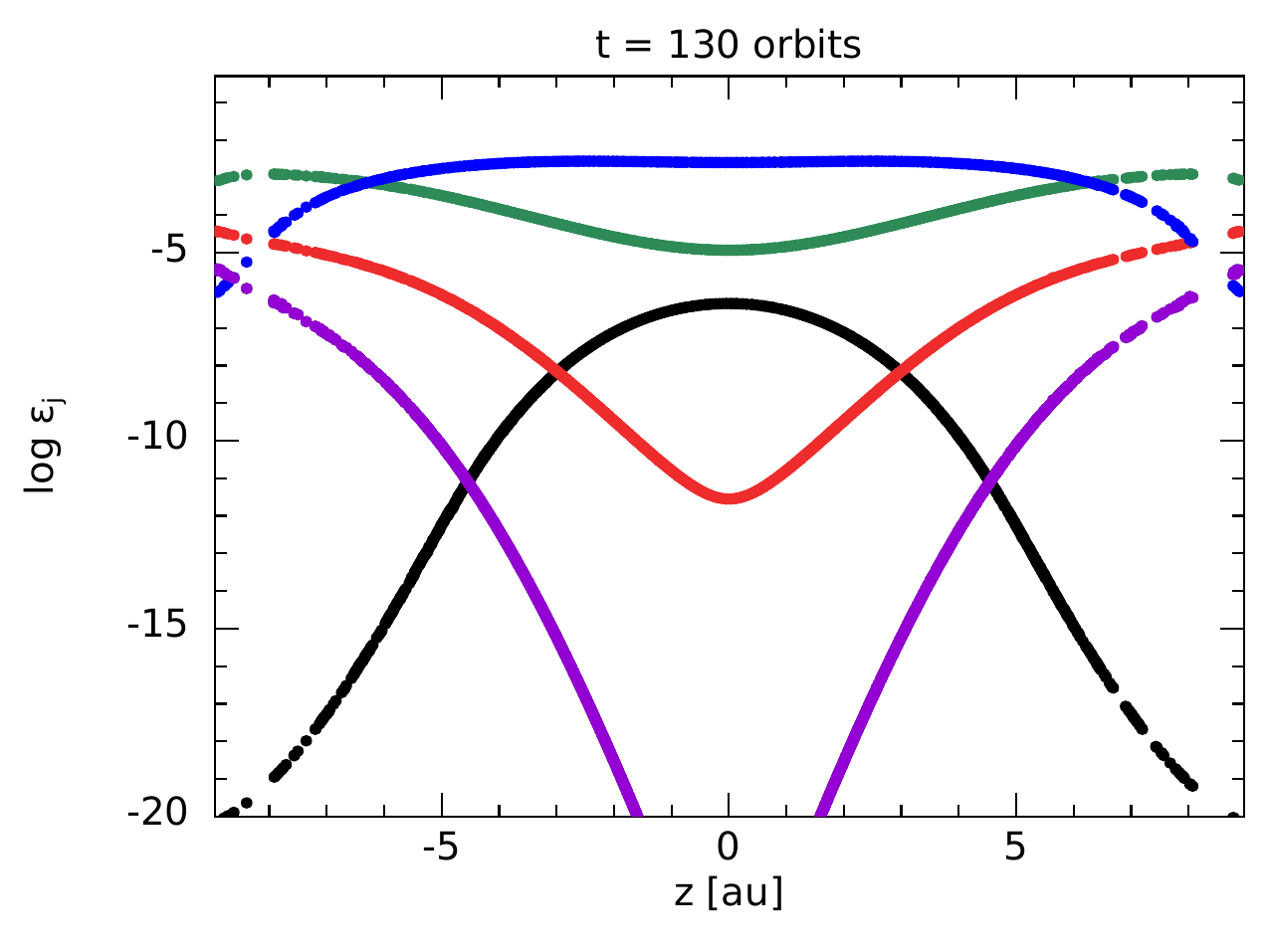}
\vspace{-0.4cm}
\caption{The vertical distributions of the dust fractions of six (out of 53) bins that have dust grain sizes of 0.036 (magenta), 0.11 (red), 0.36 (green), 1.1 (blue),  3.6 (black), and either 5.6 (second row; cyan) or 11 (first and third rows; cyan) $\mu$m after 20, 50, 130 orbits.  The calculations began with an MRN dust size distribution and included grain growth and dust settling.  In the top row, the calculation used $\alpha_{\rm SS}=0.001$ for grain growth, but did not include dust diffusion. In the other rows dust diffusion is included and the same value of $\alpha_{\rm SS}$ is used for both the dust growth and the turbulent diffusion.  The values are $\alpha_{\rm SS}=0.01$ (second row), 0.001 (third row), and $10^{-4}$ (bottom row). Generally speaking, as the turbulence is decreased the grain growth takes longer and the vertical stratification of the dust becomes stronger.  }
\label{fig:MRN_vertical}
\end{figure*}

\begin{figure}
\centering \vspace{-0.05cm} \vspace{-0.0cm}
    \includegraphics[height=6.0cm]{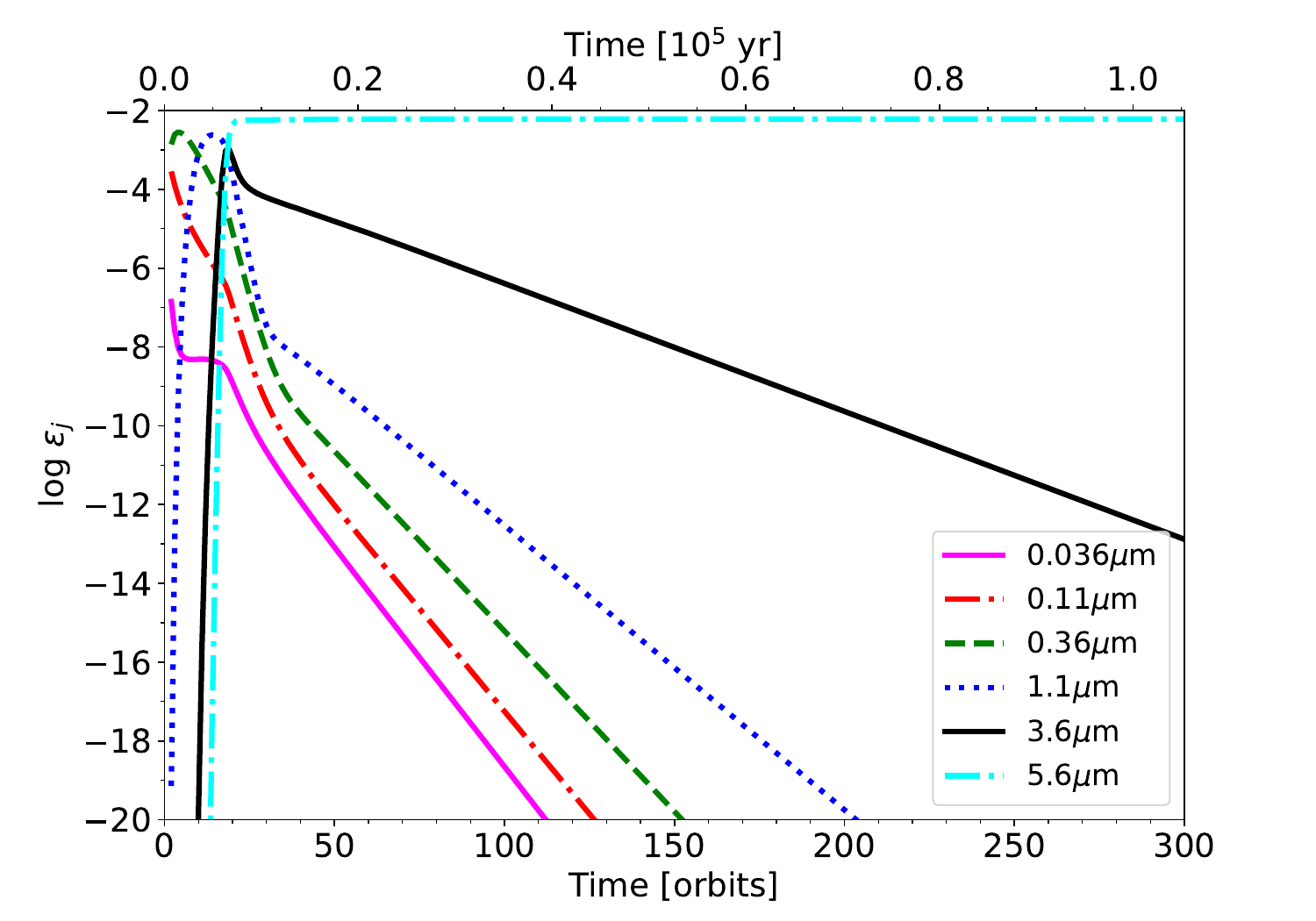}
    \includegraphics[height=6.0cm]{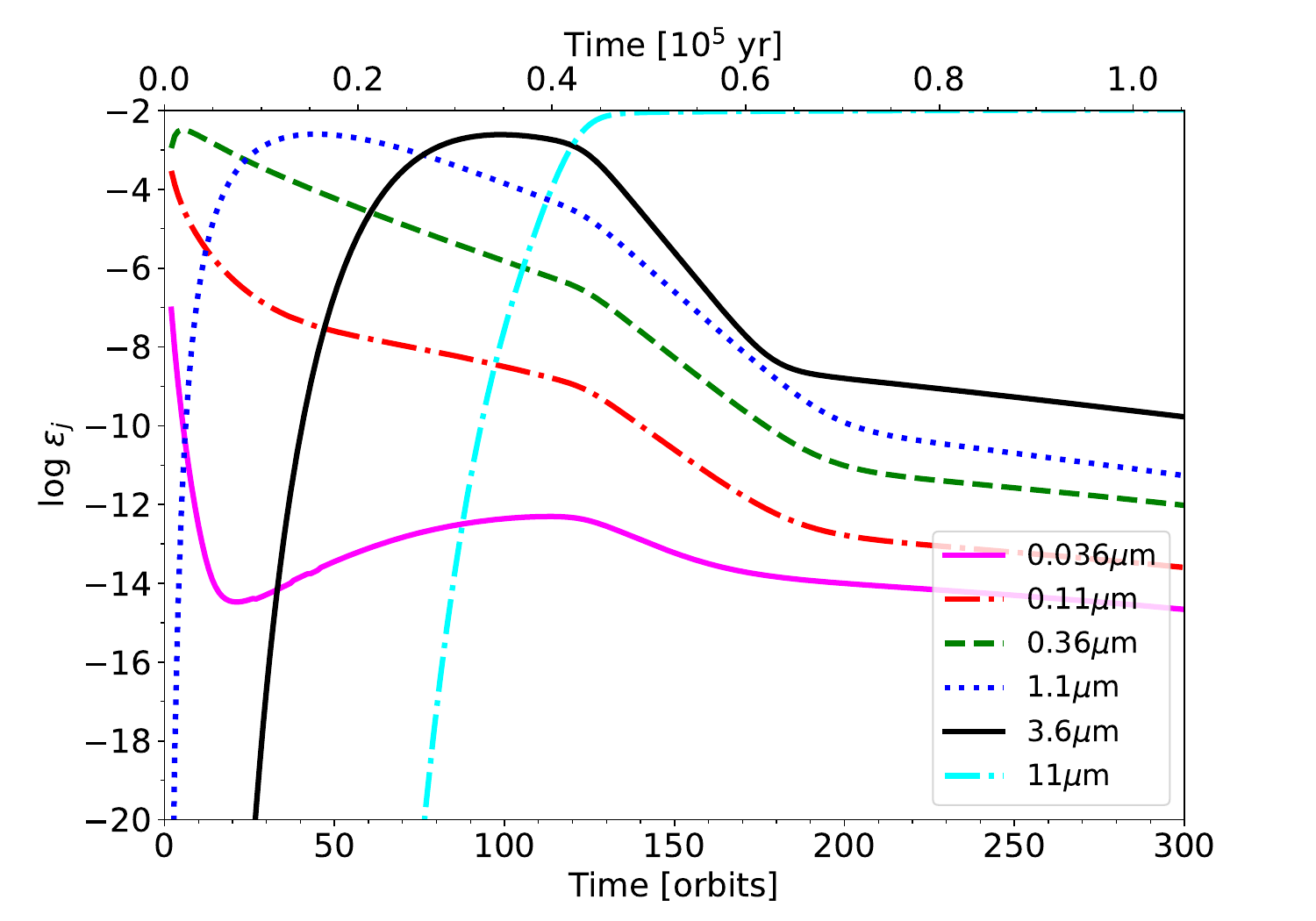}     
    \includegraphics[height=6.0cm]{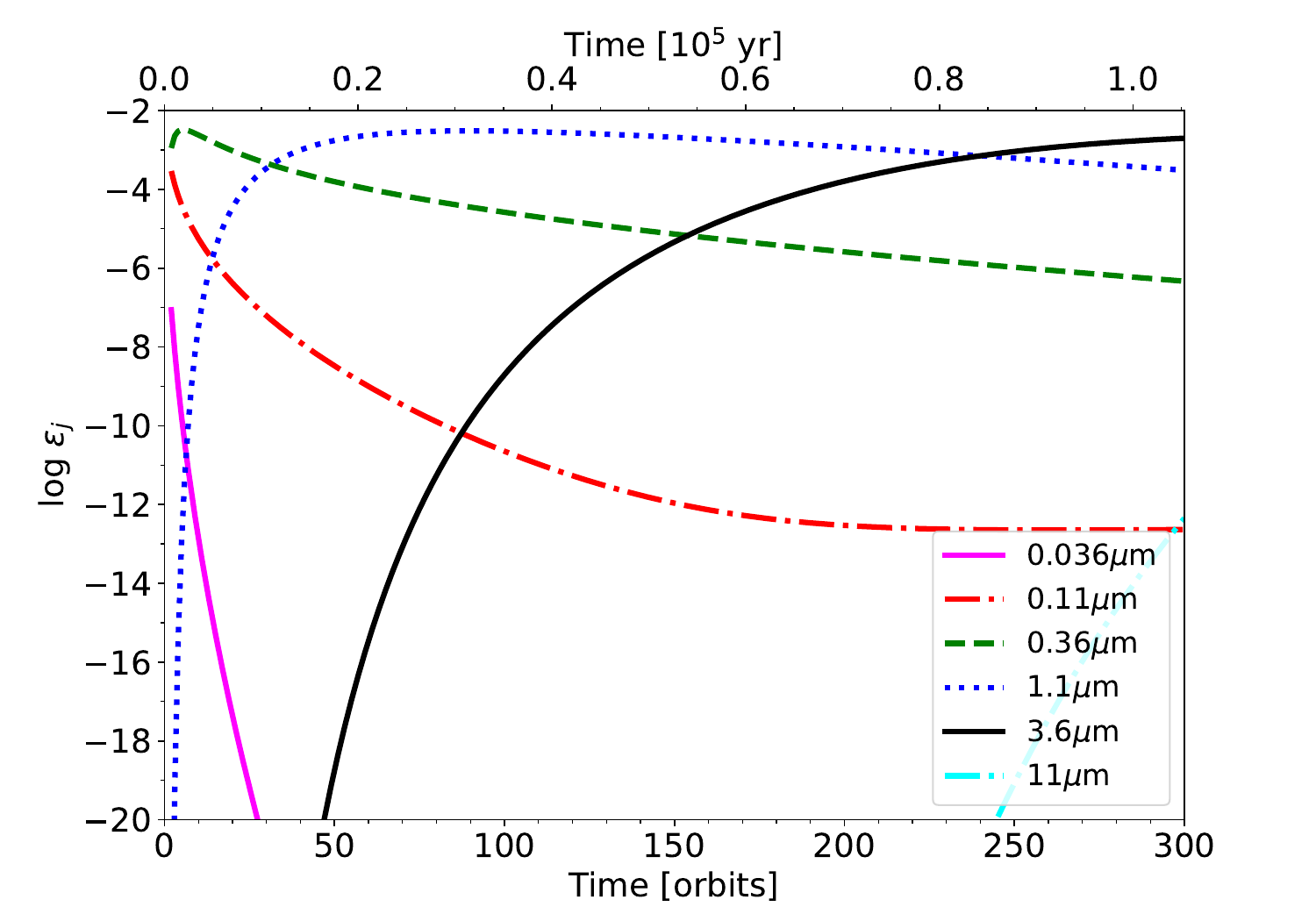} 
    \vspace{-0.5cm}
\caption{The evolution with time of the dust fractions of grains with sizes 0.036, 0.11, 0.36, 1.1, 3.6, and 11 (or 5.6, top panel) $\mu$m at the mid-plane of the disc when beginning with an MRN grain population (maximum initial size of 0.25 $\mu$m), and using the default coagulation threshold.  The three panels show, from top to bottom, the results for calculations with $\alpha_{\rm SS}=0.01, 0.001, 10^{-4}$, respectively (with the same value of $\alpha_{\rm SS}$ used for both the grain growth and turbulent diffusion).  As for the calculation without turbulent diffusion (Fig.~\ref{fig:MRNgrow_and_settle_only_mid-plane}), small grains quickly coagulate and drop in abundance, but the turbulent diffusion keeps their mid-plane abundance higher than without turbulent diffusion.  The growth of large grains depends on the level of turbulence.  With the default criteria for bouncing grains limiting the growth, the case with $\alpha_{\rm SS}=0.01$ has very rapid grain growth, but the maximum grain size is limited to $\approx 6$ $\mu$m (top panel; note that the last label in the legend differs in this panel).  With $\alpha_{\rm SS}=0.001$ the grain growth is still quite rapid, but the maximum grain size is only limited to $\approx 11$ $\mu$m (centre panel).  Finally, with $\alpha_{\rm SS}=10^{-4}$ the grain growth is very slow (bottom panel).}
\label{fig:MRNmid-plane2}
\end{figure}

\subsubsection{MRN evolution with grain growth and dust settling, but no turbulent diffusion}

For the calculation with grain growth but no dust diffusion ($\alpha_{\rm SS}=0.001$), Fig.~\ref{fig:MRNgrow_and_settle_only_mid-plane} shows the evolution with time of the dust fractions at the mid-plane of six of the 53 bins, those containing grains with mean sizes of 0.036, 0.11, 0.36, 1.1, 3.6, and 11 $\mu$m.  The top row of Fig.~\ref{fig:MRN_vertical} shows the vertical distributions of the dust fractions of the same six dust bins (using the same colours for each grain size as in Fig.~\ref{fig:MRNgrow_and_settle_only_mid-plane}) at three different times (20, 50, and 130 orbits).  

Initially only the two dust size bins that are plotted using purple and red colours (0.036, 0.11 $\mu$m, respectively) have non-zero dust abundances, since the other four size bins are for dust grains that are larger than the initial cut-off grain size (0.25 $\mu$m).  As the column of disc material evolves, the smallest grains quickly coagulate with other grains.  Grains slightly larger in size than the initial cut-off  (0.36 $\mu$m) are quickly produced throughout the vertical extent, but then as the calculation is evolved further the abundance of these grains slowly decreases too, particularly near the mid-plane, as even larger grains are produced (green dashed line in Fig.~\ref{fig:MRNgrow_and_settle_only_mid-plane}, and the green points near the mid-plane in the top row of Fig.~\ref{fig:MRN_vertical}).  By 100 orbits ($\approx 35\,000$ yrs) most of the dust near the mid-plane is contained in grains with radii of $\approx 3.6$ $\mu$m (black solid lines in Fig.~\ref{fig:MRNgrow_and_settle_only_mid-plane}, and the black points near the mid-plane in the top row of Fig.~\ref{fig:MRN_vertical}).  From about 130 orbits onward, the vast majority of the dust is contained in grains with sizes of $\sim 11 \mu$m and they do not grow further in this calculation because of the bouncing barrier.

The above description applies to the mid-plane evolution of the dust.  However, the grain growth rate depends on the square of the dust number density (all other things being equal), so the evolution of the grain populations varies with distance from the mid-plane (it is slower further from the mid-plane; top row of Fig.~\ref{fig:MRN_vertical}).  At 130 orbits and far from the mid-plane, most of the dust is contained in dust grains with sizes of only $\approx 0.4-1 \mu$m.

It should be noted that because the grains are all small to begin with in this calculation, they are well coupled to the gas even after they grow to $\approx 4$~$\mu$m in size.  Thus, it makes no significant difference to the grain growth whether or not the settling velocities are included when computing the relative grain velocities.  Similarly, there is very little settling of the dust toward the mid-plane.  Even at 300 orbits, the mid-plane dust-to-gas ratio has only increased from 0.01 to 0.0112 for these particular initial conditions.  The importance of dust settling versus dust growth depends on the exact parameters used (e.g., the minimum and maximum grain sizes, the gas density, etc). The tests in Sections \ref{sec:settle1} and this section clearly demonstrate that the dust evolution can be influence by {\it both} dust growth and dust settling, depending on the dust/gas regime. But, when starting with only small grains, the larger grains are preferentially {\it produced} near the mid-plane of the disc due to the higher grain number densities boosting the grain growth.  They do not accumulate near the mid-plane due to settling.

\subsubsection{MRN evolution with grain growth, dust settling, and turbulent diffusion}

We now present the results from three calculations that include turbulent diffusion, along with grain growth and dust settling.  For each calculation we use a different value of $\alpha_{\rm SS}$, but within each calculation we use the same value for both for the grain growth and the turbulent diffusion. Fig.~\ref{fig:MRNmid-plane2} shows equivalent graphs to Fig.~\ref{fig:MRNgrow_and_settle_only_mid-plane}, but for calculations with values of $\alpha_{\rm SS}=0.01, 0.001$, and $10^{-4}$ from top to bottom, respectively.

First we compare the middle panel of Fig.~\ref{fig:MRNmid-plane2} with Fig.~\ref{fig:MRNgrow_and_settle_only_mid-plane} (these both have the same value of $\alpha_{\rm SS}=0.001$, but the former includes turbulent dust diffusion while the latter does not). The mid-plane dust evolution is very similar, especially for the grain sizes that contain the majority of the dust at any given time.  The main difference is that including turbulent diffusion reduces the strong depletion of small dust grains at the mid-plane (where `small' means much smaller than the grain size containing most of the dust at that time). Turbulent stirring transports small grains from higher altitudes down to the mid-plane, maintaining a small but finite population there. This effect can also be seen by comparing the 3rd row of Fig.~\ref{fig:MRN_vertical} with the top row.  At any of the three particular times, the peak values of the dust fractions for any particular grain size are similar for the top and 3rd rows.  But the dispersions (i.e., the difference between the maximum and minimum values) are greatly reduced because of the mixing of the dust within the vertical column.

We now consider how this dust evolution depends on the magnitude of the turbulence (i.e., the value of $\alpha_{\rm SS}$). The top panel in Fig.~\ref{fig:MRNmid-plane2} and the second row of Fig.~\ref{fig:MRN_vertical} give the results with a higher level of turbulence: $\alpha_{\rm SS}=0.01$.  In this case the grain growth is extremely rapid near the mid-plane, and very quickly the dust becomes mono-disperse.  After only 50 orbits, essentially all the dust grains at the mid-plane have sizes $\approx 6 \mu$m.  Note that this is a smaller value than in the $\alpha_{\rm SS}=0.001$ case (in which grains stopped growing at $\approx 11 \mu$m) because the higher level of turbulence means that the bouncing barrier is reached at a smaller grain size.  Examining the second row of Fig.~\ref{fig:MRN_vertical} we see that there is also much less variation of the dust grain fractions with distance from the mid-plane due to the strong turbulent diffusion/stirring.  When the small grains get depleted, the depletion essentially happens throughout the vertical extent of the disc, rather than being confined to near the mid-plane.  Similarly, because the grains cannot grow to large sizes in this case (due to the bouncing barrier) and they are still tightly coupled to the gas, they are found throughout the vertical column (to $z>3H$) and are not confined to a thin layer.

Finally, in the bottom panel of Fig.~\ref{fig:MRNmid-plane2} and the bottom row of Fig.~\ref{fig:MRN_vertical} we give the results with a low level of turbulence: $\alpha_{\rm SS}=10^4$. With the lower turbulence the dust grows very slowly.  Even after 300 orbits most of the dust is in dust grains with sizes of only $2-4 \mu$m.  Considering vertical distributions of the different grain sizes (the bottom row of Fig.~\ref{fig:MRNmid-plane2}), substantial grain growth only occurs near the mid-plane of the disc, and because of the weak turbulent stirring the largest dust grains (i.e., the 3.6 $\mu$m grains plotted in black) tend to remain there and there is very little mixing into the atmosphere of the disc.  Thus, with low turbulence the dust population becomes strongly vertically stratified.  Again, we emphasise that this is not due to settling of the dust.  It is because the dust preferentially grows in the highest density region, which is near the disc mid-plane.

\begin{figure}
\centering \vspace{-0.05cm} \vspace{-0.0cm}
    \includegraphics[height=6.0cm]{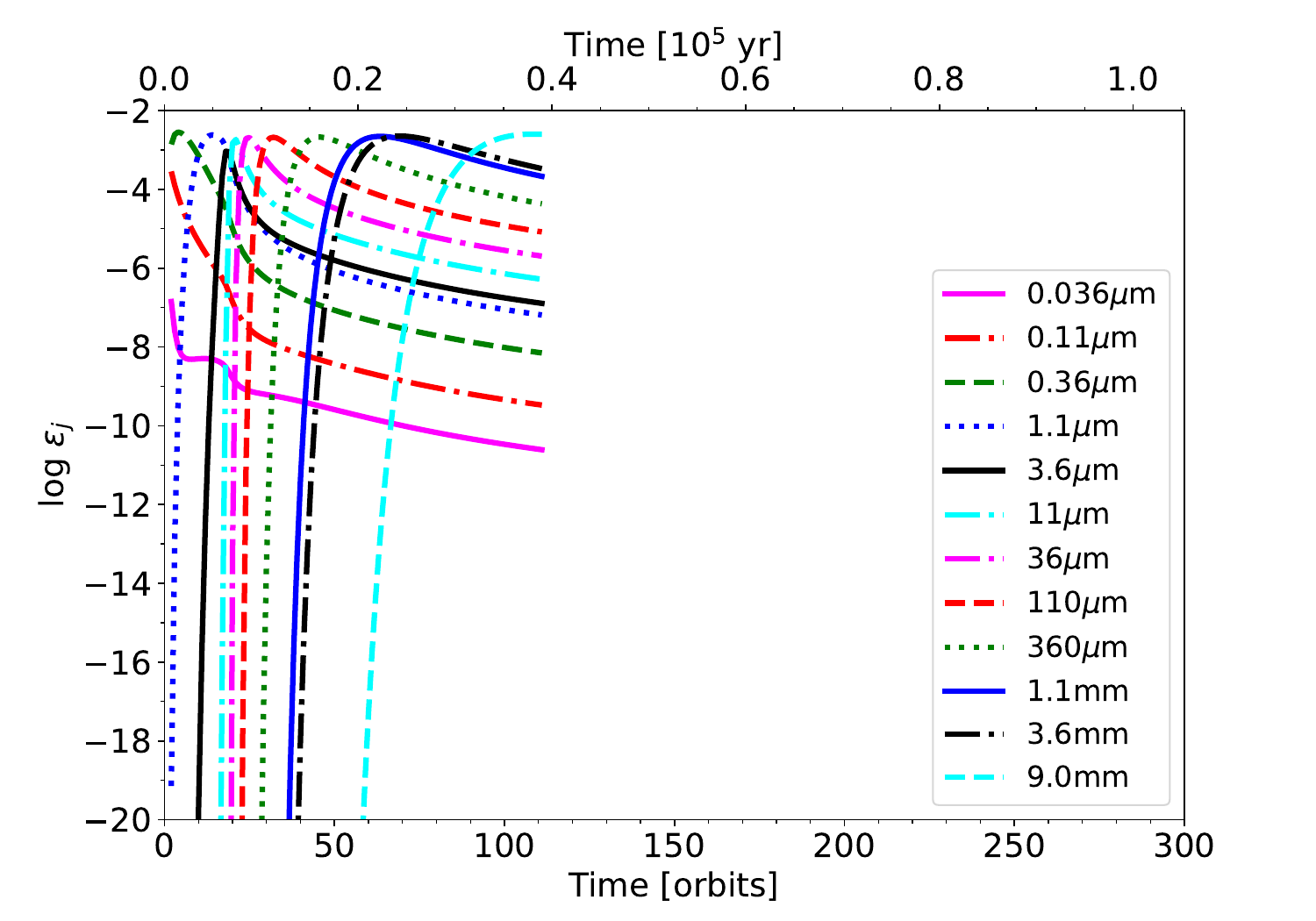}     
    \includegraphics[height=6.0cm]{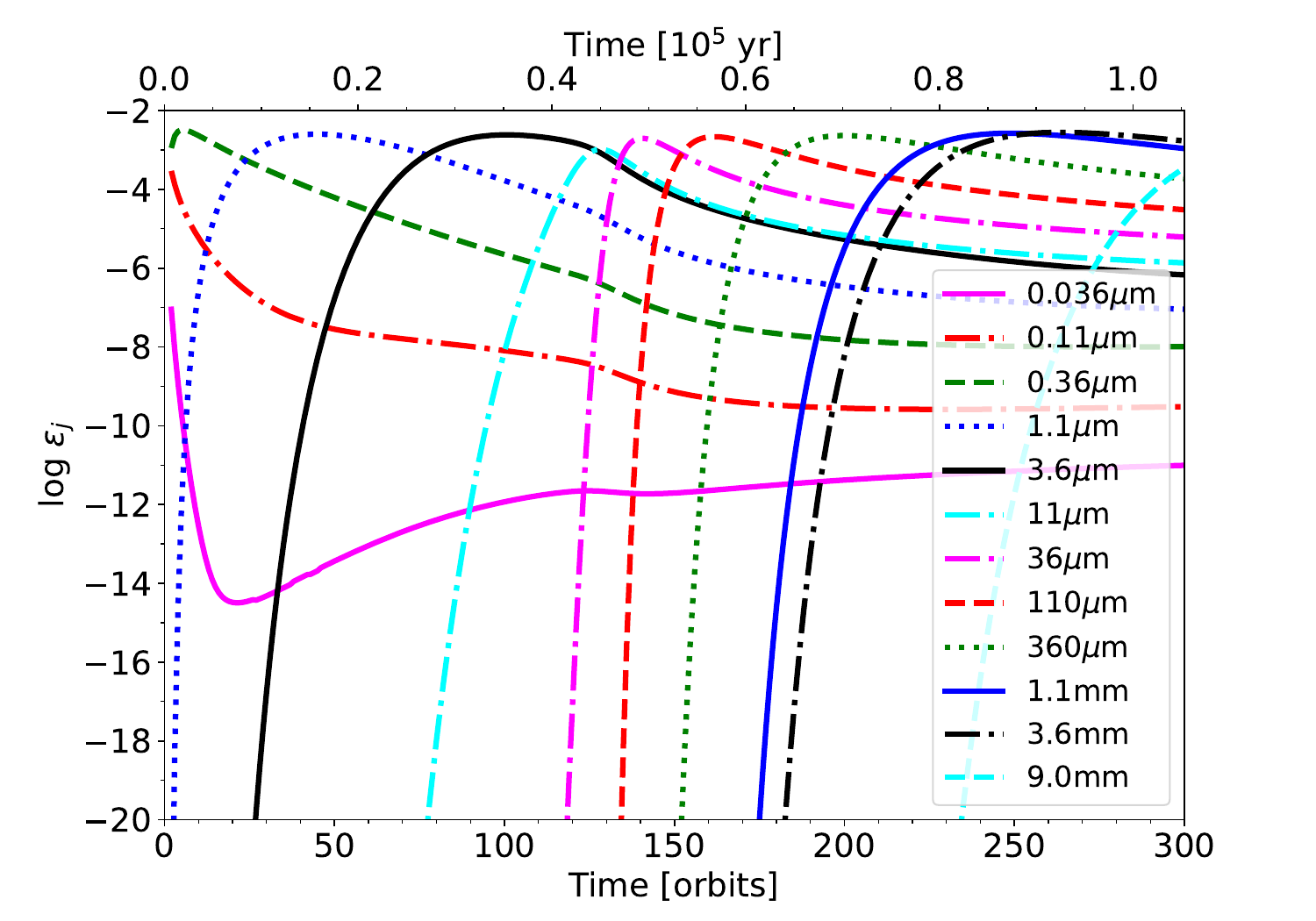} 
    \vspace{-0.0cm}
\caption{The same as the top two panels in Fig.~\ref{fig:MRNmid-plane2}, but allowing all grain collisions with relative speeds smaller than 1 km/s to stick. The top panel shows the results with $\alpha_{\rm SS}=0.01$, while the bottom panel shows the results with $\alpha_{\rm SS}=0.001$ (with the same value of $\alpha_{\rm SS}$ used for both the grain growth and turbulent diffusion). The results for $\alpha_{\rm SS}=10^{-4}$ are indistinguishable from the bottom panel of Fig.~\ref{fig:MRNmid-plane2} because with the lower turbulence and the slow grain growth the relative grain velocities remain low beyond 300 orbits.  Because the grains can grow to larger sizes in these calculations, we show the time evolution of dust fractions of grains with sizes 0.036, 0.11, 0.36, 1.1, 3.6, 11, 36, 110, 360 $\mu$m and 1.1, 3.6, 9.0 mm at the mid-plane of the disc. The calculations begin with an MRN grain population (maximum initial size of 0.25 $\mu$m).  Comparing with Fig.~\ref{fig:MRNmid-plane2}, with $\alpha_{\rm SS}=0.001$ the evolution to 100 orbits is almost identical, but after this point the grain growth no longer stalls at 11~$\mu$m. By the end of the calculation grains with sizes from 1~mm to 1~cm are abundant at the mid-plane.  Similarly, for the case with $\alpha_{\rm SS}=0.01$ (top panel), the grain growth no longer stalls at 6~$\mu$m; instead the peak grain size reaches 1~cm in only $\approx 35,000$ yrs.}
\label{fig:MRNmid-plane3}
\end{figure}

\subsubsection{Relaxing the grain bouncing condition}

Previously, the maximum grain size was limited either by slow growth at a low level of turbulence or by the coagulation threshold velocity, $v_{\rm coag}^{kj}$ (equation \ref{eq:coagulation_barrier}) which prevents sticking in high-speed collisions.

In this section, we present the results from two more calculations, with $\alpha_{\rm SS}=0.01$ and 0.001 (Fig.~\ref{fig:MRNmid-plane3}, top and bottom panels, respectively).  These calculations are identical to those from the previous section, except that we set $v_{\rm coag}^{kj}=1$ km~s$^{-1}$ for all collisions so as to allow the grains to grow much larger.  Our maximum grain size for these calculations is 10 mm.

With $\alpha_{\rm SS}=0.01$ (top panel of Fig.~\ref{fig:MRNmid-plane3}), the grain growth again occurs very quickly.  It only takes 100 orbits ($\approx 35,000$ yrs) for most of the dust at the mid-plane to be in 10 mm grains. Recall that in the previous calculation with $\alpha_{\rm SS}=0.01$ the largest grains had sizes of $\approx 6 \mu$m.  This illustrates the importance of the chosen sticking condition.

With $\alpha_{\rm SS}=0.001$ (bottom panel of Fig.~\ref{fig:MRNmid-plane3}), the grain growth is very similar to the case with the lower coagulation threshold (middle panel of Fig.~\ref{fig:MRNmid-plane2} until $\approx 110$ orbits ($\approx 39,000$ yrs) when the highest dust fraction is for grain sizes of $\approx 4 \mu$m.  However, whereas the grain growth stalled at 11 $\mu$m grain sizes soon after this time in the original calculation, in the calculation with the much greater coagulation threshold grain growth continues to sizes of several millimetre sizes in less than 300 orbits.  Again this demonstrates that the assumed sticking condition plays a decisive role in setting the maximum grain size.
 
\begin{figure}
\centering \vspace{-0.05cm}     
    \includegraphics[height=5.0cm]{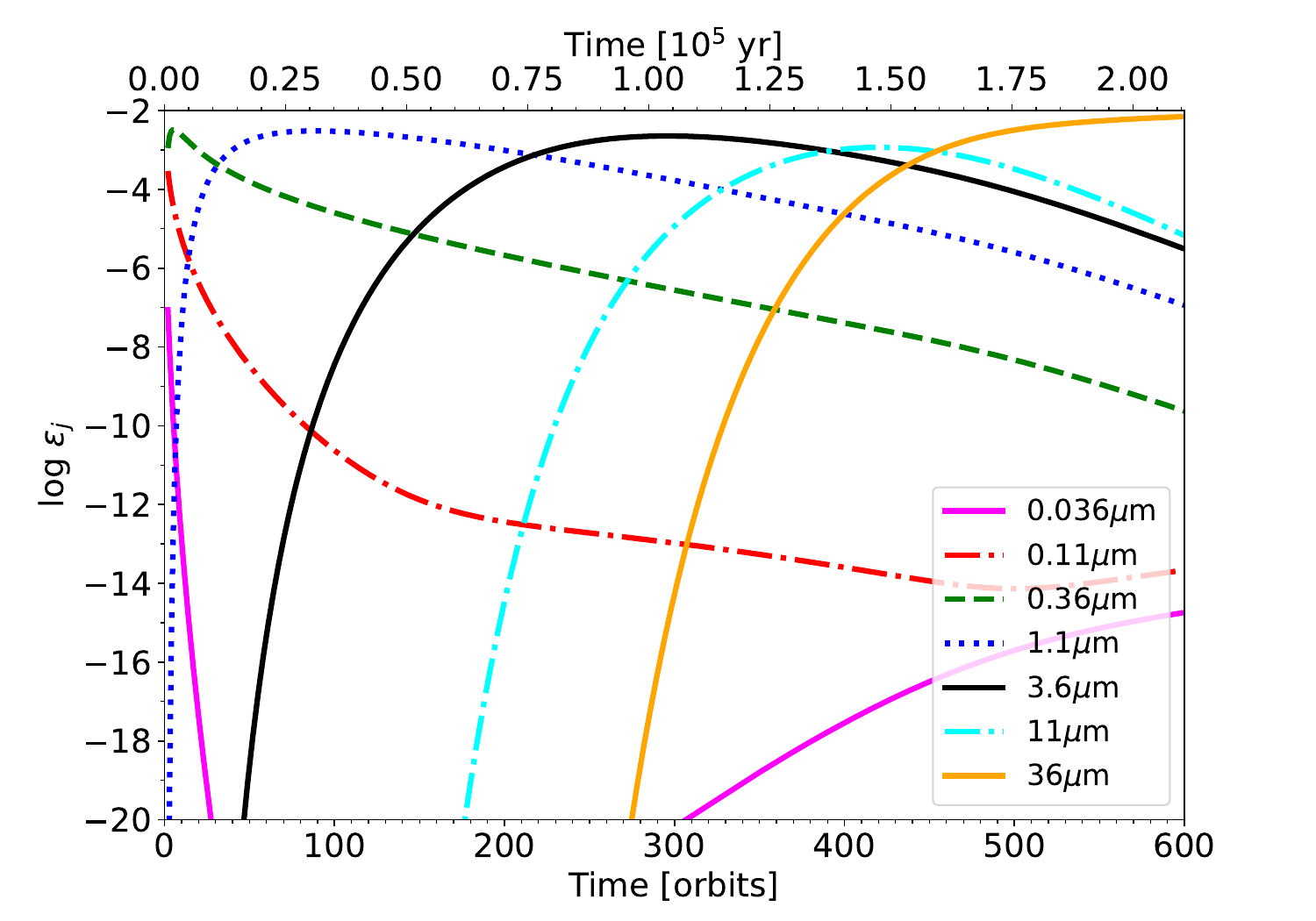}
    \includegraphics[height=5.0cm]{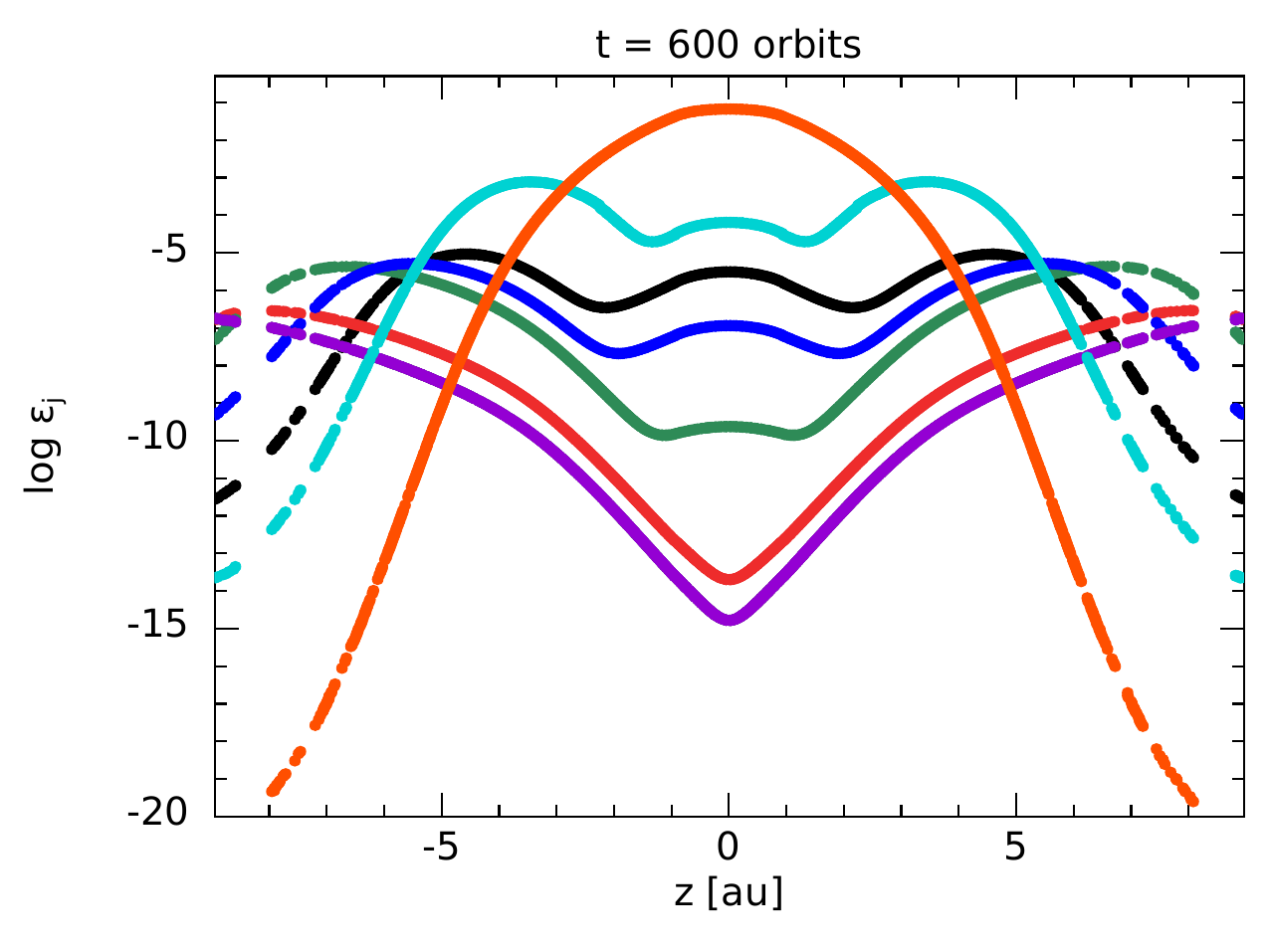} 
    \includegraphics[height=5.0cm]{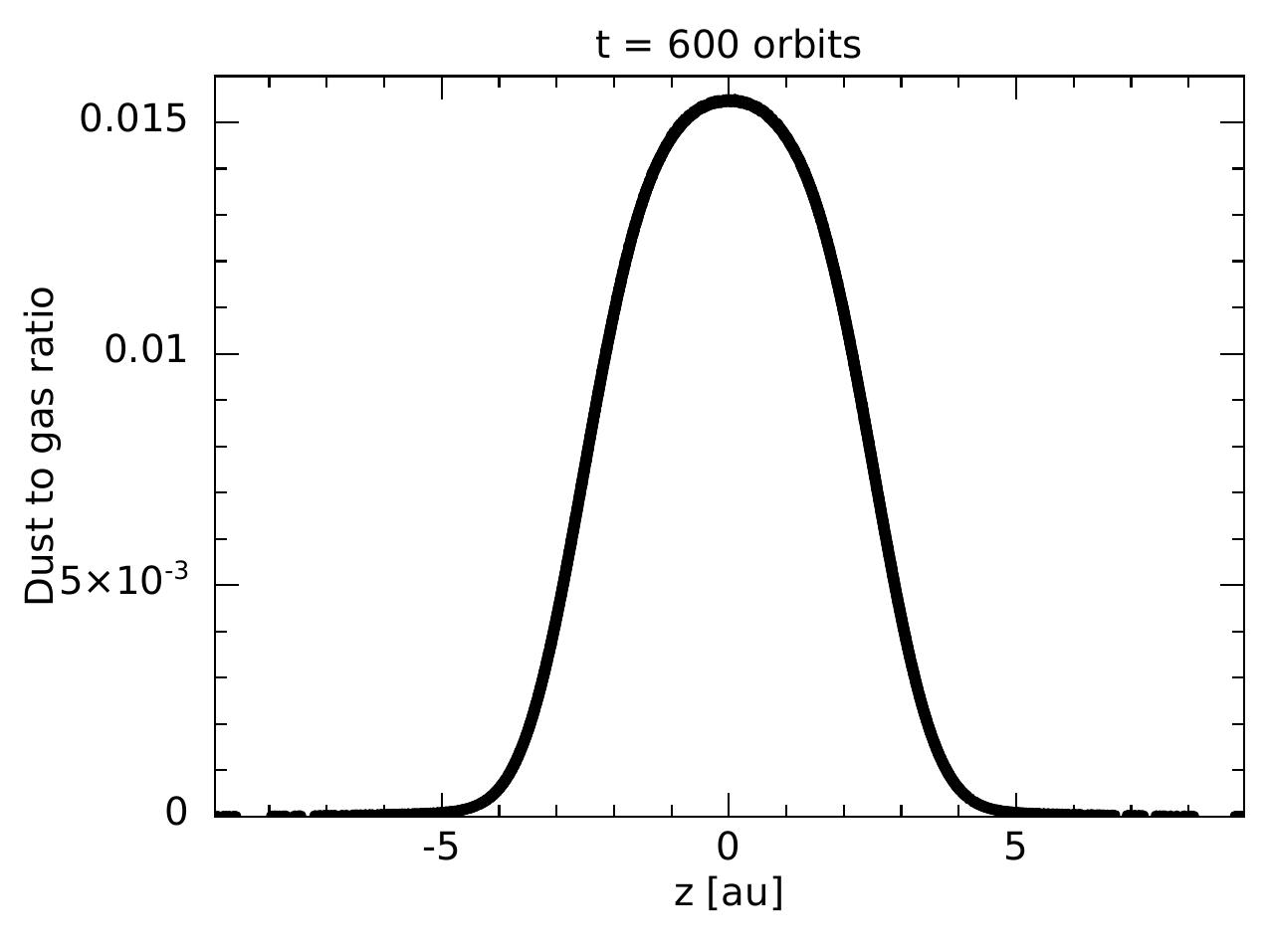} 
\caption{The results from a calculation beginning with an MRN grain population (maximum initial size of 0.25 $\mu$m) with dust growth, settling and turbulent stirring, and using $\alpha_{\rm SS}=0.001$ and the default sticking criteria for the dust growth, but using $\alpha_{\rm SS}=10^{-4}$ for the turbulent diffusion.  Top panel: the evolution with time of the dust fractions of grains with sizes 0.036, 0.11, 0.36, 1.1, 3.6, 11, and 36  $\mu$m at the mid-plane of the disc.
Middle panel: the vertical distributions of the dust fractions for the same 7 dust grain sizes (using the same colours as in the top panel) after 600 orbits. Bottom panel: the total dust-to-gas ratio (i.e., summed over all dust grains) as a function of height after 600 orbits. }
\label{fig:MRN_grow3_diff4}
\end{figure}

\subsubsection{Enhanced grain growth in a disc with weak turbulence}

Finally, we give an example of the dust evolution in a vertical column with a low turbulent diffusion/stirring ($\alpha_{\rm SS}=10^{-4}$), but in which we enhance the dust growth rate by using $\alpha_{\rm SS}=0.001$ when computing the relative grain velocities.  Real dust growth rates are poorly known, and by enhancing the dust growth rate relative to the dust stirring we can study a case in which the dust grows comparatively quickly but there is also significant settling of the dust towards the mid-plane and, thus, an increase of the mid-plane dust-to-gas ratio (Fig.~\ref{fig:MRN_grow3_diff4}).  

The calculation was evolved for 600 orbits.  At the end of the calculation the dust is mainly in the form of dust grains with sizes of $\approx 36 \mu$m (top panel of Fig.~\ref{fig:MRN_grow3_diff4}).  These relatively large grains are not that well coupled to the gas, particularly far from the mid-plane, and because we have taken the turbulent stirring to be weak these large grains are confined to a thin layer (with a scale height much smaller than the gas scale-height of $H=2.5$ au; middle panel of Fig.~\ref{fig:MRN_grow3_diff4}).  The small grains are strongly depleted near the mid-plane and largely confined to the surface layers of the disc.  Above three gas scale heights the dust population is not that different to the initial MRN population, with most of the dust contained in grains with sizes less than 0.5 $\mu$m.

In the lower panel of Fig.~\ref{fig:MRN_grow3_diff4} we give the total dust-to-gas ratio at the end of the calculation (i.e., for each SPH particle we integrate over all the dust bins to obtain the total dust mass, and compare this to the gas mass of the SPH particle).  This graph makes it clear that there has been substantial settling of the dust toward the mid-plane.  The initial dust-to-gas ratio was 0.01, whereas after 600 orbits the mid-plane dust-to-gas ratio is more than 50 percent greater (0.0155).  Moreover, although the dust population above three gas scale-heights ($|z|>7.5$ au) may be similar to the initial distribution in terms of the relative abundances of different grain sizes (e.g., most of the dust is in grains with sizes $\lsim 0.4 \mu$m), in fact there is almost no dust in this gas.

\section{Star formation calculations}
\label{sec:results}

In this penultimate section, we demonstrate the promise of this new combination of methods by presenting results from dusty radiation hydrodynamical calculations of the collapse of a molecular cloud core well past the first hydrostatic core and stellar core formation phases.  We follow the protostellar disc as it grows by accreting gas from the infalling envelope and as gas and dust are accreted by the protostar through a marginally gravitationally unstable disc.  The dust distribution is not used to set the opacities for the radiative transfer --- standard interstellar opacities are employed \citep{WhiBat2006, BatKet2015}.  Using the dust grain populations to compute local opacities is an obvious extension, but is beyond the scope of this paper.

We only consider a single set of initial conditions, as this paper is primarily a method paper.  But to demonstrate the effects of each of the processes causing the dust evolution we perform calculations with all three processes (i.e., dust coagulation, dust-gas drag, and dust turbulent diffusion) taking two different values, $\alpha_{\rm SS}=0.001$ and $10^{-4}$ (with the same value used for the dust growth and the turbulent diffusion).  We also perform calculations assuming $\alpha_{\rm SS}=0.001$ for the dust growth, but in which we turn off either the dust-gas drag, the turbulent diffusion, or both of the dynamical effects, in order to determine which processes have the greatest effect on the dust grain populations and distributions.

\subsection{Initial conditions, resolution, and sink particles}
\label{sec:initialconditions}

We use the same basic initial conditions for the molecular cloud core as those used by \cite{Bate2022}, but with only a single rotation rate.  The initial conditions consist of a 1-M$_\odot$ unstable Bonnor-Ebert sphere with an inner to outer density contrast of 20 and a radius of 5500~au.  The cloud is contained by spherical, reflective boundary conditions (using ghost particles).  The cloud is given an initial uniform rotation rate of $8.24 \times 10^{-14}$~rad~s$^{-1}$, resulting in the magnitude of the ratio of rotational to gravitational potential energy of $\beta=0.005$.  The \cite{BatKet2015} combined radiative transfer and diffuse interstellar medium method results in the initial clouds being warmer on the outside (gas temperature $\approx 15$~K) than at the centre ($\approx 6.5$~K).  The initial magnitude of the ratio of thermal energy to the gravitational potential energy is $\alpha=0.42$.  We note that the implementation of the \cite{BatKet2015} 
method used for these calculations includes the energy terms associated with equations \ref{eq:energytot} and \ref{eq:dudt_drag}, but omits the energy term (equation \ref{eq:dudt_diffusion}) associated with the dust diffusion.  We do not expect this
to have a significant effect on the results.

The Bonnor-Ebert density distribution is set up using $10^6$ equal-mass SPH particles that are placed on a uniform cubic lattice that is radially deformed to produce the required density profile.  The resolution is an order of magnitude larger than the number of particles required to resolve the local Jeans mass throughout the calculation ($\approx 10^5$ SPH particles per solar mass are required to marginally resolve the minimum Jeans mass; \citealt{BatBur1997, Trueloveetal1997, Whitworth1998}; \citealt*{HubGooWhi2006}).  

We take an initial dust-to-gas ratio of 1:100 ($\varepsilon = 1/101$), with a MRN dust grain size distribution $n(a) \propto a^{-3.5}$ with $a_{\rm min}=5$~nm and $a_{\rm cutoff}=0.25$~$\mu$m.  The calculations use 53 grain size bins with logarithmic spacing from $a_{\rm min}$ to $a_{\rm max}=1$~mm (i.e., 10 bins per decade in size).  This is the same dust size resolution as used by \cite{Bate2022}, who showed that using either 10 bins or 20 bins per decade gave similar results.  See also Appendix \ref{appendixA} in which we compare the performance of the coagulation method with analytic solutions of the Smoluchowski equation.  We assume spherical grains with an intrinsic density of 3~g~cm$^{-3}$, and we assume that the Epstein drag force is valid for all our calculations.  This requires that $s < (9/4) \lambda$, where $\lambda$ is the gas mean free path, and that the relative velocity between the gas and the dust is much less than the mean thermal velocity of the gas, i.e. $|  \mbox{\boldmath{$v$}}_{\rm D} - \mbox{\boldmath{$v$}}_{\rm G} | \ll v_{\rm th}$ \citep{Weidenschilling1977}.  Both conditions are easily satisfied throughout all of the calculations considered in this paper because, as will be seen below, the dust grains do not grow beyond a few tenths of a millimetre during the calculations even in the densest regions of gas and the dust-gas relative velocities remain low.

To follow the calculations significantly beyond stellar core formation, we use a sink particle \citep{BatBonPri1995} with an accretion radius, $r_{\rm acc} =0.5$ au.  The sink particle is inserted when the maximum density reaches $\rho = 10^{-5}$~g~cm$^{-3}$ which occurs during the second collapse phase, when the first hydrostatic core collapses from the inside due to the dissociation of molecular hydrogen.  After the sink particle is formed, any SPH particles (which model both gas and dust) that pass inside the accretion radius of the sink particle are accreted by the sink particle, if their specific angular momentum is smaller than that required for a circular orbit of radius $r_{\rm acc}$ around the sink particle.  The only force acting between sink and SPH particles is gravity (without any softening).  No other boundary conditions are applied.

\begin{figure*}
\centering \vspace{-0.2cm} 
    \includegraphics[height=22cm]{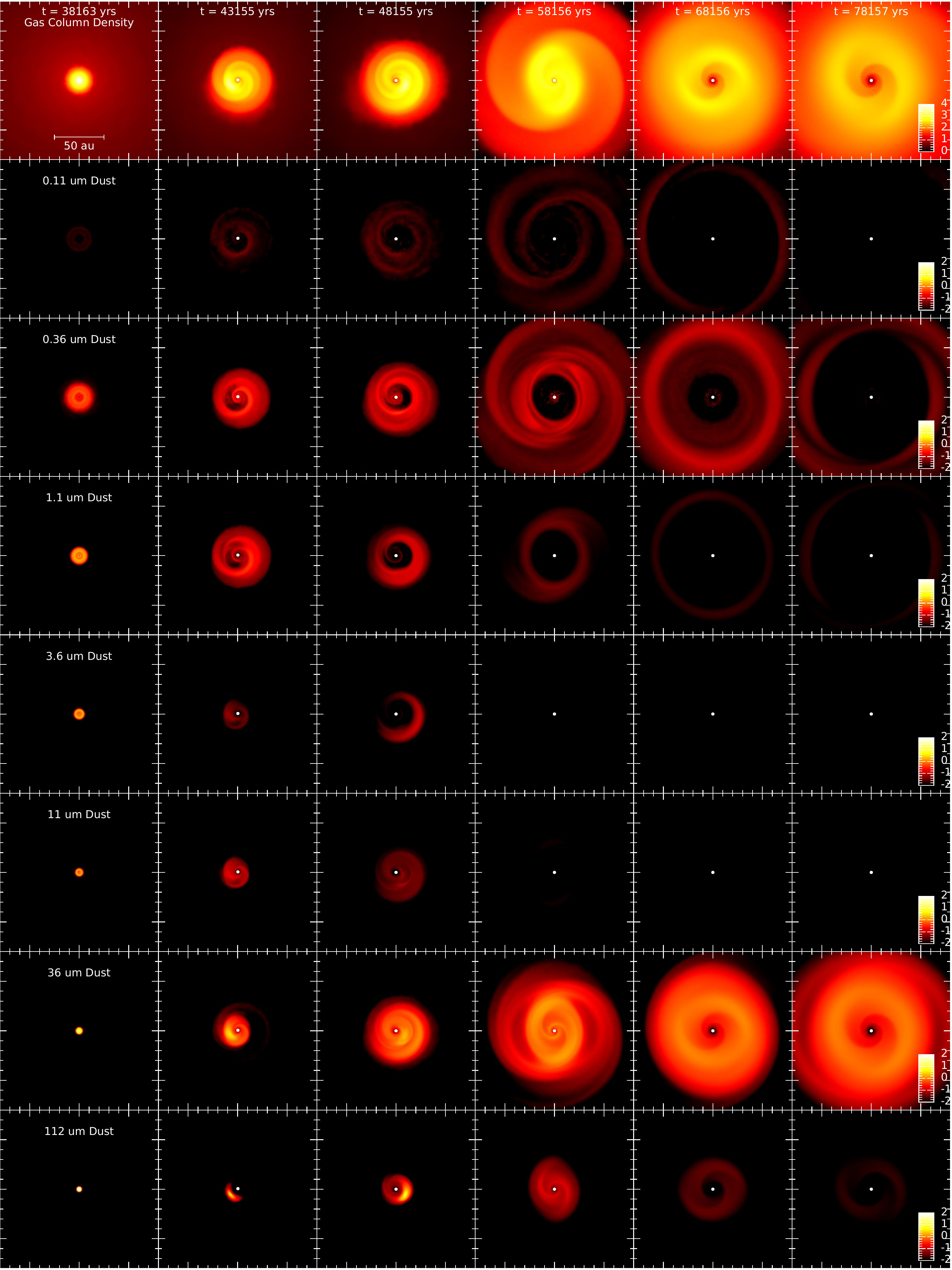}
\vspace{-0.2cm}
\caption{The protostellar disc produced from the collapse of a rotating Bonnor-Ebert molecular cloud core, using all three dust evolutionary processes (i.e., dust coagulation, dust-gas drag, and turbulent diffusion) and assuming $\alpha_{\rm SS}=0.001$.  Each panel shows the logarithm of the column density (in g~cm$^{-2}$), viewed down the rotation axis, of either the gas (top row) or the dust contained in a particular dust bin (the other rows, with dust sizes 0.11, 0.36, 1.1, 3.6, 11, 36 or 112 $\mu$m from the second row down to the bottom row).  Each column shows the state of calculation at a different time.  The first column shows the calculation just before the sink particle is inserted (i.e., just before stellar core formation).  Subsequent columns show the state 5000, 10,000, 20,000, 30,000 and 40,000 years later, with the sink particle plotted as a white dot. Note how the dust mass in various dust size bins varies greatly with the evolutionary phase. }
\label{fig:SF3}
\end{figure*}

\subsection{Results}
\label{sec:SFresults}

\subsubsection{Star formation with $\alpha_{\rm SS}=0.001$}

In Fig.~\ref{fig:SF3} we show snapshots of the calculation that includes all three dust processes (dust growth, dust-gas drag, and turbulent diffusion), with $\alpha_{\rm SS}=0.001$.  Each panel shows the column density of either the gas (top row) or the dust contained in a particular dust bin (other rows), viewed looking down the rotation axis.  Each column shows a different time.  The first column shows the state of the calculation just before the sink particle is inserted (i.e., just before stellar core formation).  Subsequent rows show the state of the disc at intervals of 5000 or 10,000 years, with the last column showing the state of the disc 40,000 years after stellar core formation.  At this latter time, the mass of the sink particle (protostar) is 0.47 M$_\odot$ (a little less than half the mass of the original cloud core) and the disc is almost as massive, at $\approx $0.40 M$_\odot$.

\cite{Bate2022} conducted a thorough study of the dust growth during the formation and evolution of the first hydrostatic core, for a variety of initial conditions.  The reader particularly interested in this phase is referred to that paper \citep[see also][]{Marchand_etal2023}.  With this particular set of initial conditions the first hydrostatic core is a rotating, oblate disc with a radius of $\approx 15$ au, just before stellar core formation.  The first column of Fig.~\ref{fig:SF3} clearly illustrates some of the main findings of \cite{Bate2022}, namely that the typical dust grain size increases rapidly with decreasing radius within the first hydrostatic core, and that large dust grains (in excess of 100 $\mu$m) are produced deep inside the first hydrostatic core (radii less than a few au), before the stellar core is formed. These dust populations are mono-disperse --- at a given location within the first core, most of the dust mass is composed of grains with a narrow range of sizes.  This can even been seen in the dust column density plots in Fig.~\ref{fig:SF3}.  For example, near the centre of the first hydrostatic core, most of the dust column density is contained in dust bins with sizes $36-112$ $\mu$m, and the column density plots for dust bin sizes 0.36--11 $\mu$m have central `holes'.  Note that very little dust mass is found in dust bins with sizes smaller than 0.25 $\mu$m (e.g., the 0.11 $\mu$m panels in the second row).  The smallest dust grains are very rapidly depleted by coagulation.  Note also that these fundamental results for the dust grain distributions of the first hydrostatic core are not substantially affected by the inclusion of dust-gas drag or dust turbulent diffusion.  The former is expected --- \cite{Bate2022} showed that the dust should not migrate significantly relative to the gas on the evolutionary timescales of the first hydrostatic core because it remains well coupled to the gas and the timescales are short.  \cite{Bate2022} did not consider the possible effects of turbulent dust diffusion, but from the results in the first column of Fig.~\ref{fig:SF3} the evolutionary timescales of the first hydrostatic core are so short that there is no substantial impact of turbulent dust diffusion on the radial grain segregation within the first hydrostatic core.

By 5000 yrs after stellar core formation (the second column in Fig.~\ref{fig:SF3}), the oblate first hydrostatic core has become a fully-developed disc with a radius of $\approx 30$ au (second panel of the top row). It is marginally gravitationally unstable, with the gas column density displaying prominent spiral arms.  Spiral structure persists in the disc for as long as this calculation was evolved, and the gravitational torques exerted on the disc material transfer angular momentum outward and mass onto the sink particle \citep{LauBod1994}.  By the end of the calculation the disc has grown to approximately 150 au in radius.

Note how different the images look for different grain sizes.  The dust column density renderings all use the same surface density scales, which are 100 times lower than the gas column density rendering. Thus, the relative intensities of the images show the dust sizes that contain most of the dust mass.  As mentioned above, there is almost no dust mass in small grains ($<0.25~\mu$m).  Significant dust with sizes $\approx 0.36~\mu$m exists, but as the calculation evolves most of this small dust is found in the outskirts of the disc (e.g., radii $>50$ au by 30,000 yrs after stellar core formation). Much of the gas and dust in the outskirts of the disc has recently arrived from infall of the envelope, and with the densities typically being lower in the outskirts than at smaller disc radii not much grain growth has occurred yet.

Dust with intermediate sizes of $\approx 1-15~\mu$m (i.e., significantly larger than the MRN cutoff of 0.25 $\mu$m, but smaller than $\approx 20~\mu$m) exists in significant quantities until around 10,000 years after stellar core formation, but it is essentially absent after from 20,000 years onward (columns 4--6 in Fig.~\ref{fig:SF3}).  The dust in the bulk of the disc grows rapidly and is mono-disperse with typical grain sizes of $\approx 30-45~\mu$m (columns 6 and 7), extending out to a radius of $\approx 60$ au by the end of the calculation.  Some dust grows to sizes $\approx 110~\mu$m in the very densest parts of the disc (disc radii $\lsim 20$ au, bottom row, second, third and fourth panels).  
These large dust grains form in the densest regions of the spiral arms, producing an asymmetric distribution, reminiscent of the trends found by \cite{Bate2022} in their rapidly rotating pre-stellar discs. However, these large grains disperse radially and few persist by the end of the calculation. Note that the disc surface density within the inner $\approx 10-20$ au is affected by the sink particle towards the end of the calculation. Better sink particle boundaries \citep[e.g.,][]{BatBonPri1995} are required to study the evolution of the inner disc.

\begin{figure*}
\centering \vspace{-0.2cm} 
    \includegraphics[height=22cm]{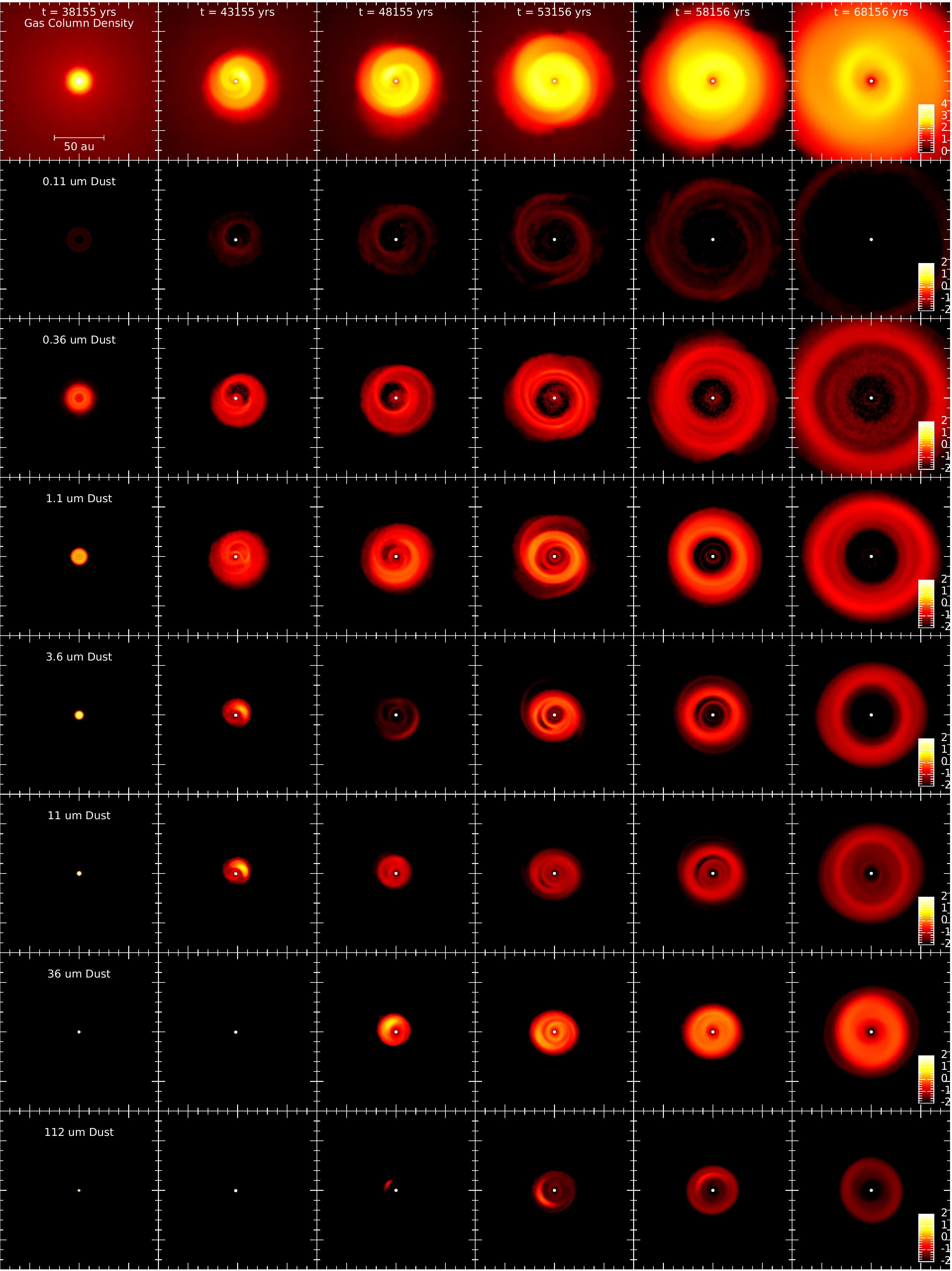}
\vspace{-0.2cm}
\caption{The equivalent of Fig.~\ref{fig:SF3}, except taking $\alpha_{\rm SS}=10^{-4}$ for the dust coagulation and dust turbulent diffusion (note that the times differ for columns 4--6).  All three dust evolutionary processes are included (i.e., dust coagulation, dust-gas drag, and turbulent diffusion).  Each panel shows the logarithm of the column density (in g~cm$^{-2}$), viewed down the rotation axis, of either the gas (top row) or the dust contained in a particular dust bin (the other rows, with dust sizes 0.11, 0.36, 1.1, 3.6, 11, 36, or 112 $\mu$m from the second row down to the bottom row).  Each column shows the state of calculation at a different time.  The first column shows the calculation just before the sink particle is inserted (i.e., just before stellar core formation).  Subsequent columns show the states 5000, 10,000, 15,000, 20,000 and 30,000 years later, with the sink particle plotted as a white dot (note the different times for columns 4-6 compared to Fig.~\ref{fig:SF3}). Note how the dust is less mono-disperse with the lower value of $\alpha_{\rm SS}$, and the radial extent of the large, 36 $\mu$m, grains is much smaller than in Fig.~\ref{fig:SF3}. }
\label{fig:SF4}
\end{figure*}

\subsubsection{Star formation with $\alpha_{\rm SS}=10^{-4}$}

In Fig.~\ref{fig:SF4} we show the results from an identical calculation, but in which we set $\alpha_{\rm SS}=10^{-4}$ for the computation of the relative grain velocities and the dust diffusion due to turbulence.  This calculation was followed to 30,000 years after stellar core formation, at which time, the mass of the sink particle (protostar) is 0.43 M$_\odot$ and the mass of the disc is $\approx $0.36 M$_\odot$.  Note that columns $4-6$ are at different times than columns $4-6$ in Fig.~\ref{fig:SF3} (column 4 is 15,000 years after stellar core formation, while columns 5 and 6 are at the same times as columns 4 and 5 in Fig.~\ref{fig:SF3}, respectively).

Comparing Figs.~\ref{fig:SF3} and \ref{fig:SF4} we notice several important differences.  First, in the first column (just before stellar core formation), the 0.36 $\mu$m and 1.1 $\mu$m renderings are similar, but with the lower turbulence the radial extents of the regions dominated by $11-112~\mu$m grains are smaller in Fig.~\ref{fig:SF4} than in Fig.~\ref{fig:SF3}.  With the lower relative grain velocities, higher densities (which are found at smaller radii within the first hydrostatic core) are required to produce large grains in the same amount of time.

Second, at later times, the dust in the disc with $\alpha_{\rm SS}=10^{-4}$ shows less of a tendency to become mono-disperse.  For example, 15,000 yrs after stellar core formation (Fig.~\ref{fig:SF4}, column 4), significant dust mass is contained in each of the 1.1, 3.6, 11, and 36 $\mu$m dust bins (rows 4--7 in Fig.~\ref{fig:SF4}) at disc radii $\lsim 20$ au, whereas even at 10,000 years after stellar core formation in Fig.~\ref{fig:SF3} there is little dust in bins with sizes 3.6--11 $\mu$m and most of the dust mass is in grains with sizes $\approx 20-110~\mu$m.  Nevertheless, the dust is slowly becoming mono-disperse even in the $\alpha_{\rm SS}=10^{-4}$ calculation, since at disc radii $\lsim 25$ au the dust column densities for grain sizes $\lsim 11$~$\mu$m dust bin are much lower than for the 36~$\mu$m dust bin at 30,000 yrs after stellar core formation (the rightmost column).

Third, the growth of large grains takes noticeably longer.  The smallest dust grains ($\lsim 0.11~\mu$m) are rapidly and strongly depleted in both calculations, because Brownian motion provides significant relative grain velocities for the smallest grains.  But with the lower value of $\alpha_{\rm SS}$, larger grains coagulate more slowly.  For example, 5000 yrs after stellar core formation there is essentially no dust in the disc with grain sizes $\gsim 30~\mu$m (column 2, rows 7--8 of Fig.~\ref{fig:SF4}), whereas with the higher level of turbulence most of the dust at disc radii $\lsim 15$ au already had sizes $\approx 36~\mu$m and there was even some dust with sizes $\approx 112~\mu$m.  Similarly, at later times, the radius in the disc out to which 36~$\mu$m dust has become abundant is smaller.  For example, 20,000 yrs after stellar core formation the the 36~$\mu$m dust extends out to only $\approx 25$ au in the disc with $\alpha_{\rm SS}=10^{-4}$ (column 4, row 7 in Fig.~\ref{fig:SF4}), whereas in the calculation with $\alpha_{\rm SS}=0.001$ the 36~$\mu$m dust extends out to $\approx 60$ au (column 4, row 7 in Fig.~\ref{fig:SF3}).

Thus, as expected, a lower level of hydrodynamic turbulence leads to the slower growth of large dust grains, due to the smaller relative grain velocities.  Note, however, that even this lower level of turbulent dust diffusion is sufficient to produce relatively smooth distributions of dust column density for each of the individual dust size bins, compared to the calculations in the next section that were performed without turbulent dust diffusion.

\begin{figure*}
\centering \vspace{-0.2cm} 
    \includegraphics[height=21.5cm]{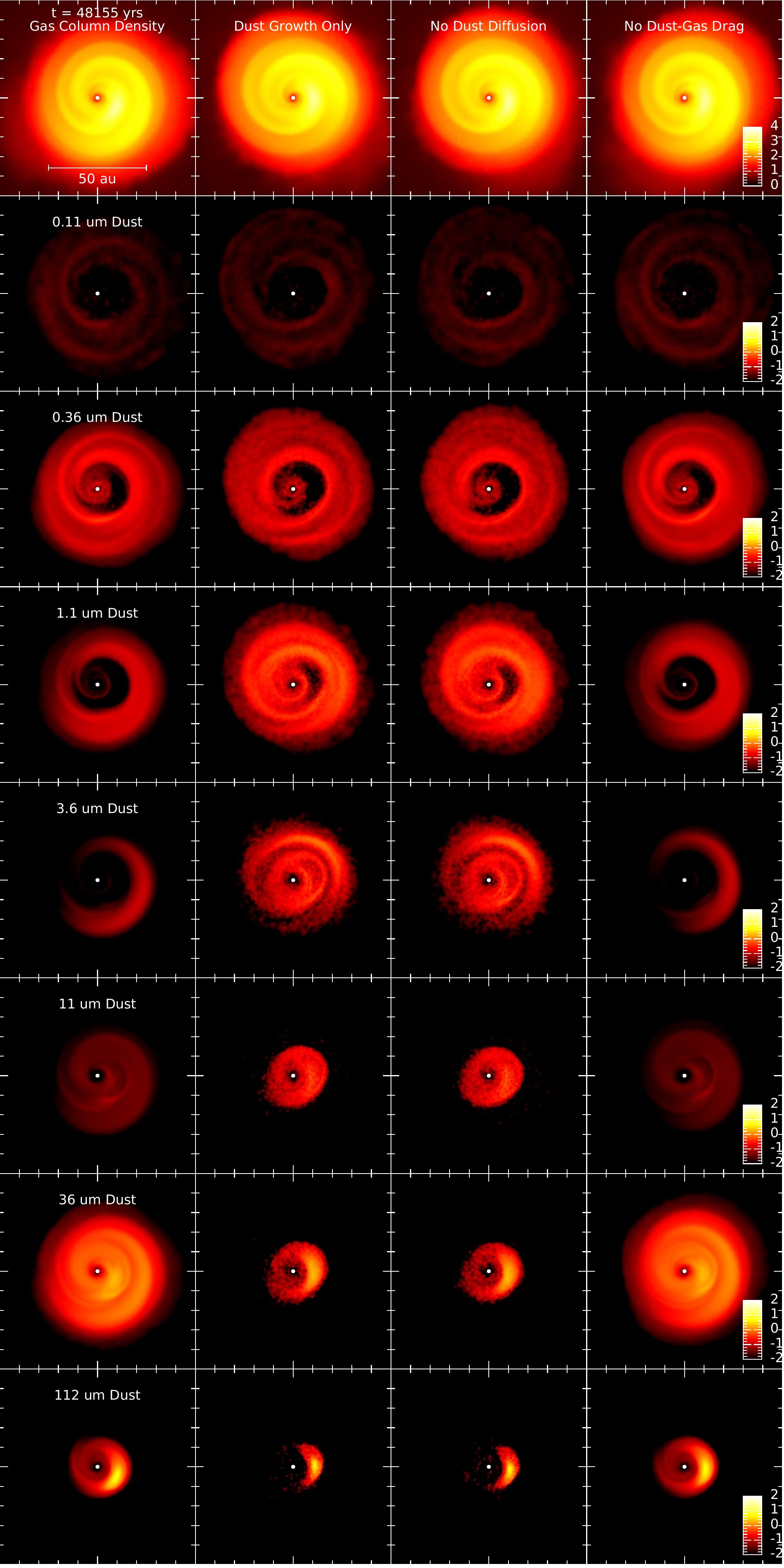}
\vspace{-0.2cm}
\caption{Comparison of four separate calculations that have different mixtures of dust evolutionary processes included.  All panels show the protostellar discs 10,000 years after stellar core formation.  All four calculations assume $\alpha_{\rm SS}=0.001$.  The first column includes all three processes: dust coagulation, dust-gas drag, and turbulent diffusion (it is the same as the second column of Fig.~\ref{fig:SF3}).  The second column only includes dust coagulation.  The third column includes dust coagulation and dust-gas drag, but no turbulent diffusion. The rightmost column includes dust coagulation and turbulent diffusion, but no dust-gas drag.  Each panel shows the column density, viewed down the rotation axis, of either the gas (top row) or the dust contained in a particular dust bin (the other rows, with dust sizes 0.11, 0.36, 1.1, 3.6, 11, 36, or 112 $\mu$m from the second row down to the bottom row).  Note that the leftmost and rightmost columns are similar, and the two middle columns are similar. This shows that the dust-gas drag has little effect, but that the inclusion of dust turbulent diffusion leads to a much greater rate of large dust grain production than when it is excluded. }
\label{fig:SFroles}
\end{figure*}

\subsubsection{The roles of turbulent diffusion and gas-dust drag}
\label{sec:roles}

In Fig.~\ref{fig:SFroles} we show the results from four calculations all at the same time, 10,000 years after stellar core formation (i.e., after the insertion of the sink particle).  All the calculations use the same method for dust coagulation, with $\alpha_{\rm SS}=0.001$.  The first column shows the same results as the second column of Fig.~\ref{fig:SF3}.  The second column shows the results from an identical calculation, except that the dust-gas drag and dust turbulent diffusion were turned off (i.e., as in \citealt{Bate2022}).  The third column shows the results from a calculation that included the dust-gas drag, but not the dust turbulent diffusion, while the fourth column shows with results obtained from a calculation that includes dust turbulent diffusion, but not dust-gas drag.

By comparing each of the columns it is clear that the dust-gas drag has little effect on the distributions of dust in these calculations.  Most of the dust is still too small to undergo significant radial or azimuthal migration relative to the gas (or relative to other dust grains) on these spatial scales due to the effects of dust-gas drag, at least on these timescales.

By contrast, the inclusion of turbulent dust diffusion has an enormous effect, both on the distributions of the dust grains, and also on the dust coagulation itself.  Recall that in the calculation which only includes dust coagulation (i.e., using only the method of \citealt{Bate2022}), the dust population within each fluid element evolves independently of all the other fluid elements.  There is no mixing of the dust between fluid elements.  Adding the effects of dust turbulent diffusion clearly allows substantial mixing of the dust between fluid elements (particularly with $\alpha_{\rm SS}=0.001$, but also with $\alpha_{\rm SS}=10^{-4}$ by inspection of Fig.~\ref{fig:SF4}).  The somewhat `speckled' distributions of dust in the second and third columns of Fig.~\ref{fig:SFroles} are symptomatic of the complete absence of dust mixing between fluid elements in the second column, and the severe lack of mixing in the third column.  Adding the 
turbulent dust diffusion produces much smoother (less `speckled') dust distributions.

However, by comparing the results with and without turbulent dust diffusion it is also clear that the diffusion doesn't just smooth out the dust distributions --- it also accelerates the dust growth.  Although the distributions of small dust ($\lsim 0.36 \mu$m) are similar in all columns, in the leftmost and rightmost columns of Fig.~\ref{fig:SFroles} there is much more dust in the 36 and 112 $\mu$m size bins than in centre two columns.  Moreover, even at this early time, the calculations depicted in the leftmost and rightmost columns are clearly producing mono-disperse dust distributions in the inner 15 au of the discs (there is a lack of dust in the bins with sizes 3.6 and 11 $\mu$m), whereas the dust is distributed more evenly distributed across the bins from 3.6--36~$\mu$m in the centre two columns.  We attribute this accelerated growth of large dust grains to the continual mixing of fresh small and intermediate-sized grains into the regions that have already managed to produce large grains, thus allowing them to grow further (something that cannot occur if the dust population of each fluid element is evolved separately, as in \citealt{Bate2022} and the second column of Fig.~\ref{fig:SFroles}).

The importance of the dust turbulent diffusion for both the distribution of dust grains and the grain growth itself means that the results of future calculations will depend on the choice of the diffusion model.  For example, \cite{FroNel2009} studied the turbulent diffusion of dust particles in MHD turbulent discs.  They found that the vertical profile of the dust was quite different to that obtained from simple turbulent diffusion models such as that used here.  In particular, they found that the abundances of small dust particles (a few microns in size) in the surfaces of discs tended to be greater with MHD turbulence than that predicted using a constant vertical diffusion coefficient because MHD turbulence is not homogeneous and tends to be stronger in the surface layers.  They provide an alternative dust diffusion model based on their MHD results.  In the future it will be important to study how different turbulence models affect the dust grain growth rates and distributions.

\section{Conclusions}
\label{sec:conclusions}

We have extended an existing three-dimensional smoothed particle hydrodynamics code, {\sc sphNG}, that has been previously used to model aspects of star and planet formation, to enable it to model the evolution of populations of dust grains that are subject to coagulation, dust-gas dynamical drag, and turbulent dust diffusion.  To accomplish this, we combined the dust coagulation method of \cite{HirYan2009}, \cite{HirOmu2009} and \cite{Bate2022}, with the {\sc multigrain} method of \cite{HutPriLai2018}, which models the dust-gas drag of multiple grain sizes, including the back-reaction on the gas.  In addition, we have developed a new implicit method to model dust stirring and mixing by unresolved hydrodynamic turbulence (e.g., in protoplanetary discs).  Furthermore, to solve the dust-gas drag equations, we have adapted the implicit method developed by \cite{ElsBat2024} to deal with distributions of dust grain sizes, as opposed to a single dust species.

We have presented the results of various test calculations of dust settling, dust growth, and turbulent dust diffusion in a vertical column of a protoplanetary disc, by extending a test case originally proposed by \cite{PriLai2015} and used by \cite{HutPriLai2018} to test the {\sc multigrain} method.

Using this code we have presented, for the first time, calculations that follow the growth, mixing and dynamical evolution of dust grain distributions during the formation of a star and its protostellar disc.  The three-dimensional, radiation hydrodynamical calculations follow the collapse of gravitationally-unstable molecular cloud cores, through the first hydrostatic core phase, and beyond stellar core formation, until most of the molecular cloud core has fallen into the protostar (modelled using a sink particle) or its protoplanetary disc. 

Although this paper is primarily a method paper, we make several important findings from our dusty star formation calculations, albeit using a single set of initial conditions:

\begin{enumerate}
\item The inclusion of dust-gas dynamical drag and turbulent dust diffusion does not substantially alter the main findings of \cite{Bate2022}.  In particular, rapid dust growth occurs within the first hydrostatic core, before the formation of a stellar core, and it produces dust grain populations that are locally mono-disperse (peaked around a single grain size) with a typical grain size that increases with decreasing radius within the first hydrostatic core (i.e., a radially-segregated dust grain population).  Dust grains with sizes $>100~\mu$m can form deep within the first hydrostatic core.
\item Small dust grains (sizes $\lsim 0.25~\mu$m) are strongly depleted in the disc surrounding the young star. With reasonably high levels of disc turbulence (i.e., $\alpha_{\rm SS}=0.001$), we find that dust grains may quickly grow to mono-disperse populations with typical grain sizes of $\approx 30-100~\mu$m throughout much of the disc (e.g., disc radii of up to 50 au only 20,000 years after stellar core formation).  However, the growth rates depend strongly on the level of disc turbulence, with $\alpha_{\rm SS}=10^{-4}$ giving significantly slower dust growth rates and it takes longer to develop a mono-disperse grain size distribution. 
\item In general, the radial and azimuthal distributions of dust grains with different sizes within the protostellar discs are extremely variable, both spatially and temporally.
\item As opposed to the dust evolution during the first hydrostatic core phase, turbulent dust diffusion within the disc provides a substantial enhancement of the rate of dust grain coagulation compared to calculations that do not include turbulent dust diffusion.  We attribute this to the fact that turbulent dust diffusion provides a continual source of small and intermediate dust grains to the high-density regions (e.g., near the disc mid-plane) in which the largest dust grains are growing.  However, it should be noted that, since both dust distribution and growth are governed by turbulent diffusion, the results inevitably depend on the adopted turbulence model.  This sensitivity should be explored in future calculations. Furthermore, we consider dust coagulation but we do not consider dust fragmentation.  Our chosen grain coagulation threshold means that the largest grains in our star formation calculations remain relatively small ($a_{\max} \lsim 100$ micron), quite well coupled to the gas, and the grain distributions evolve towards monodisperse populations in the long-term.  Employing different coagulation thresholds may produce
larger grains, that have higher relative velocities, thereby leading to grain fragmentation that may repopulate the small grain population.  Again, such effects should be explored in future calculations.
\end{enumerate}

Finally, there are several obvious extensions of this work in the future, although they are beyond the scope of this paper.  First, since these calculations provide full grain size distributions for each fluid element, this information could be used to compute local dust opacities which could be used when computing the radiation transport (currently the calculations still assume standard interstellar opacities).  Second, similarly, the local grain size distributions could be used when determining the local ionisation rate of the disc for calculations that include magnetic fields \citep[e.g.,][]{Guillet_etal2020, Marchand_etal2023, ValLebHen2024}.

\section*{Acknowledgments}

The authors thank the anonymous referee for several suggestions that helped to improve this paper. MRB thanks Sebastiaan Krijt for several useful conversations. MAH acknowledges support from the DFG program {\it Closing the Loop--Using Synthetic Observations of Simulated Star-forming Regions to Test Observational Properties} (Project No. 426714422). He also gratefully acknowledges earlier support from the Swiss National Science Foundation within the framework of the National Centre for Competence in Research PlanetS. DE acknowledges the support of a Science and Technology Facilities Council (STFC) studentship (ST/V506679/1). 

Some of the graphs and all of the renderings of the calculations were produced using SPLASH \citep{Price2007}, an SPH visualization tool publicly available at http://users.monash.edu.au/$\sim$dprice/splash.

The calculations performed for this paper used the University of Exeter Supercomputer, Isca, and the Science and Technology Facilities Council (STFC) DiRAC High Performance Computing (HPC) facilities (www.dirac.ac.uk).  This work used the DiRAC Memory Intensive service Cosma8 at Durham University, managed by the Institute for Computational Cosmology on behalf of the STFC DiRAC HPC Facility. The DiRAC service at Durham was funded by BEIS, UKRI and STFC capital funding, Durham University and STFC operations grants. This work also used the DiRAC Data Intensive service (DIaL-3) at the University of Leicester, managed by the University of Leicester Research Computing Service on behalf of the STFC DiRAC HPC Facility (www.dirac.ac.uk). The DiRAC service at Leicester was funded by BEIS, UKRI and STFC capital funding and STFC operations grants. DiRAC is part of the UKRI Digital Research Infrastructure. 

For the purpose of open access, the author has applied a Creative Commons Attribution (CC BY) licence to any Author Accepted Manuscript version arising.

\section*{Data Availability}

The data from the calculations discussed in this paper will be made available by the corresponding author on receipt of a reasonable request.

\bibliography{/Users/mbate/Tex/mbate}

\appendix

\section{Testing the grain growth scheme}
\label{appendixA}

The growth of a population of dust grains through binary collisions can be described
by the Smoluchowski equation \citep{Smoluchowski1916}.  Several explicit analytic solutions
exist that can used to test numerical grain growth prescriptions.  In this Appendix
we demonstrate the performance of our chosen grain growth method 
(Section \ref{sec:coagulation}), which was originally developed by \cite{HirYan2009}
and subsequently applied to star formation by \cite{HirOmu2009} and modified
by \cite{Bate2022}.

\cite{LomLai2021} discuss the historical study of the Smoluchowski equation and they 
provide a clear description of the analytic solutions, the main elements of which are
reproduced here for the convenience of the reader.  The continuous 
Smoluchowski equation can be written
\begin{equation}
\begin{split}
\frac{\partial n(m,t)}{\partial t} = & \frac{1}{2} \int_0^m K(m-m^\prime, m^\prime) n(m-m^\prime, t) n(m^\prime, t) {\rm d}m^\prime    \\
 & \displaystyle  - n(m, t)  \int_0^\infty K (m, m^\prime) n(m^\prime, t) {\rm d}m^\prime,
\end{split}
\label{eq:Smoluchowski}
\end{equation}
where $m$ and $m^\prime$ are the masses of two colliding particles and the coagulation kernel $K(m, m^\prime)$ is a symmetric function of $m$ and $m^\prime$ for binary collisions that is related to the probability of collision.  For the kernel, for physical collisions between particles this is often written
\begin{equation}
K(m,m^\prime) =  \beta(m,m^\prime,\Delta v) \Delta v(m, m^\prime) \sigma(m,m^\prime)
\label{eq:kernel}
\end{equation}
where $\Delta v$ is the mean relative velocity between two particles, $\sigma$ is the mean effective collision cross-section, and $\beta$ denotes the mean sticking probability of the grains.

Our discretised form of equation \ref{eq:Smoluchowski} is equation \ref{eq:coag}, which is written in terms of the mass density of grains, $\tilde{\rho}$, rather than the number density, $n$.  Our equivalent of $K(m, m^\prime)$ is $\alpha_{kj}$, given by equation \ref{eq:alpha}.

The Smoluchowski equation can be written in dimensionless form \cite{Scott1968, Drake1972, LomLai2021} by defining
\begin{equation}
\begin{split}
& x \equiv m/m_0, y \equiv m^\prime/m_0, {\cal K}(x,y) = K(m,m^\prime)/K_0, \\
& \tau = (K_0 N_0)t, f(x,t) = m_0 n(m,t)/N_0,
\end{split}
\end{equation}
where $N_0$ is the initial total number density of particles of mean mass, $m_0$, and $K_0$ is a normalising constant with dimensions [length]$^3$/time.  Equation \ref{eq:Smoluchowski} can then be written as
\begin{equation}
\begin{split}
\frac{\partial f(x,y)}{\partial \tau} = & \frac{1}{2} \int_0^x {\cal K}(y, x-y) f(y, \tau) f(x-y, \tau) {\rm d}y    \\
 & \displaystyle  - f(x, \tau)  \int_0^\infty {\cal K} (y, x) f(y, \tau) {\rm d}y,
\end{split}
\label{eq:Smoluchowski_nondim}
\end{equation}

Explicit analytic solutions to this equation exist for three types of coagulation kernel: constant, additive, and multiplicative.  The constant kernel solution can be represented by ${\cal K}(x,y) = 1$, the additive kernel as ${\cal K}(x,y) = x+y$, and the multiplicative by ${\cal K}(x,y) = xy$.  For physical collisions of spherical particles, the ballistic kernel can be written ${\cal K}(x,y) = \pi(x^{1/3}+y^{1/3})\Delta v$ where $\Delta v$ is the mean relative velocity between particles. However, no known analytic solution exists for the ballistic kernel.  The constant and additive kernels are the most similar to the ballistic kernel; the multiplicative kernel represents an explosive increase of the collisional frequencies with grain size which is not the type of growth expected for small dust grains within a protoplanetary disc.  Therefore, in this Appendix we only consider the constant and additive kernels.

For the constant kernel ${\cal K}(x,y) = 1$ an analytic solution of equation \ref{eq:Smoluchowski_nondim} can be derived \citep{Muller1928, Schumann1940, Melzak1957, Rajagopal1959, Scott1968, SilTak1979} for the initial condition $f(x,0) = \exp(-x)$ such that
\begin{equation}
f(x,\tau) = \frac{4}{(2+\tau)^2} \exp \left[ -  \frac{2 x}{(2+\tau)} \right].
\label{eq:constantsolution}
\end{equation}
Physically, this kernel implies that the frequency of collisions between two particles is independent of their mass (or size).

For the additive kernel ${\cal K}(x,y) = x+y$ an analytic solution of equation \ref{eq:Smoluchowski_nondim} can be derived \citep{Golovin1963, Scott1968} for the initial condition $f(x,0) = \exp(-x)$ such that
\begin{equation}
\begin{split}
& T \equiv 1- \exp(-\tau), \\
& f(x,\tau) = \frac{(1-T \exp(-x[1+T])}{xT^{1/2}} I_1(2x T^{1/2})
\end{split}
\label{eq:additivesolution}
\end{equation}
where $I_1$ is the upper modified Bessel function of first kind.
Physically, this kernel implies that the frequency of collisions increases as the mass (or size) of the grains increases.

The dimensionless grain mass density is given by $g(x, \tau) = x f(x,\tau)$.  This is the quantity that is plotted in \cite{LomLai2021} and below.

\subsection{The initial conditions for the SPH code}

In what follows, we compare our grain growth algorithm with the above analytic solutions by making the minimum number of changes to the SPH code.  The code uses dimensionless units defined by a unit of length and a unit of mass, which we set to $L=10^{16}$ cm and $M=1$ M$_\odot$, respectively, and in which the gravitational constant $G=1$.  These are typical values for star formation calculations.  We set up a cubic box with size length, $L$, containing a total mass of 1 M$_\odot$, represented by 8000 SPH particles on a uniform Cartesian lattice (i.e., $20\times 20\times 20$ particles), and we employ periodic boundary conditions.  This gives a mean mass density of unity in code units, or $1.991\times 10^{-15}$~g~cm$^{-3}$ (this is the total density, including both gas and dust).  The code unit of time is $t_0 = (L^3/(G M))^{1/2} = 8.6763 \times 10^{10}$ seconds.  We set the temperature to be very cold (so that the particles do not move significantly during the test, because we only wish to test the grain growth algorithm).

Our grain growth algorithm is built around grain mass densities, $\tilde{\rho}$, and grain sizes, $a$, rather than grain number densities, $n$, and grain masses, $m$.  Thus, some care is required in comparing the above analytic solutions with the results provided by the SPH code.  Specifically, the code works with dust fractions, $\epsilon_i$, in multiple grain size bins, rather than dust number densities in multiple dust mass bins (i.e., $\epsilon_i(a)$, rather than $n(m)$).  Moreover, the dust fractions are in logarithmically-spaced bins (i.e., d$\log{a}$ rather than d$m$).  The dust fractions per bin decrease as the number of bins is increased, so when plotting the evolution of the dust with time we need to use $\epsilon_i/\log \delta$ where $\delta = \log(a_{\rm max} /a_{\rm min})/N$ (see Section \ref{sec:coagulation}).  Furthermore, because $m \propto a^3$ (for spherical grains), we need to plot $f(x,\tau)$ versus $a^3$ rather than $x$.

The SPH code specifies grain sizes in units of microns (i.e. $10^{-4}$ cm).  Therefore, we define the mass, $m_0$ to be the mass of a grain with size $a_0=1$ micron (equation \ref{eq:mass_grain}) such that $m_0 = 4 \pi \times 10^{-12}$~g~cm$^{-3}$ (using $\rho_{\rm gr} = 3.0$~g~cm$^{-3}$).  This gives $N_0$ to be $\rho_d/m_0 \approx 4.56\times 10^{-6}$~cm$^{-3}$.  For the kernel, equation \ref{eq:kernel} for $K(m,m^\prime)$ must be compared with equation \ref{eq:alpha}, for $\alpha_{kj}$, the latter of which includes the same quantities in the numerator, but also the product $m_j m_k$ in the denominator due to the use of grain mass densities rather than number densities.  Thus, for example, choosing the constant kernel ${\cal K}(x,y)$, corresponds to setting $\alpha_{kj} \propto 1/(m_j m_k)$. 

The equivalent of the setting the dimensionless initial grain mass distribution to $g(x,0) = x \exp(-x)$ is to set
the initial dust mass density in bin $i$ to be 
\begin{equation}
\epsilon_i \propto (m/m_0)^2 \exp( - m/m_0), 
\label{eq:testdustfrac}
\end{equation}
in which the first term is squared due to the fact that the SPH code considers the dust mass fractions within logarithmically-spaced bins (i.e., $m$~d$\log m$ = d$m$).  These dust fractions are then normalised so that $\epsilon = \sum_i \epsilon_i$ is the required the total dust fraction (i.e., the total dust-to-gas ratio is $\epsilon/(1-\epsilon)$).

Finally, in order to plot the equivalent of $g(x,0) = x \exp(-x)$ versus $x$ (and the equivalent of $g(x,\tau)$ at time $\tau$), we plot ${\cal N} \epsilon_i/\log(\delta) \times a^3$ versus $a^3$, since $x=m/m_0=a^3/a_0^3$ and $a_0=1$ (in units of microns).  The factor, ${\cal N} = 1/(3 \epsilon)$, is a normalisation factor such that the graph of ${\cal N} \epsilon_i/\log(\delta) \times a_0^3$ versus $a_0^3$ has a peak of $1/e$ at $a_0^3=1$ (since $g(x,0) = x \exp(-x)$ has its peak value of $1/e$ at $x=1$).  For the calculations below, we choose to set the initial total dust-to-gas ratio to be 1/20, which gives ${\cal N} = 7$.  This also gives $N_0 = \epsilon M/L^3 / m_0 \approx 7.545 \times 10^{-6}$ cm$^{-3}$.

We use the same dust size range as for the star formation simulations (Section \ref{sec:initialconditions}), i.e. $a_{\rm min}=5$~nm and $a_{\rm max}=1$~mm.  We perform simulations with either 26, 53, 106, or 212 logarithmically-spaced bins (corresponding to 5, 10, 20, or 40 bins per decade in grain size, respectively).  This is also equivalent to approximately 1.7, 3.3, 6.7, or 13.3 bins per decade in grain mass (since $m \propto a^3$).

Using the above initial conditions, the only changes we need to make to the SPH code to run the idealised test case rather than a physically realistic calculation (e.g., Section \ref{sec:results}), are to set the initial grain mass distribution using equation \ref{eq:testdustfrac} and to change the equation for $\alpha_{kj}$ (see below).  Being able to reproduce the analytic solutions numerically with only these minimal changes give us confidence that the grain growth algorithm in the SPH code is valid.

\begin{figure}
\centering \vspace{-0.25cm} \vspace{0cm}
    \includegraphics[width=8.8cm]{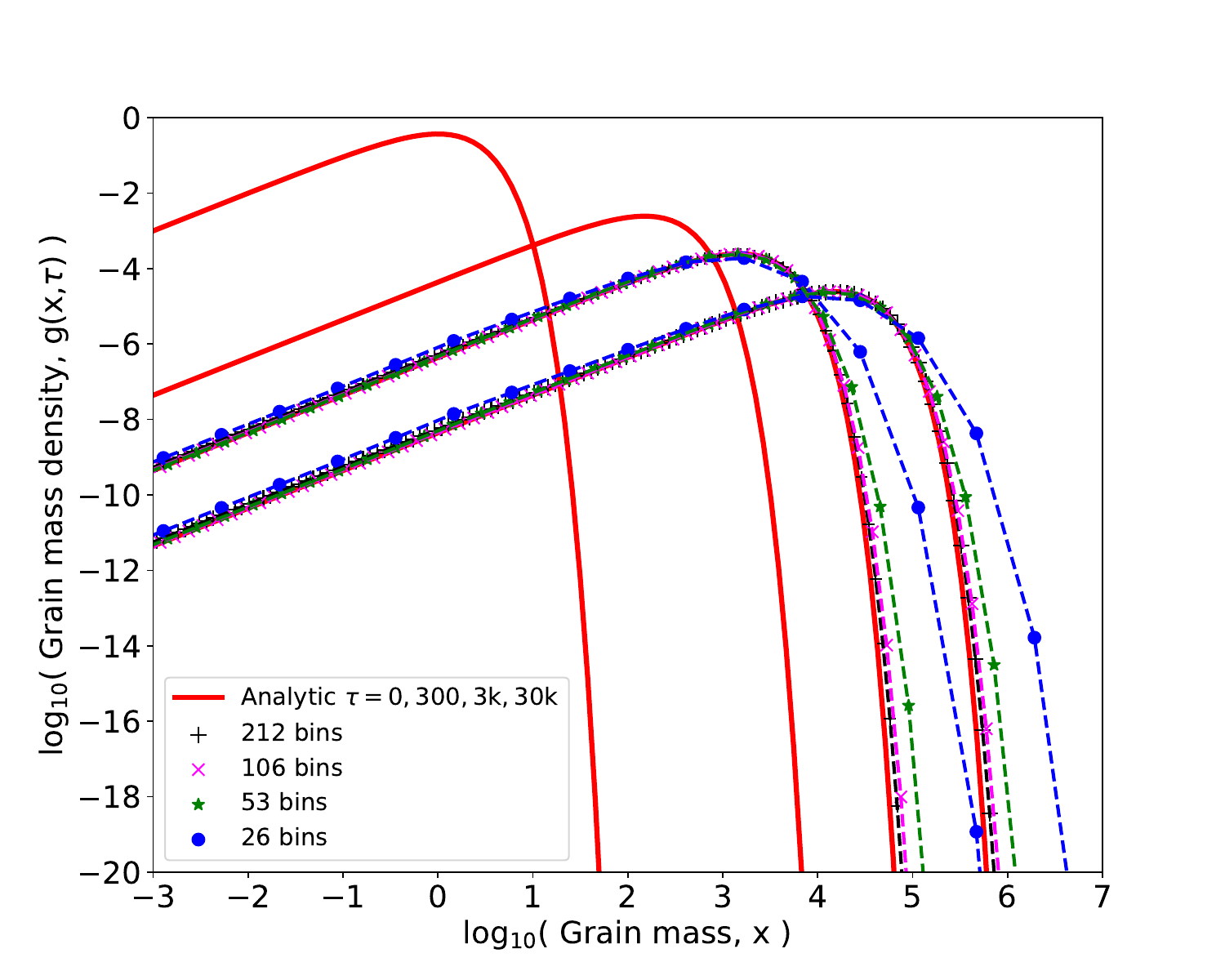}         \vspace{-0.5cm}
\caption{Comparison of the numerical grain growth method with the analytic solution when using a constant coagulation kernel.  The red solid lines show the analytic solutions for the dimensionless grain mass densities, $g(x,\tau)$, at dimensionless times $\tau = 0, 300, 3\,000, 3\times 10^4$.  We show the analytic initial conditions and the distributions at three other times to make it clear how the grain mass density distribution evolves with time.  The various symbols, connected by dashed lines, show the dimensionless grain mass densities in the dust bins obtained from the SPH calculations, at dimensionless times $\tau = 3\,000$ and $3\times 10^4$, obtained using approximately 5, 10, 20, and 40 bins per decade in grain size. We show the numerical distributions at two times to show how the numerical distributions evolve.  As is typical for numerical grain growth schemes, the formation of large grains tends to be overestimated at low resolution, but the results converge towards the analytic result as the number of dust bins is increased. }
\label{A1}
\end{figure}

\subsection{Numerical results with the constant kernel}

\cite{LomLai2021} performed calculations that evolved the initial grain distribution under the action of a constant kernel to dimensionless time $\tau = 3 \times 10^4$, for which the analytic solution is given by equation \ref{eq:constantsolution}.  We also evolve our test calculations to this time.
To run SPH calculations using a constant kernel, we set $\alpha_{kj} = A/(m_j m_k)$ with $A=10^{-4}$.  The time measured in code units is $t_{\rm code} = \tau m_0 / (A N_0 t_0) \approx 0.015277 \times  \tau$. For example, for $\tau=3 \times 10^4$ the corresponding time in code units is $t_{\rm code} \approx 458.3$.

The results from four calculations with different numbers of dust bins are shown in Fig.~\ref{A1}.  The red solid lines show the analytic solutions for the dimensionless grain mass densities, $g(x,\tau)$, at dimensionless times $\tau = 0, 300, 3\,000, 3\times 10^4$.  The symbols show the dimensionless grain mass densities in the dust bins obtained from the SPH calculations, at dimensionless times $\tau = 3\,000$ and $3\times 10^4$, obtained using 26, 53, 106, or 212 bins that are logarithmically-spaced in grain size between 5 nm and 1000 mm.  

As expected, using a small number of bins overestimates the high-mass tail of the grain mass density distribution.  This is the well-known over-diffusion problem in which numerical schemes tend to overestimate the formation of large grains at low resolution. However, our numerical solutions for this test case are of a similar accuracy to the methods of \cite{LomLai2021} and \cite*{LuiGroWam2019} for a similar number of bins per decade in grain mass.

Using as few as 5 bins per decade in grain size also over-estimates the low-mass tail and underestimates the peak.  Although the total dust mass is conserved, this is the lowest possible number of bins before the scheme completely breaks down (see Section \ref{sec:coagulation}), and should never be used in practice.  Using 10 or more bins per decade in grain size provides reasonably accurate solutions and the numerical solutions converge towards the analytic solution as expected as the number of bins is increased (Fig.~\ref{A1}).  Nevertheless, for production calculations it would be wise to test that any conclusions are robust to changes in the number of bins.

\begin{figure}
\centering \vspace{-0.25cm} \vspace{0cm}
    \includegraphics[width=8.8cm]{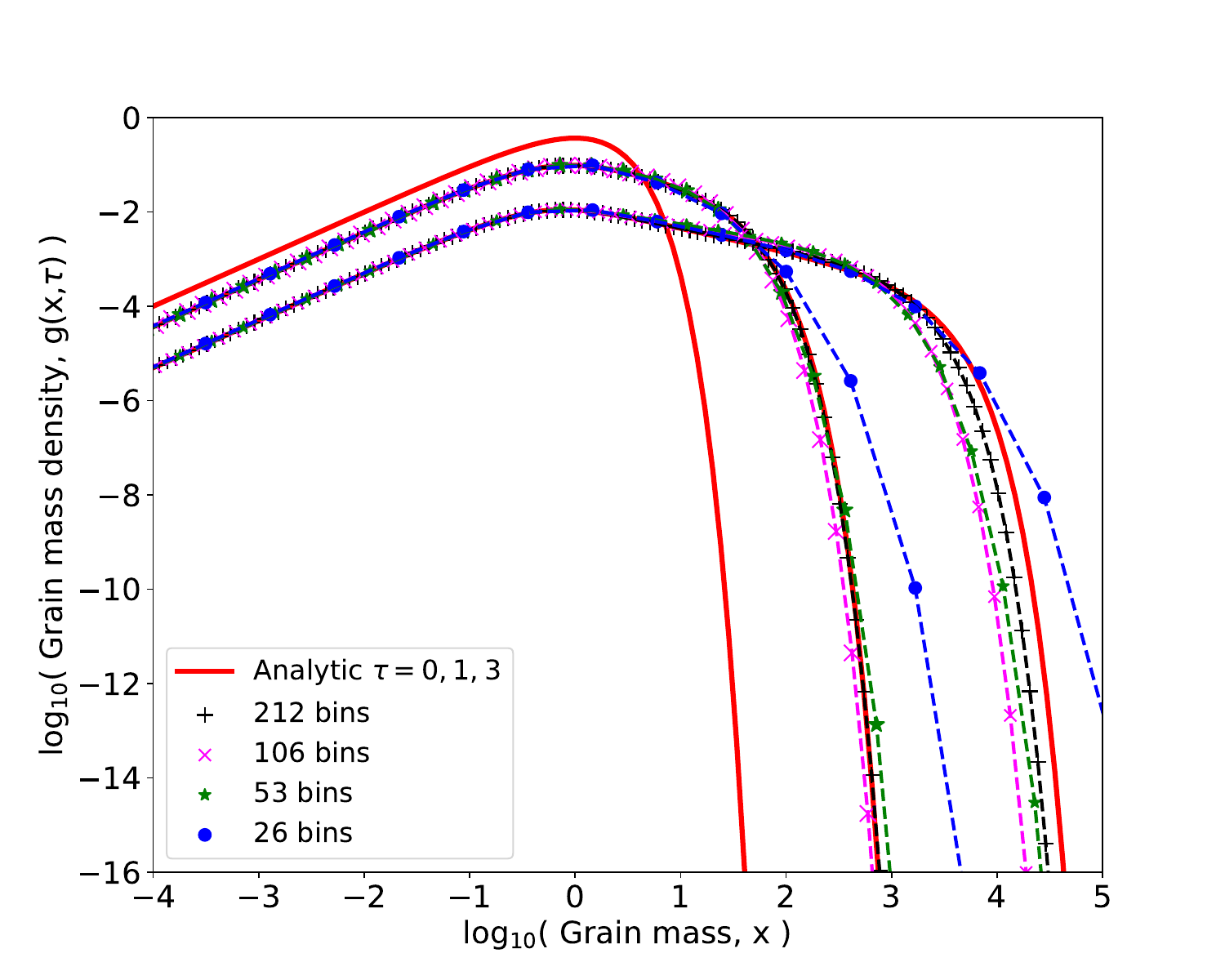}        
    \includegraphics[width=8.8cm]{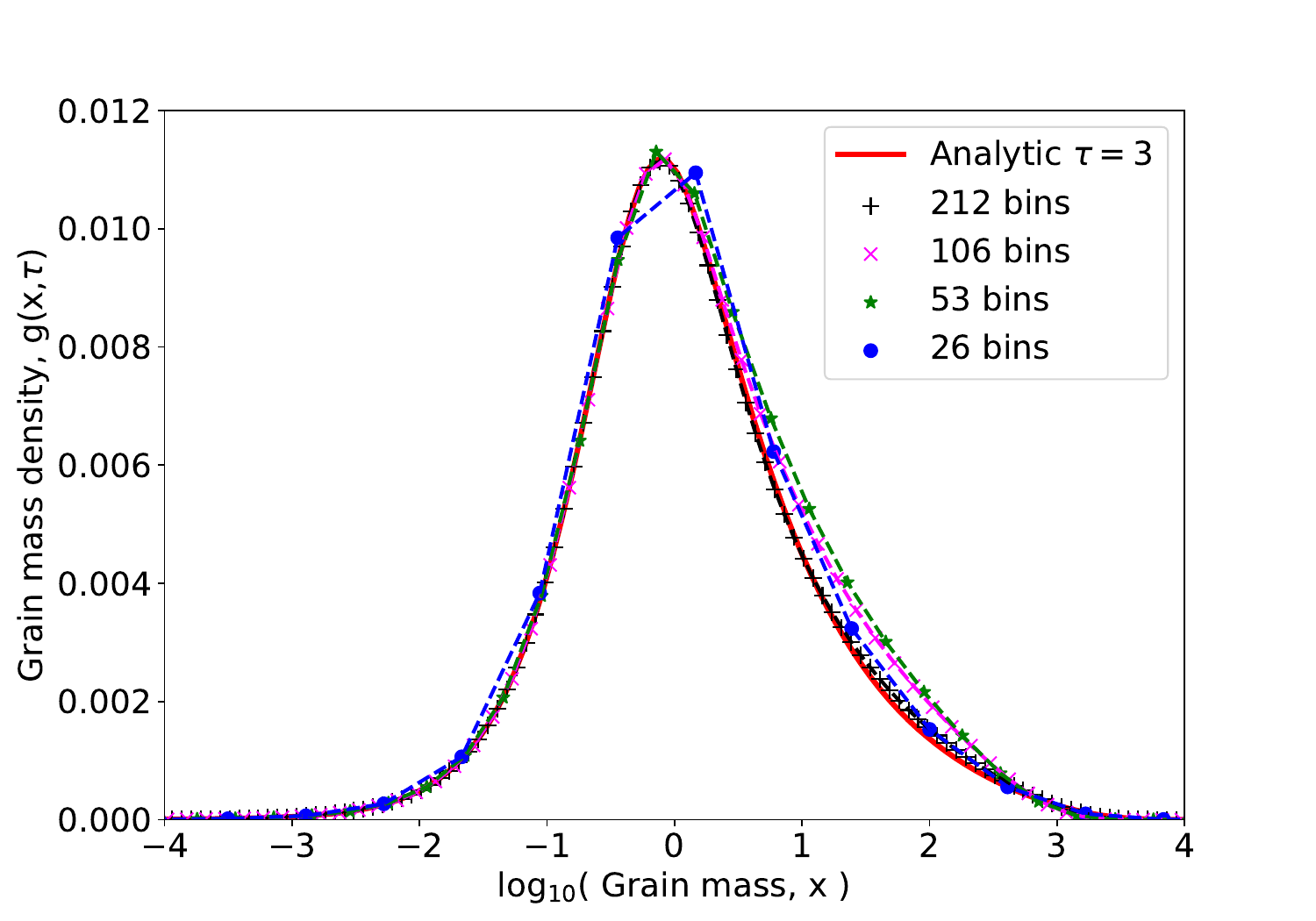}         \vspace{-0.5cm}
\caption{Comparison of the numerical grain growth method with the analytic solution when using an additive coagulation kernel.  In the top panel, the red solid lines show the analytic solutions for the dimensionless grain mass densities, $g(x,\tau)$, at dimensionless times $\tau = 0, 1, 3$.  The various symbols, connected by dashed lines, show the dimensionless grain mass densities in the dust bins obtained from the SPH calculations, at dimensionless times $\tau = 1$ and $\tau = 3$, obtained using approximately 5, 10, 20, and 40 bins per decade in grain size. In the lower panel, we plot the analytic solution and numerical results at $\tau=3$ using a linear vertical scale so that the grain mass densities for medium-mass grains ($\log_{10}(x)=[0,3]$ at $\tau=3$) can be more easily seen.  In contrast to the test case using the constant kernel, for resolutions $\geq 10$ bins per decade in grain size the formation of large grains is slightly {\it underestimated} (and the densities of medium-mass grains are slightly overestimated), but the numerical solutions again converge towards the analytic solution for large numbers of bins.}
\label{A2}
\end{figure}

\subsection{Numerical results with the additive kernel}

\cite{LomLai2021} performed calculations that evolved the initial grain distribution under the action of the additive kernel to dimensionless time $\tau = 3$, for which the analytic solution is given by equation \ref{eq:additivesolution}.  We also evolve our test calculations to this time.
To run SPH calculations using the additive kernel, we set $\alpha_{kj} = B((m_j) + (m_k))/(m_j m_k)$ in the SPH code with $B=100$.  In this case, the time measured in code units is $t_{\rm code} = \tau / (B N_0 t_0) \approx 1215.66 \times  \tau$ (note the missing factor of $m_0$ compared to the previous section; an alternative would be to divide the masses in the numerator of the equation for $\alpha_{kj}$ by $m_0$, but this would require an additional change to the SPH code and we have deliberately kept the changes to a minimum). For example, for $\tau=3$ the corresponding time in code units is $t_{\rm code} \approx 3647.0$.

The results from four calculations using the additive kernel with different numbers of dust bins are shown in Fig.~\ref{A2}.  In the top panel, the red solid lines show the analytic solutions for the dimensionless grain mass densities, $g(x,\tau)$, at dimensionless times $\tau = 0, 1, 3$.  The symbols show the dimensionless grain mass densities in the dust bins obtained from the SPH calculations, at dimensionless times $\tau = 1$ and $\tau = 3$, obtained using 26, 53, 106, or 212 bins that are logarithmically-spaced in grain size between 5 nm and 1000 mm.   In the lower panel, we use a linear vertical scale and only plot the results at $\tau=3$.

The behaviour for this test case is interesting.  Whereas using the constant kernel overestimates the formation of large grains, when using the additive kernel this only happens at the lowest resolution (26 bins, or 5 bins per decade in grain size). All of the higher resolution calculations slightly {\it underestimate} the grain mass densities of large grains.  The underestimate is strongest between 10 and 20 bins per decade in grain size (Fig.~\ref{A2}).  Increasing the resolution further, for example from 20 to 40 bins per grain size, the numerical results converge toward the analytic solution.  The total dust mass is still conserved, so when the large grain mass density is underestimated, the density of medium-mass grains ($\log_{10}(x)=[0,3]$ at $\tau=3$) is slightly overestimated (see the lower panel of Fig.~\ref{A2}).

The behaviour of our chosen numerical method for modelling dust growth when increasing the numerical resolution is quite different to some other methods, such as those discussed in \cite{LomLai2021}.  As with the constant kernel, other methods tend to overestimate the formation of large grains.  This is usually considered to be problematic and a numerical effect that should be minimised, since the purpose of studying grain growth is often specifically to study the formation and behaviour of large grains.  Our chosen method for modelling grain growth avoids this problem of over-producing large grains when using the additive kernel and $\gsim 10$ bins per decade in grain size.

\end{document}